%% file: modeling_high_dimensional_multichannel_brain_signals_revise.tex
\documentclass{article}

\input{definitions}

\usepackage{bm}
\usepackage{xr-hyper}
\usepackage{dsfont}
\usepackage{color}
\usepackage{amsmath}
\usepackage{algorithm}
\usepackage[noend]{algpseudocode}
\usepackage{csquotes}
\makeatletter
\def\BState{\State\hskip-\ALG@thistlm}
\makeatother

\numberwithin{equation}{section}
\DeclareMathOperator{\sgn}{sgn}

\title{Modeling High Dimensional Multichannel Brain Signals}
\date{}
\author{Lechuan Hu$^1$, Norbert J. Fortin$^2$ and Hernando Ombao$^{1,3,4}$}
\footnotetext[1]{Department of Statistics, University of California, Irvine, USA}
\footnotetext[2]{Department of Neurobiology and Behavior, University of California, Irvine, USA}
\footnotetext[3]{Statistics Program, King Abdullah University of Science and Technology (KAUST), Saudi Arabia}
\footnotetext[4]{Corresponding author: hernando.ombao@kaust.edu.sa}

\begin{document}
\maketitle

\abstract{
Our goal is to model and measure functional and effective (directional) connectivity in multichannel brain physiological signals (e.g., electroencephalograms, local field potentials). The difficulties from analyzing these data mainly come from two aspects: first, there are major statistical and computational challenges for modeling and analyzing high dimensional multichannel brain signals; second, there is no set of universally-agreed measures for characterizing connectivity. To model multichannel brain signals, our approach is to fit a vector autoregressive (VAR) model with potentially high lag order so that complex lead-lag temporal dynamics between the channels can be captured. Estimates of the VAR model will be obtained by our proposed hybrid LASSLE (LASSO+LSE) method which combines regularization (to control for sparsity) and least squares estimation (to improve bias and mean-squared error). Then we employ some measures of connectivity but put an emphasis on partial directed coherence (PDC) which can capture the directional connectivity between channels. PDC is a frequency-specific measure that explains the extent to which the present oscillatory activity in a sender channel influences the future oscillatory activity in a specific receiver channel relative to all possible receivers in the network. The proposed modeling approach provided key insights into potential functional relationships among simultaneously recorded sites during performance of a complex memory task. Specifically, this novel method was successful in quantifying patterns of effective connectivity across electrode locations, and in capturing how these patterns varied across trial epochs and trial types.
}

\vspace{0.1in}


\baselineskip=24pt

\newpage

\input{Introduction_revise}

\input{Method_revise}
\input{Simulation_revise}
\input{Application_revise}
\input{Conclusion}

\newpage
\input{Acknowledgement}
\bibliographystyle{plainnat}
\bibliography{reference}

\end{document}

%% file: definitions.tex
\usepackage{setspace}
\usepackage{multirow}
\usepackage{subfigure}
\usepackage{amsmath,amssymb,amsfonts,mathrsfs}
\usepackage{natbib}
\usepackage{graphicx, epstopdf}
\usepackage{fullpage}
\usepackage[top=1in, bottom=1in, left=1in, right=1in]{geometry}

\usepackage{graphicx}
\usepackage{graphics}
\usepackage{multimedia}
\usepackage[english]{babel}
\usepackage[latin1]{inputenc}
\usepackage{times}
\usepackage[T1]{fontenc}
\usepackage{pgf,pgfarrows,pgfnodes,pgfautomata,pgfheaps,pgfshade}
\usepackage[latin1]{inputenc}
\usepackage{colortbl}
\usepackage{color}
\usepackage{xcolor}
\usepackage[english]{babel}
\usepackage{bm}
\usepackage{bbm}
\usepackage{multimedia}
\usepackage{subfig}
\usepackage{multicol}
\usepackage{pdfpages}
\usepackage{enumerate}
 \usepackage{epstopdf}

\renewcommand{\Pr}{{\mathbb P}}

\xdefinecolor{DarkBlue}{rgb}{0,0,0.65}


%% file: Introduction_revise.tex
\section{Introduction}

Connectivity between populations of neurons is crucial to fully characterize brain processes during cognition (e.g., memory and learning) and even during resting-state. Moreover, alterations in brain connectivity is widely believed to be implicated in a number of neurological and mental diseases such as obsessive compulsive disorder and Alzheimer's disease. However, the underlying mechanisms of brain connectivity remain elusive. First, there is no set of universally-agreed measures for characterizing connectivity. Second, there are major statistical and computational challenges for modeling and analyzing multichannel brain signals -- especially when the number of parameters is large which often happens when the number of channels is large and/or the temporal lag for parametric models such as vector autoregressive (VAR) is high. Our contribution in this paper is a scalable approach to estimate connectivity in multichannel brain physiological signals modeled with high dimensional parameters.

The work is motivated by our current collaborations with the Fortin Laboratory (UC Irvine) whose research requires developing a systematic statistical framework to quantify functional and effective connectivity among multi-site neural activity signals recorded in rats performing complex memory tasks. The electrophysiological data recorded from rats include local field potentials (LFPs) and an example of a recording for one epoch (here an epoch is 1 second time block) is given in Figure~\ref{fig:LFP}. LFP signals have excellent temporal resolution (here $1000$ observations per second). It is comparable to electroencephalograms (EEGs) in terms of temporal resolution and both capture electrical activity of the neurons. However, LFPs are recorded invasively since these are obtained from electrodes that are chronically implanted inside the brain. Because LFPs are obtained from implanted electrodes, they have lower contamination compared to scalp EEGs. They contain less non-neuronal physiological activity (e.g., muscular activity) and therefore possess a higher signal-to-physiological-noise ratio. One disadvantage of LFPs, however, is its limited utility in humans due to its invasive nature. However, these will continue to be a valuable tool for investigating brain function in animals which can then provide useful information for modeling brain function in humans. One of the challenges to fitting statistical models to LFPs is that the parameter space can be high dimensional. The number of recording tetrodes ($P$) in LFPs can range from 8-100; and the temporal order ($d$) of parametric models such as vector autoregressive (VAR) models needs to be sufficiently large in order to accurately capture the dynamics in these complex processes. In this setting the number of parameters in a VAR model is $P^2d$, which can be large.

\begin{figure}[h!]\centering
\includegraphics[width=0.6\textwidth,height=0.5\textwidth]{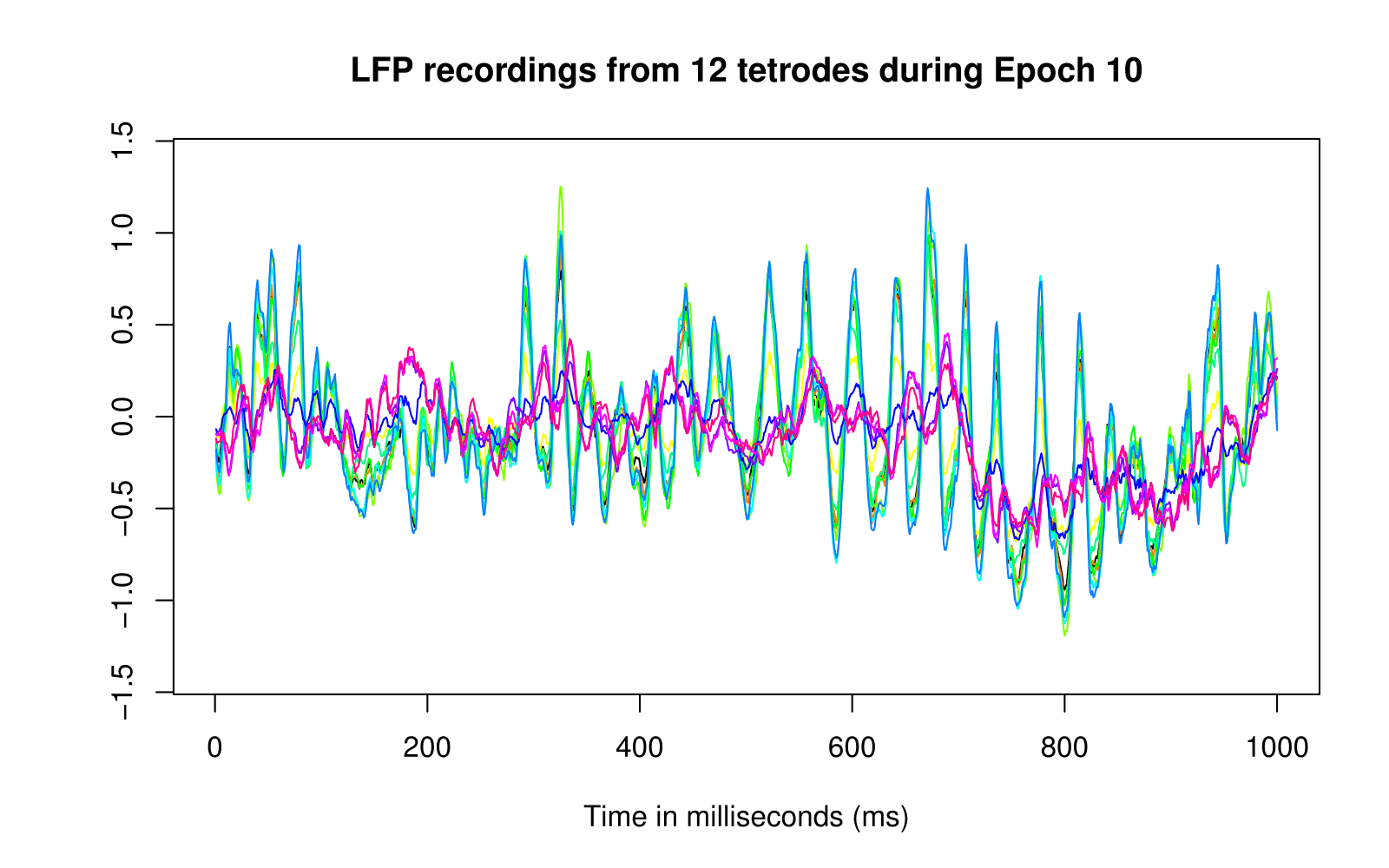} \\
\caption{Local field potential (LFP) recordings from 12 tetrodes during one epoch ($1000$ milliseconds; $T = 1000$). Each time series with color indicates the LFP recording from one tetrode.}
\end{figure}\label{fig:LFP}

In this paper we will develop a computationally scalable method for fitting high dimensional complex models that addresses two important goals in brain science: (1.) To identify the connectivity structure between channels in a brain network and (2.) To quantify both the strength and directionality of connectivity between these channels. Our approach is to fit a VAR model with potentially high temporal lag in order to more accurately capture complex lead-lag temporal dynamics between the channels or leads. Estimates of the VAR model will be obtained by a combination of regularization to maintain high specificity and least squares estimation to reduce bias and mean-squared error. The method will be applied to LFPs obtained from a rat performing an odor sequence memory task, in which he is required to identify each odor as being presented in the correct or incorrect sequence position.

To characterize connectivity in a multichannel LFP signal we shall use the vector autoregressive (VAR) model (\cite{shumway2006time}). A $P$-dimensional brain signal
${\mathbf X}_t$ is said to follow a VAR model of order $d$, denoted VAR($d$), if it has the representation
\begin{equation} \label{var}
{\mathbf X}_t \ = \ {\Phi}_1{\mathbf X}_{t-1}+ \ldots +  {\Phi}_d{\mathbf X}_{t-d} + \varepsilon_t \quad t=d+1,...,T
\end{equation}
where ${\Phi}_\ell$'s $\in {\mathfrak R}^{P \times P}$ are the autoregressive coefficient matrices and $\varepsilon_t \overset{iid}{\sim}N_P(\overrightarrow{0}, \Sigma)$. The interconnectivity between channels is determined by the autoregressive coefficient matrices $\{\Phi_\ell\}_{\ell=1}^d$ and spatial covariance matrix $\Sigma$. Thus, the VAR model provides a broad framework for capturing complex temporal and cross-sectional interrelationship among the time series (in particular, directionality of frequency-specific connectivity). Consequently it can be applied to model the Granger-causal relation between channels (\cite{kaminski2001evaluating}).

\begin{figure}[h!]\centering
	\begin{tabular}{cc}
		\includegraphics[width=2.5in,height=2in]{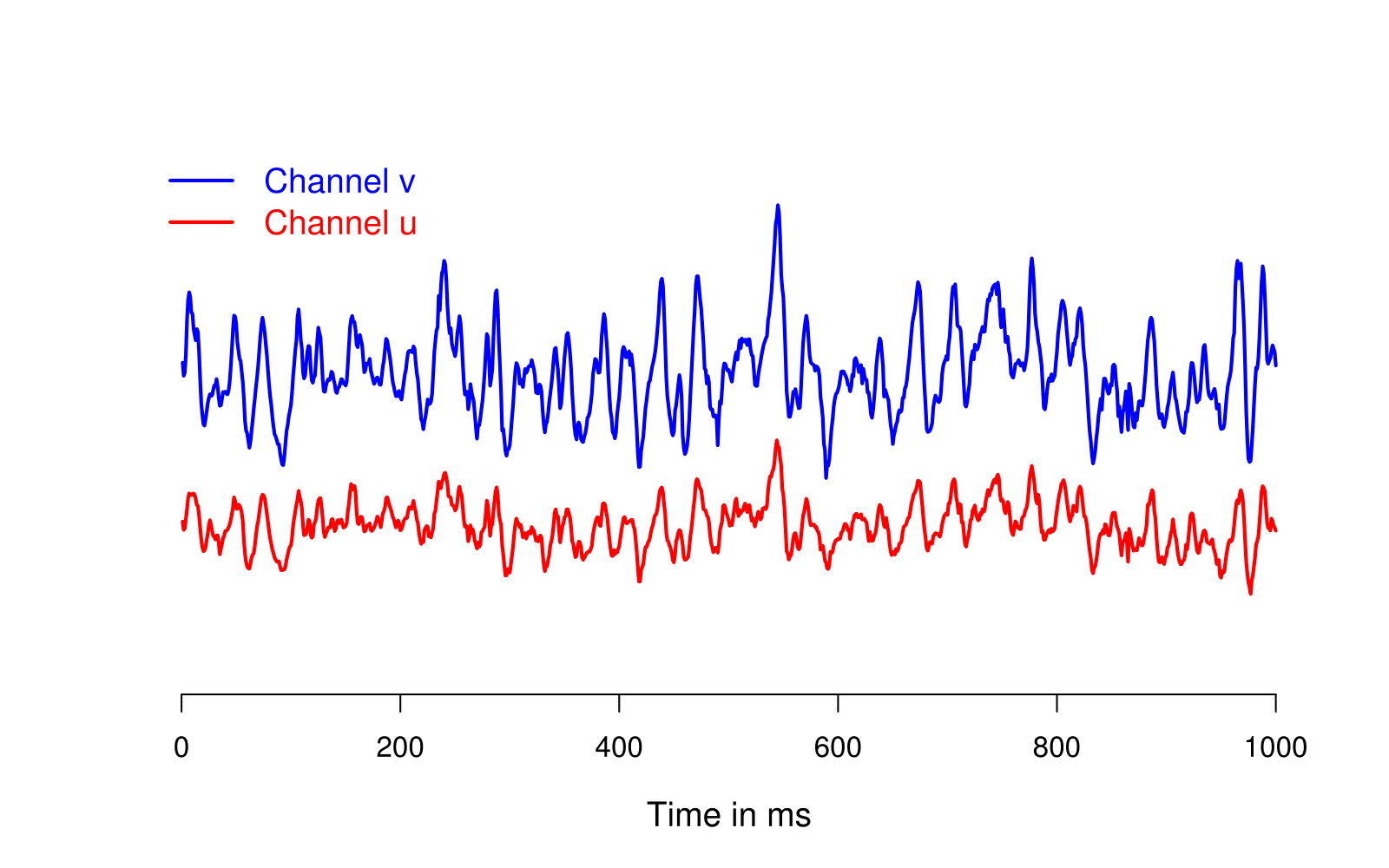} &
		\includegraphics[width=2.25in,height=1.75in]{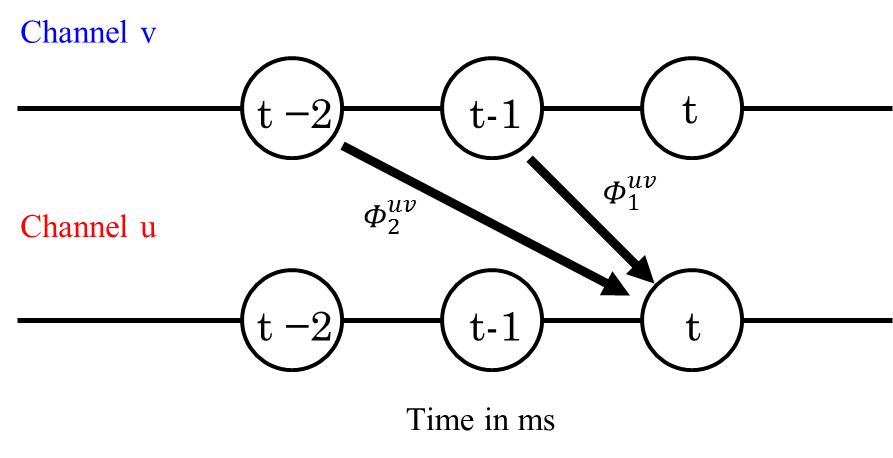}\\
		(a) LFP traces&(b) Explicit lagged cross-dependence in VAR
	\end{tabular}
	\caption{LFP traces and VAR. $\Phi_{\ell}^{uv}$ ($\ell=1,2$) captures the impact of the input from $v$-th channel at time $t-\ell$ to brain activity at $u$-th channel at the current time $t$.}
	\label{fig:VARExample}
\end{figure}

To illustrate connectivity via the VAR matrix, consider Figure \ref{fig:VARExample} and denote the LFP traces of brain region to be $u$-th and $v$-th channel. Then the entry $\Phi_{\ell}^{uv}$ shows the impact of the input from $v$-th channel at time $t-\ell$ to brain activity at $u$-th channel at the current time $t$. If $\Phi_{\ell}^{uv} = 0$ and $\Phi_{\ell}^{vu} = 0$ for all lags $\ell$ then, there is no connectivity between these two channels as determined by VAR model. A positive value indicates that the signal of $v$-th channel at time $t-\ell$, conditional on LFP values at other times, has positive linear dependence with $u$-th channel at time $t$. That is, a marginal increase in activity in $v$-th channel leads to a increased future activity in $u$-th channel. Thus, the entries of $\{\Phi_\ell\}_{\ell=1}^d$ contain the information of brain connectivity between channels. In this paper, we shall use partial directed coherence (PDC) (\cite{baccala2001partial} and \cite{baccala2014partial}) to characterize effective (directed) connectivity. This measure is more specific and provides more information, in particular frequency-specific directionality,  than simply the coefficients of the VAR matrices. PDC is frequency-specific: it measures how an oscillatory activity (at a particular frequency band) at a present time in one channel may impact oscillatory activity of the same frequency band at another channel at a future time point.

As noted above effective connectivity between channels will be characterized by the VAR coefficient matrices. This is challenging because the parameter space of a VAR model for brain signals is usually high. For example, if we fit a VAR$(10)$ model to $12$ leads or channels, there are $10 \times 12^2 = 1440$ parameters in total to estimate, which subsequently requires intensive computation. One could suggest fitting a model with low temporal lag in order to reduce the number of parameters. The problem with this suggestion, unfortunately, is that a low temporal order might miss potentially important features of the data such as multiple peaks in the
spectra. 

One classic estimation approach is via least squares which (as long as there are sufficient data points) provides unbiased estimator for the elements of the VAR coefficient matrices but at the cost of high demand of computing. The least squares estimate (LSE) does not possess the specificity for coefficients with true value of zero. Hence it cannot provide an adequate answer to the first question about identifying functionally connected regions in brain network. Another common estimation approach is the LASSO (least absolute shrinkage and selection operator) method which is a particular representative of the penalized regression family (\cite{tibshirani1996regression}, \cite{fu1998penalized}, \cite{zhao2009composite}, \cite{hesterberg2008least}). Compared with LSE, the LASSO approach requires smaller computation time (\cite{mairal2012complexity}). Most importantly, LASSO has higher specificity of zero-coefficients. The main limitation of the LASSO (and most regularization methods) is that the estimators of the non-zero coefficients are biased. Thus, it could lead to misleading results when investigating strength of brain effective connectivity. Inspired by the strengths of each of the two classical approaches (i.e., LSE and the LASSO), we propose to combine these in a two-step estimation procedure which we call the LASSLE method. We demonstrate that LASSLE has inherited low bias for non-zero estimates and high specificity for zero-estimates from LSE and the LASSO separately. As a result, the proposed two-step method has higher specificity and significantly lower mean squared error (MSE) in the simulation study. At this stage, the full theoretical justification is being developed but the numerical experiments are encouraging.

A natural question to ask is whether or not the LASSO method is appropriate for fitting VAR models to brain signals. The answer lies in whether or not brain signals such as LFPs and EEGs indeed exhibit sparse connectivity structure. Due to volume conduction for EEGs, it is not likely that the connectivity structure between channels is sparse. However, though LASSO aims to shrink many of the VAR coefficients to zero but this does not necessarily lead to a sparse connectivity structure. Keep in mind that a pair of channels are functionally disconnected only if {\it all} of the its corresponding VAR coefficients at {\it all} lags are estimated to be zero. Thus, imposing sparsity on the VAR coefficient matrices helps to weed out the less important parameters in the VAR model but does not oversimplify the connectivity structure.

The remainder of this paper is arranged as follows. In Section 2, we present the proposed hybrid LASSLE (LASSO+LSE) method followed by finite sample simulation studies in Section 3 and analysis of LFP signals in Section 4 and the Conclusion in Section 5.

%% file: Method_revise.tex
\section{A proposed two-step LASSO+LSE procedure for fitting a VAR model}

First, we note  that the VAR($d$) model can be alternatively written in a form
\begin{equation} \label{altform}
\underbrace{
  \begin{bmatrix}
  ({\mathbf X}_T)' \\
  \vdots \\
  ({\mathbf X}_{d+1})'
  \end{bmatrix}
}_{Y}=
\underbrace{
  \begin{bmatrix}
  ({\mathbf X}_{T-1})'& \cdots &({\mathbf X}_{T-d})'\\
  \vdots & \ddots & \vdots\\
  ({\mathbf X}_d)'& \cdots & ({\mathbf X}_1)'
  \end{bmatrix}
}_{\mathbb X}
\underbrace{
  \begin{bmatrix}
  (\Phi_1)' \\
  \vdots \\
  (\Phi_d)'
  \end{bmatrix}
}_{B}+
\underbrace{
  \begin{bmatrix}
  (\varepsilon_T)' \\
  \vdots \\
  (\varepsilon_{d+1})'
  \end{bmatrix}
}_{E}.
\end{equation}
Next, denote $Y = [y_1, y_2, ... , y_P]$, $B = [b_1, b_2, ... , b_P]$, $E = [e_1, e_2, ... , e_P]$.
Denote the $k^{th}$ column vector of the matrices $Y$, $B$ and $E$ $(k=1,2,...,P)$
to be $y_k$, $b_k$, $e_k$.
Then we have
\begin{equation} \label{sub}
\underbrace{y_k}_{m\times 1}= {\mathbb X}\underbrace{b_k}_{q \times 1}+
\underbrace{e_k}_{m \times 1},\quad e_k \overset{indep}{\sim}N_m(\overrightarrow{0}, \sigma_{kk}I_m)
\end{equation}
where $m = T - d$, $q = P \times d$, and $\sigma_{kk}$ is the $k^{th}$ diagonal element of the covariance matrix
$\Sigma$. Note that Equation~(\ref{altform})
is finally decomposed into many sub-linear regression problems of
estimating $\{b_k\}_{k=1}^P$ in a parallel manner and all the entries of connectivity matrices
are included in $\{b_k\}_{k=1}^P$.

\subsection{Least squares estimation (LSE)}

To fit a linear regression model, the most common approach is via least squares estimation so that the least squares estimator $\widehat{b}_k$ satisfies
\begin{equation}
\widehat{b}_k = \underset{b_k \in {\mathfrak R}^q} {\mathrm{argmin}} {\| y_k - {\mathbb X}b_k\|}^2
\end{equation}
which gives the unbiased estimator $\widehat{b}_k = ({\mathbb X}'{\mathbb X})^{-1} {\mathbb X}'y_k$. Some papers (\cite{han2013direct}) argue that in high dimensional case the number of parameters $q$ can be larger than the number of observations $m$, thus this method has limitations due to the nonsingular matrix ${\mathbb X}'{\mathbb X}$. However, we do not worry about this when analyzing the LFP data since normally we have replicated measurements from multiple epochs. The biggest problem here is that LSE has poor specificity for coefficients with true value of zero. It always produces estimates that are very close to zero rather than exactly zero, which reflects non-connectivity between channels. Indeed when LFP channels are not effectively connected with each other, then an excess non-zero estimate could lead to incorrect characterizations of connectivity through partial directed coherence. Moreover, even a trivial amount of bias for one coefficient, when added across thousands of coefficients, can produce large mean squared error (as demonstrated in the simulation study).

\subsection{LASSO family estimation}

In order to overcome the problem of non-specificity by the LSE method, recent attention has been focused on the family of penalized regression models as viable solutions to this problem. One of the well known methods of this family is LASSO regression (with $L_1$ penalty term). The estimates given by LASSO are the solution to the minimization of the criterion
\begin{equation} \label{lasso}
\widetilde{b}_k = \underset{b_k \in {\mathfrak R}^q} {\mathrm{argmin}}{\| y_k - {\mathbb X}b_k\|}^2 + {\lambda \|b_k\|_1}
\end{equation}
The penalty term will force a lot of excess non-zero estimates to exact zero, which provides good estimate for the sparsity of the VAR coefficient matrices $\{\Phi_\ell\}_{\ell=1}^d$ and could consequently greatly simplify the calculation of connectivity measures (e.g., PDC) by focusing only on the more important coefficients. In the implementation of the algorithms for LASSO, we take advantage of the results demonstrated by \cite{friedman2010regularization} where estimation of generalized linear models with convex penalties can be handled by cyclical coordinate descent and computed along a regularization path. The price of LASSO is that the non-zero estimates are biased of true values which leads to incorrect estimates of the strength of connectivity between channels (PDC). 

\subsection{LASSLE: proposed two-step estimation method}
Motivated by both the advantages and limitation of each of the previous approaches, we propose a two-step procedure to estimate VAR model parameters. Our method consists of these two steps:

\begin{itemize}

\item[\textbf{Step 1.}] Apply LASSO to identify entries in $\{\Phi_\ell\}_{\ell=1}^d$ whose estimates
are not set to 0.
\begin{equation}
\widehat {S}_k = \{j \in \{1,...,q\}: \widehat{b}_k^j \neq 0\}
\end{equation}
\item[\textbf{Step 2.}] Fit LSE with the constraint that \enquote{zero} entries estimates from Step 1 are
fixed to 0
\begin{equation}
\widetilde{b_k}_{LAS} = \underset{b_k : b_k^j = 0, j \in \widehat{S}_k^c} {\mathrm{argmin}}{\| y_k - {\mathbb X}b_k\|}_2^2
\end{equation}

\end{itemize}

\begin{algorithm}
\caption*{LASSLE Algorithm}
\begin{algorithmic}[1]
\Procedure{Two-step Estimation}{}
\BState \emph{Step 1}:
\State {Generate a sequence of $(d,\lambda)$ and randomly divide data to K folds}
\State {For a possible choice of $(d,\lambda)$, leave one fold as test data at each time}
\State {Train \ref{sub} with LASSO method on other folds and compute $\{\widehat{\Phi}_\ell\}_{\ell=1}^d$ for $\{\Phi_\ell\}_{\ell=1}^d$}
\State {Based on $\{\widehat{\Phi}_\ell\}_{\ell=1}^d$, calculate prediction error on test set and finally take average}
\State {Select $(d,\lambda)$ with the lowest average prediction error}
\State {Obtain estimate $\{\widehat{b}_k\}_{k=1}^P$ for $\{b_k\}_{k=1}^P$ in Equation~(\ref{sub}) of lag $d$ using LASSO method with $\lambda$}
\BState \emph{Step 2}:
\If {$\widehat{b}_k^j = 0$}
\State {Set $b_k^j = 0$}.
\EndIf
\If {$\widehat{b}_k^j \neq 0$}
\State {Keep $b_k^j$}.
\EndIf
\State {Obtain estimate $\{\widetilde{b_k}_{LAS}\}_{k=1}^P$ for $\{{b_k}\}_{k=1}^P$ in Equation~(\ref{sub}) with LSE under above constriction}
\State {Obtain estimate $\{\widetilde{\Phi_\ell}_{LAS}\}_{\ell=1}^d$ for $\{\Phi_\ell\}_{\ell=1}^d$ by arranging $\{\widetilde{b_k}_{LAS}\}_{k=1}^P$}
\EndProcedure
\end{algorithmic}
\end{algorithm}

To obtain the optimal tuning parameter $\lambda$, we employ a $K$-fold cross-validation test in Step 1. A sequence of candidates of $\lambda$ will be pre-specified and the optimal value is selected such that the average of prediction error on test data is minimized.

\subsection{Theoretical consideration}
For linear regression, under Irrepresentable Condition\footnote{Assume $b_k = (b_k^1,...,b_k^J,b_k^{J+1},...,b_k^q)^T$, where $b_k^j \neq 0$ for $j = 1,...,J$ and
$b_k^j = 0$ for $j = J+1,...,q$. Let $b_k^{(1)} = (b_k^1,...,b_k^J)^T$ and $b_k^{(2)} = (b_k^{J+1},...,b_k^q)^T$.
Denote Gram matrix $\Psi = \frac{1}{n}{\mathbb X}'{\mathbb X} =
\begin{pmatrix}
\Psi_{11} & \Psi_{12}\\
\Psi_{21} & \Psi_{22}
\end{pmatrix}$, then Irrepresentable Condition is satisfied if there exists a positive constant vector $\eta$, such that $|\Psi_{21}(\Psi_{11})^{-1}\sgn(b_k^{(1)})|\leq 1-\eta$.}, 
$\{\widehat{b}_k\}_{k=1}^P$ have sign consistency assured by LASSO
estimator (\cite{zhao2006model}), which means for sufficient large sample size $T-d$
\begin{equation}
 \Pr(\sgn(\widehat{b}_k) = \sgn(b_k)) \rightarrow 1
\end{equation}
where $\sgn(b_k)$ is the sign function with value of 1, 0 or -1 corresponding to $b_k > 0$, $b_k = 0$ or $b_k < 0$ respectively. Therefore, ${\textbf P}(\widehat{S}_k \neq S_k) \rightarrow 0$, which implies high specificity of true zero VAR coefficients. Then our inaccurate non-zero estimate will be updated in Step 2. Since we put a constraint for LSE in Step 2, the computing is much simplified compared with merely LSE. Moreover, the bias and mean squared error of LASSLE estimator will be bounded (\cite{liu2013asymptotic})
\begin{equation}
||E(\widetilde {b_k}_{LAS}) - b_k||_2^2 \leq 2{\textbf P}(\widehat{S}_k \neq S_k)\{O(\frac{1}{m})+||b_k||_2^2 + \tau \sigma_{kk}\}
\end{equation}
\begin{equation}
E||\widetilde {b_k}_{LAS} - b_k||_2^2 \leq 2 \frac{\sigma_{kk}}{m}tr(\Psi_{11}^{-1})+\sqrt {{\textbf P}(\widehat{S}_k \neq S_k)}\{O(\frac{1}{m})+||b_k||_2^2 + \tau \sigma_{kk}\}
\end{equation}
Thus, our non-zero estimates are almost unbiased, which is significantly improved from LASSO. Final estimates given by LASSLE in simulation study have substantially lower general mean squared error. Thus our approach is able to both indicate the most important effective connectivity and give a more precise estimate of the strength of connectivity.

\subsection{Measure of dependence}

In this section, we enumerate the different measures of dependence between components of a multivariate time series (or between different channels) using the VAR model. First, a $P$-channel time series, denoted $\{\mathbf X_t=(\mathbf X_t^1,...,\mathbf X_t^P)', t=1,2,...\}$, is weakly stationary if the following are satisfied:
\begin{itemize}
\item[(a.)] $E(\mathbf X_t)$ is constant over all time $t$, and
\item[(b.)] the autocovariance function matrix
\begin{center}
$cov(X_t,X_{t+h}) = \Gamma(h) = \left( \begin{array}{cccc}
\gamma_{11}(h) & \gamma_{12}(h)&\dots & \gamma_{1P}(h)\\
\gamma_{21}(h) & \gamma_{22}(h)&\dots& \gamma_{2P}(h)\\
\vdots & \vdots & \ddots & \vdots \\
\gamma_{P1}(h) & \gamma_{P2}(h)&\dots& \gamma_{PP}(h)\\
\end{array} \right )$
\end{center}
depends only on the lag $h$, where $\gamma_{uv}(h) = cov(X_t^u,X_{t+h}^v)$ for all  pairs of
channels $u,v=1,...,P$.
\end{itemize}
Moreover, if the sequence of auto- and cross-covariance between any pair of channels
$u$ and $v$ is absolutely summable, i.e.,
$\sum_{h=-\infty}^{\infty}|\gamma_{uv}(h)| < \infty$, then we define the spectral density
matrix of $\{\mathbf X_t\}$ to be
\begin{equation}
f(\omega) = \sum_{h=-\infty}^{\infty}\Gamma(h)e^{-2\pi i \omega h}, \quad -1/2 \leq \omega \leq 1/2.
\end{equation}
The spectral matrix has dimension $P \times P$ whose diagonal elements $f_{uu}(\omega)$ are the auto-spectra of the channels at frequency $\omega$ and the off-diagonal elements $f_{uv}(\omega)$ are the cross-spectra of channels $u$ and $v$ at frequency $\omega$.

The first dependency measure that we will consider is coherency. 
Coherency between the $u$-th and $v$-th channels at frequency $\omega$, is defined as
\begin{equation}
\rho_{uv}(\omega)=\frac{f_{uv}(\omega)}{\sqrt{f_{uu}(\omega)}\sqrt{f_{vv}(\omega)}}.
\end{equation}
One can interpret coherency as the cross-correlation between the $\omega$-oscillatory component in channel $u$ and the $\omega$-oscillatory component in channel $v$ (\cite{ombao2008evolutionary}).

The second dependency measure is coherence. Coherence between the $u$-th and $v$-th channels at frequency $\omega$, is defined as
\begin{equation}
\rho_{uv}^2(\omega)=\frac{\vert f_{uv}(\omega) \vert^2}{f_{uu}(\omega)f_{vv}(\omega)}.
\end{equation}
When $\rho_{uv}^2(\omega)$ is close to $1$ then both channels $u$ and $v$ share a common $\omega$-oscillatory activity. Moreover, when the cross-correlation between the $u$ and $v$ channels is $0$ at all time lags, then the coherency (and coherence) between these channels at {\it all} frequencies is 0. A large coherence value between channels $u$ and $v$ could be due to direct connectivity between these two channels or could be indirectly due to the intervening effect of other channel(s). To measure the strength of connectivity between a pair of channels -- with the effect of all intervening channels removed -- we shall use partial coherence.

The third dependency measure is partial coherence. Define the matrix $g(\omega)=f^{-1}(\omega)$ and denote the diagonal elements as $g_{pp}(\omega)$. Let $h(\omega)$ be a diagonal matrix whose elements are $g_{pp}^{-1/2}(\omega)$. Define the matrix $C(\omega)$ to be
\begin{equation}
C(\omega) = -g(\omega)h(\omega)g(\omega)
\end{equation}
Then, the partial coherence between the $u$-th and $v$-th channels is the modulus squared of the $(u,v)$-th element of $C(\omega)$ (\cite{fiecas2010functional}, \cite{fiecas2011generalized})
\begin{equation}
\zeta_{uv}^2(\omega) = |C_{uv}(\omega)|^2
\end{equation}

We now present the fourth dependency measure which is partial directed coherence developed in \cite{baccala2001partial} and refined in \cite{baccala2014partial}. Consider a VAR($d$) model given by Equation~(\ref{var}), define
\begin{equation}\label{fourier}
A(\omega) = I-\sum_{\ell=1}^{d}\Phi_\ell{\exp(-i2\pi \omega \ell/\Omega)}
\end{equation}
be the transform of sequence $\{\Phi_\ell\}_{\ell=1}^d$ at frequency $\omega$, where $\Omega$ is the sampling frequency. The partial directed coherence from channel $v$ to channel $u$ at frequency $\omega$ is defined as
\begin{equation}\label{pdc}
\pi_{uv}^2(\omega)= \frac{|A_{uv}(\omega)|^2}{\sum_{m=1}^{P}|A_{mv}(\omega)|^2}
\end{equation}
which measures the direct influence from channel $v$ to channel $u$ conditional on all the
outflow from channel $v$. PDC gives an indication on the extent to which present frequency-specific oscillatory activity from a sender channel explains future oscillatory activity in a specific receiver channel relative to all channels in the
network. 

\subsection{Model selection}
To determine the best order $\widehat{d}$ of VAR, we first use least squares estimation to obtain $\{\widehat{\Phi}_\ell\}_{\ell=1}^{d_j}$ for each candidate order in the set $\{d_j\}_{j=1}^J$.
We search among a class of reasonable temporal lag orders. From our analysis of LFPs where there are usually less than 4 peaks in the spectrum,
it would be reasonable to use an upper bound of $12$ as the temporal lag order.
Then we calculate the sum of squared errors
\begin{equation} \label{SSE}
SSE(d_j) = \sum_{t=d_j+1}^{T}(\mathbf X_t - \sum_{\ell=1}^{d_j}\widehat{\Phi}_\ell\mathbf X_{t-\ell})(\mathbf X_t - \sum_{\ell=1}^{d_j}\widehat{\Phi}_\ell\mathbf X_{t-\ell})'
\end{equation}
Consequently the conditional MLE of the error covariance matrix $\Sigma$ for a candidate order $d_j$ is
\begin{equation} \label{sigmahat}
\widehat{\Sigma}(d_j) = SSE(d_j)/(T-d_j)
\end{equation}
which is analogous to univariate regression case. To choose the optimal lag, we compute three information criteria - the Akaike Information criterion (AIC),
Bayesian Information criterion (BIC) and Hannan-Quinn information criterion (HQC), respectively, for each candidate order $d_j$. 
\begin{equation} \label{aic}
AIC(d_j) = \log|\widehat{\Sigma}(d_j)|+2/T(P^2d_j)
\end{equation}
\begin{equation} \label{bic}
BIC(d_j) = \log|\widehat{\Sigma}(d_j)|+\log T/T(P^2d_j)
\end{equation}
\begin{equation} \label{hqc}
HQC(d_j) = \log|\widehat{\Sigma}(d_j)|+2\log\log T/T(P^2d_j)
\end{equation}
The optimal order for each criterion, denoted $\widehat{d}$ is the minimizer of the cost functions and thus gives the optimal balance between fit (as measured by SSE) and model complexity (as expressed by the penalty terms). It has been noted that $\widehat{d}^{BIC} \leq \widehat{d}^{HQC} \leq \widehat{d}^{AIC}$ when $T \geq 16$ (\cite{ivanov2005practitioner}). In the analysis of LFPs, the difference between $\widehat{d}^{BIC}$ and $\widehat{d}^{AIC}$ is at most 1, therefore we choose $\widehat{d}^{AIC}$ to capture more temporal correlation by fitting a VAR of slightly higher lag order.

\subsection{Bootstrap-based inference}
To conduct inference on the VAR parameters, a general idea is to derive the asymptotic property of the estimated VAR coefficient matrices. However, this is not trivial and is still under investigation given the high dimensionality of the VAR parameter space. An alternative is to develop a bootstrap-based inference, which has been used in time series (\cite{paparoditis2002frequency}, \cite{politis2003impact}, \cite{kirch2007resampling}, \cite{shao2010dependent}, \cite{kreiss1992bootstrap}). After obtaining the estimates of the VAR($d$) coefficient matrices, $\{\widetilde {\Phi}_\ell\}_{\ell=1}^d$, we use these estimates and corresponding residuals to generate new bootstrapped trials. Denote

\begin{equation}
R_t = \mathbf X_t - \widetilde\Phi_1\mathbf X_{t-1}-...-\widetilde\Phi_d\mathbf X_{t-d}, \quad t=d+1,...,T.
\end{equation}
by residual at time $t$. Then, to generate a bootstrapped trial $\{\mathbf X_t^{(b)}\}_{t=1}^T$, we shall use the following bootstrap algorithm. Define the bootstrapped residuals to be $\{R_t^{(b)}\}_{t=d+1}^{T}$, which are selected with replacement from $\{R_t\}_{t=d+1}^{T}$. Let ${\mathbf X}_{t}^{(b)}= \mathbf X_t$ when $t=1,...,d$, then ${\mathbf X}_{t}^{(b)}= \sum_{\ell=1}^{d}{\widetilde \Phi_\ell}{\mathbf X}_{t-\ell}^{(b)}+R_{t}^{(b)}$ are bootstrapped data at time $t$ when $t=d+1,...,T$.

\begin{algorithm} \label{boot}
	\caption*{Bootstrap Algorithm}
	\begin{algorithmic}[1]
		\State {For $b=1,...,B$}
		\BState \emph{Step 1}:
		\State {Let ${\mathbf X}_{t}^{(b)}= {\mathbf X}_{t}$ for $t=1,2,...,d$}
		\BState \emph{Step 2}:
		\State {Randomly sample bootstrapped residuals $\{R_t^{(b)}\}_{t=d+1}^{T}$ from $\{R_t\}_{t=d+1}^{T}$} with replacement
		\BState \emph{Step 3}:
		\State {Let ${\mathbf X}_{t}^{(b)}= \sum_{\ell=1}^{d}{\widetilde \Phi_\ell}{\mathbf X}_{t-\ell}^{(b)}+R_{t}^{(b)}$ be bootstrapped data at time $t$ when $t=d+1,...,T$}
	\end{algorithmic}
\end{algorithm}

Given the $b$-th bootstrap time series $\{{\mathbf X}_{t}^{(b)}\}_{t=1}^T$, we compute the VAR coefficient estimates which we denote by $\{\widetilde \Phi_\ell^{(b)}\}_{\ell=1}^d$ using the LASSLE method and then compute partial directed coherence estimate. We repeat this procedure a sufficient large number of times, then we can find the empirical distribution and obtain the 95\% bootstrap confidence interval of both VAR parameters and PDCs.

%% file: Simulation_revise.tex
\section{Simulation study}
\subsection{Simulation design}

\begin{figure}[h!]\centering
\includegraphics[width=0.5\textwidth]{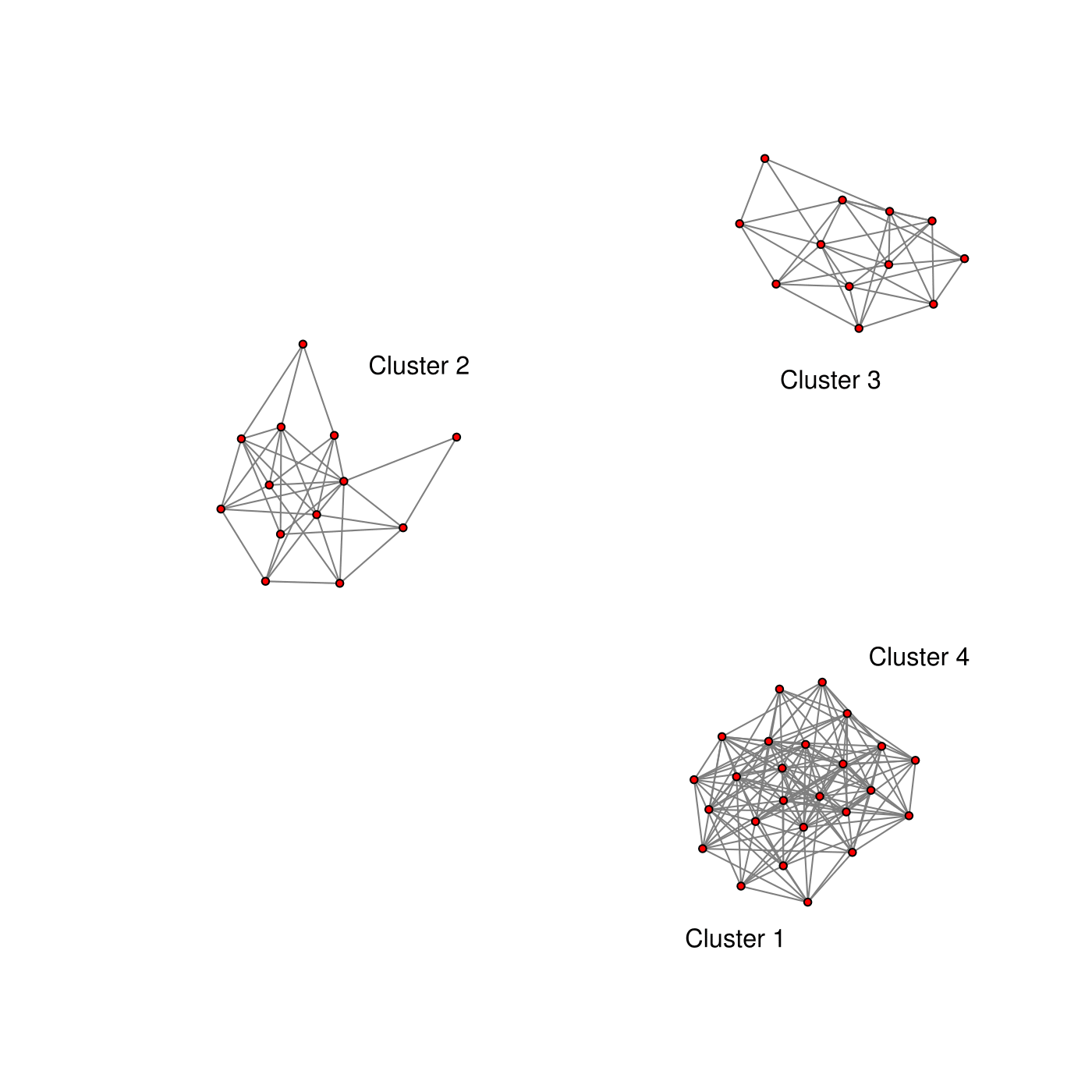}\\
\caption{50 brain channels of \enquote{Cluster} type}
\label{cluster_demo}
\end{figure}

To compare the performance of the proposed LASSLE approach with the classical methods (i.e., LSE only and LASSO only), we conducted a simulation study of VAR($d$) model for two different brain network types. The first is \enquote{Cluster}, which is a type of network that has high level local and global connectivity efficiency. In Figure \ref{cluster_demo}, channels (red nodes) are located in four brain regions, while the edge between two red nodes indicates connectivity at channel level. Auto-connectivity inside each region makes channels from the same region connect like a cluster, and cross-connectivity between brain regions determines whether these clusters are connected with each other. For example, Cluster 2 is independent from other regions, but Cluster 1 and Cluster 4 are connected due to the cross-connectivity at region level.

\begin{figure}[h!]\centering
\includegraphics[width=0.5\textwidth]{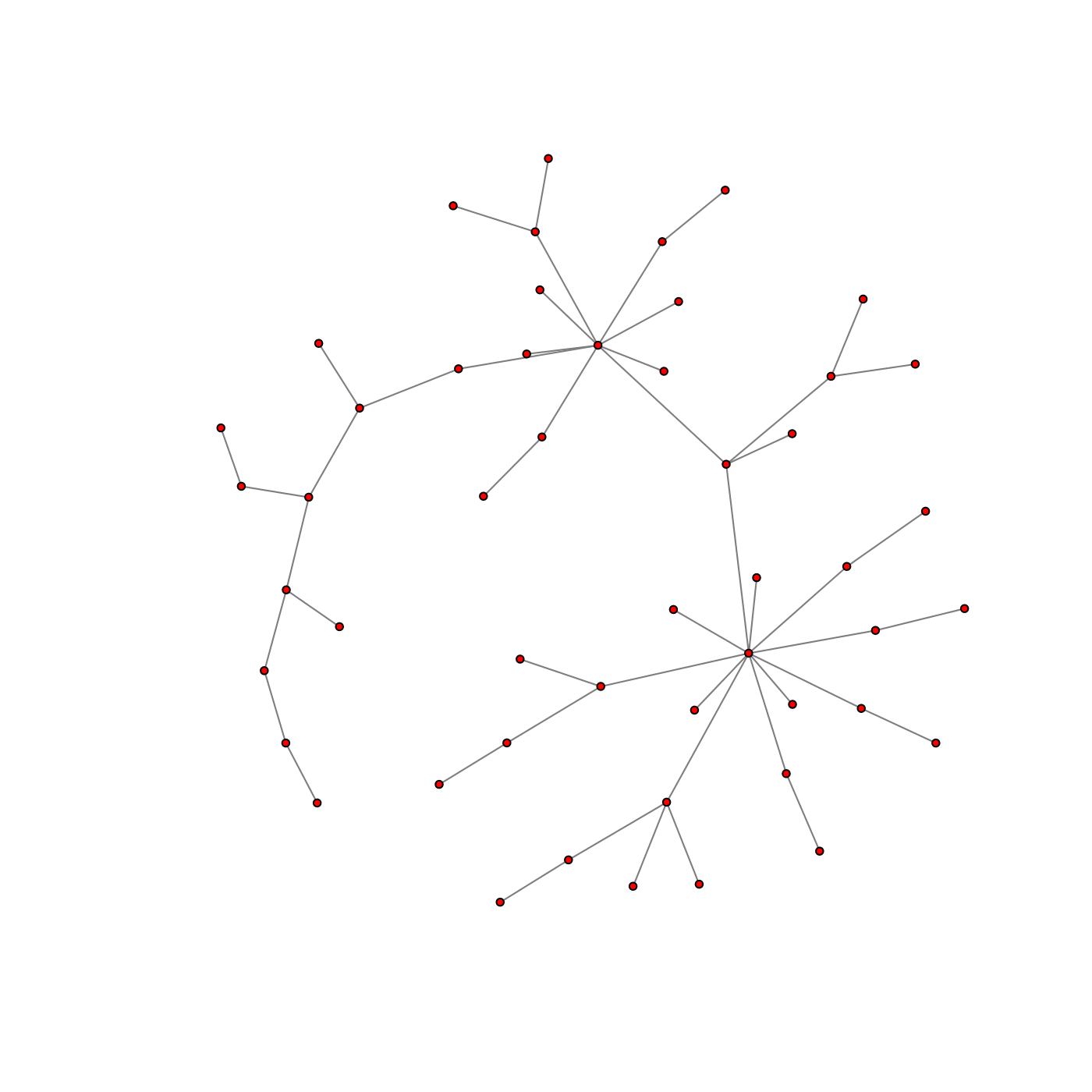}\\
\caption{50 brain channels of \enquote{Scale-free} type}
\label{scale_demo}
\end{figure}

In the second type \enquote{Scale-free}, shown in Figure \ref{scale_demo}, there is no significant auto-connectivity or cross-connectivity at the region level, but all the brain channels are connected within the network. Most of the channels have several connections with other channels, with the exception that a few channels are heavily connected. The idea is that these channels play a central role in the organization of entire brain network, as they are mostly responsible for the connectivity efficiency.

For both network types, we use Equation~(\ref{var}) to generate time series data sets. The VAR matrix $\Phi_1$ of setting $\{P=50$, $d=1\}$ is visualized in Figure \ref{cluster_compare_normal}(a) and Figure \ref{scale_compare_normal}(a). Each small square represents the non-zero entry of $\Phi_1$ and different colors indicate different values according to the color bar. The blank part of coefficient matrix are the zero entries. In addition, $\varepsilon_t$ follows a Gaussian distribution and the covariance matrix is not necessary to be diagonal. We run $N = 1,000$ simulations for each VAR setting respectively and the time series data of each channel contains $T=10,000$ time points. Then we apply LSE, LASSO and our LASSLE method to estimate coefficient matrices, and compare their results with two important criteria. The first one evaluates how successful the estimate identifies the specific entries with true value of zero, as shown by the visualization of absolute difference between true coefficient matrix and estimated one. The second criteria is their mean squared error,
defined as
\begin{equation}
MSE = \frac{\sum_{n,\ell,i,j}(\Phi_\ell^{ij}-\widehat{\Phi}_\ell^{ij})^2}{N}
\end{equation}
where $\{\Phi_\ell^{ij}\}_{\ell=1}^d$, $\{\widehat{\Phi}_\ell^{ij}\}_{\ell=1}^d$ represent entries of $\{\Phi_\ell\}_{l=1}^d$ and $\{\widehat{\Phi}_\ell\}_{\ell=1}^d$ respectively. Lower MSE indicates better centering at true connectivity matrix.

\subsection{Simulation results}
Due to the display limit of high dimensional matrix, we only demonstrate visualized results of VAR setting $\{P=50$, $d=1\}$.

\subsubsection{Results from the \enquote{Cluster} setting}

\begin{figure}[H] \centering
	\begin{tabular}{cc}
		\includegraphics[width=0.4 \textwidth]{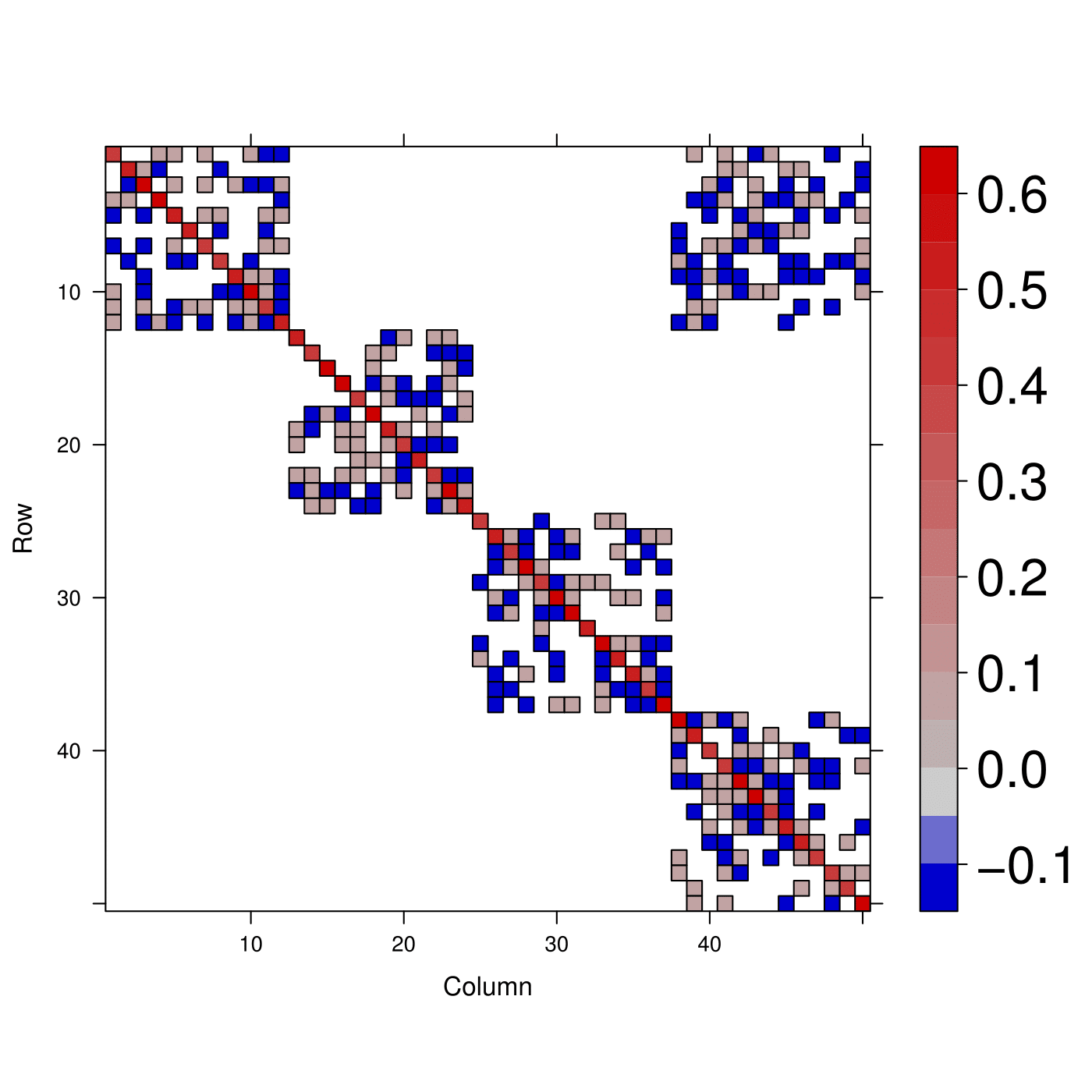}&
		\includegraphics[width=0.4 \textwidth]{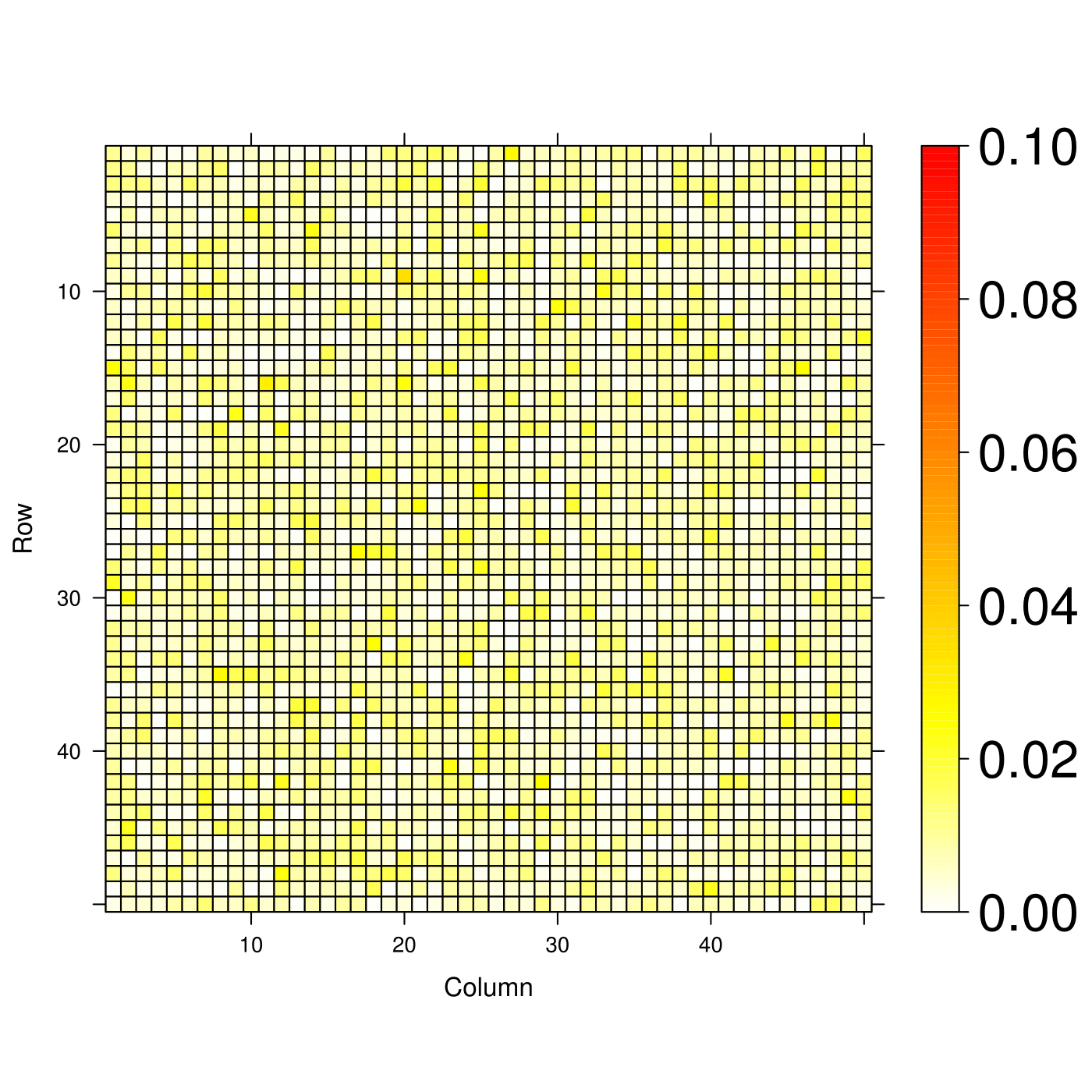} \\
		(a) \enquote{Cluster} type VAR matrix &(b) Absolute difference of LSE\\
		\includegraphics[width=0.4 \textwidth]{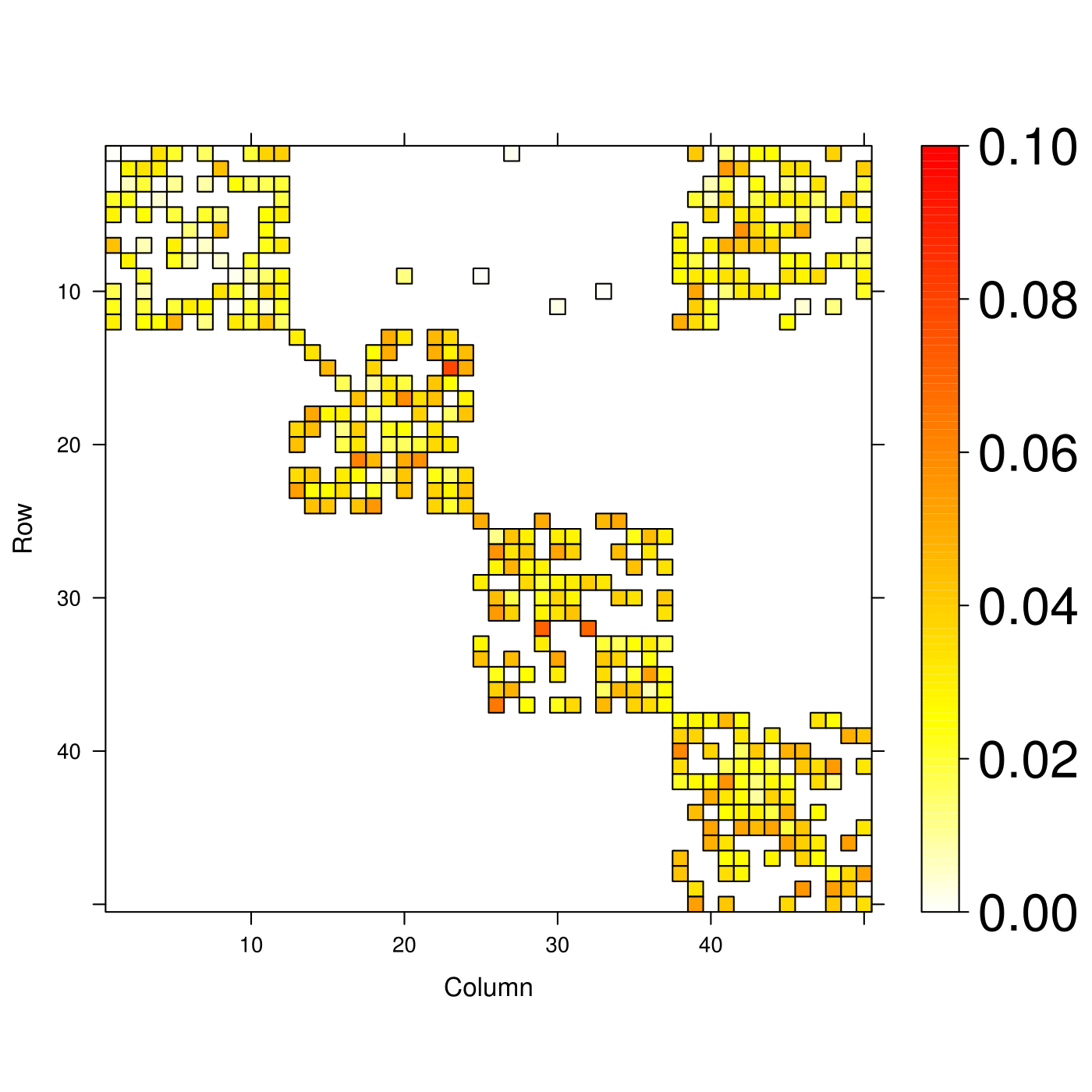} &
		\includegraphics[width=0.4 \textwidth]{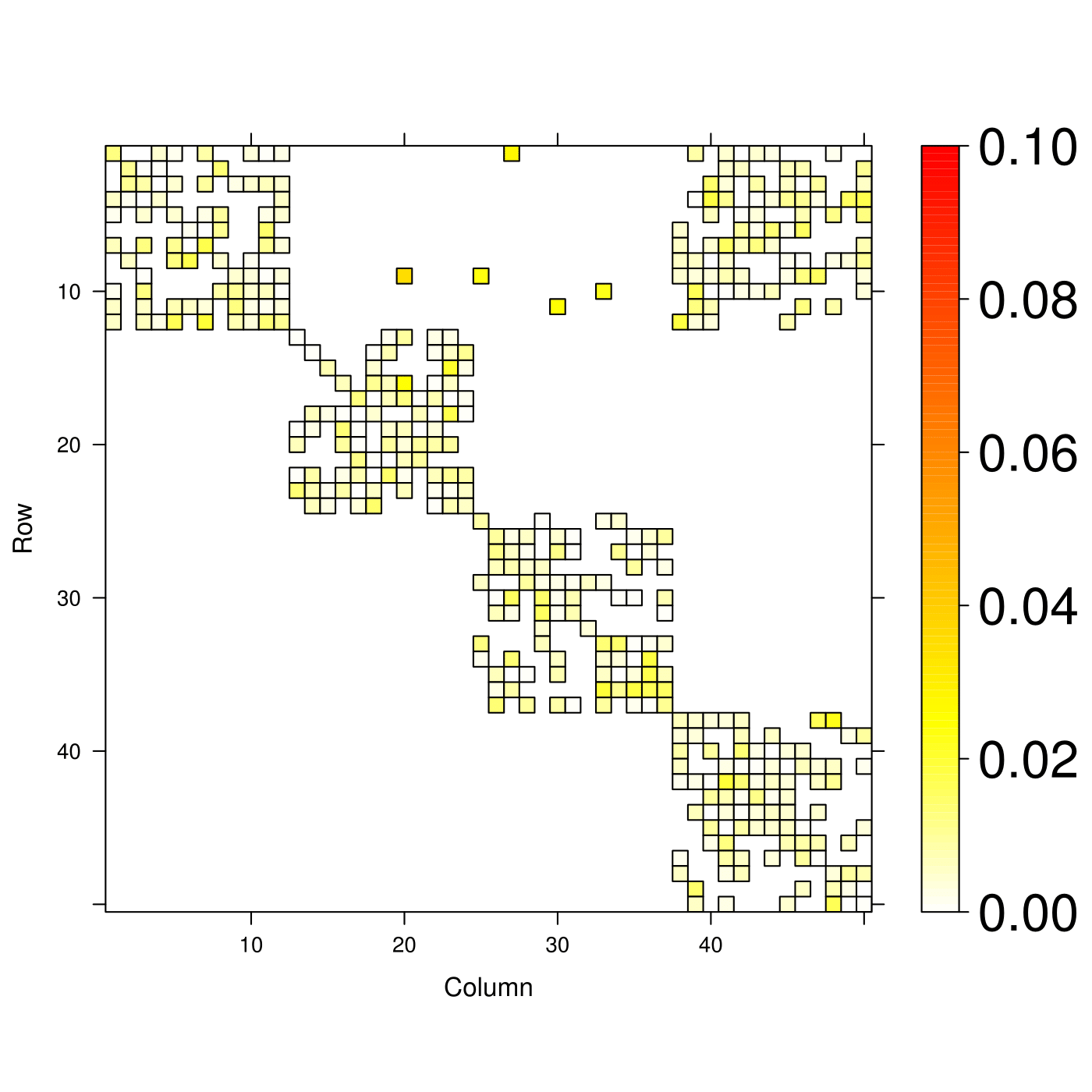} \\
		(c) Absolute difference of LASSO &(d) Absolute difference of LASSLE
	\end{tabular}
	\caption{Comparison of specificity of true zero on \enquote{Cluster} type data with Gaussian noise. Figure \ref{cluster_compare_normal}(a) demonstrates the true VAR(1) coefficient matrix with $P=50$. Figure \ref{cluster_compare_normal}(b),(c),(d) yield the absolute difference between true matrix and estimated matrix by LSE, LASSO and LASSLE method respectively.}
	\label{cluster_compare_normal}
\end{figure}

In this setting, 50 channels represent measurements in four brain regions and only region 1 and region 4 have cross-connectivity. In the coefficient matrix, all non-zero entries are first randomly assigned either 0.1 or -0.1, then 0.5 is added to all diagonal entries (shown in Figure \ref{cluster_compare_normal}(a)). Figure \ref{cluster_compare_normal}(b),(c),(d) yield the absolute difference between true connectivity matrix and estimated one by LSE, LASSO and LASSLE method. The color of small squares ranging from white to red indicates the value of absolute difference of each entry. The blank part of the matrix implies that the estimate has given correct zero-estimate for those true zero entries so that there is no need to distinguish the difference with color. Table 1 demonstrates the MSE results of all three methods under different VAR parameter setting $\{P,d\}$.

\begin{table}[H]\centering
\small
\begin{tabular}{|c| c| c| c| c| c|}\hline\hline

\multicolumn{3}{|c|}{VAR Parameter Setting} & \multicolumn{3}{c|}{MSE $\times 10^{-3}$} \\
\hline
Number & $P$ & $d$ & LSE & LASSO & LASSLE \\
\hline
100 & 10 & 1 & 8 & 93 & 4 * \\
\hline
500 & 10 & 5 & 58 & 326 & 36 * \\
\hline
1,000 & 10 & 10 & 134 & 412 & 82 * \\
\hline
2,500 & 50 & 1 & 176 & 464 & 24 * \\
\hline
5,000 & 50 & 2 & 457 & 739 & 93 * \\
\hline
10,000& 100 & 1 & 697 & 1016 & 65 * \\
\hline

\end{tabular}
\caption {Comparison of MSE between three methods on \enquote{Cluster} type data} 
\end{table}

\subsubsection{Results from the \enquote{Scale-free} setting}
\begin{figure}[H]\centering
	\begin{tabular}{cc}
		\includegraphics[width=0.4 \textwidth]{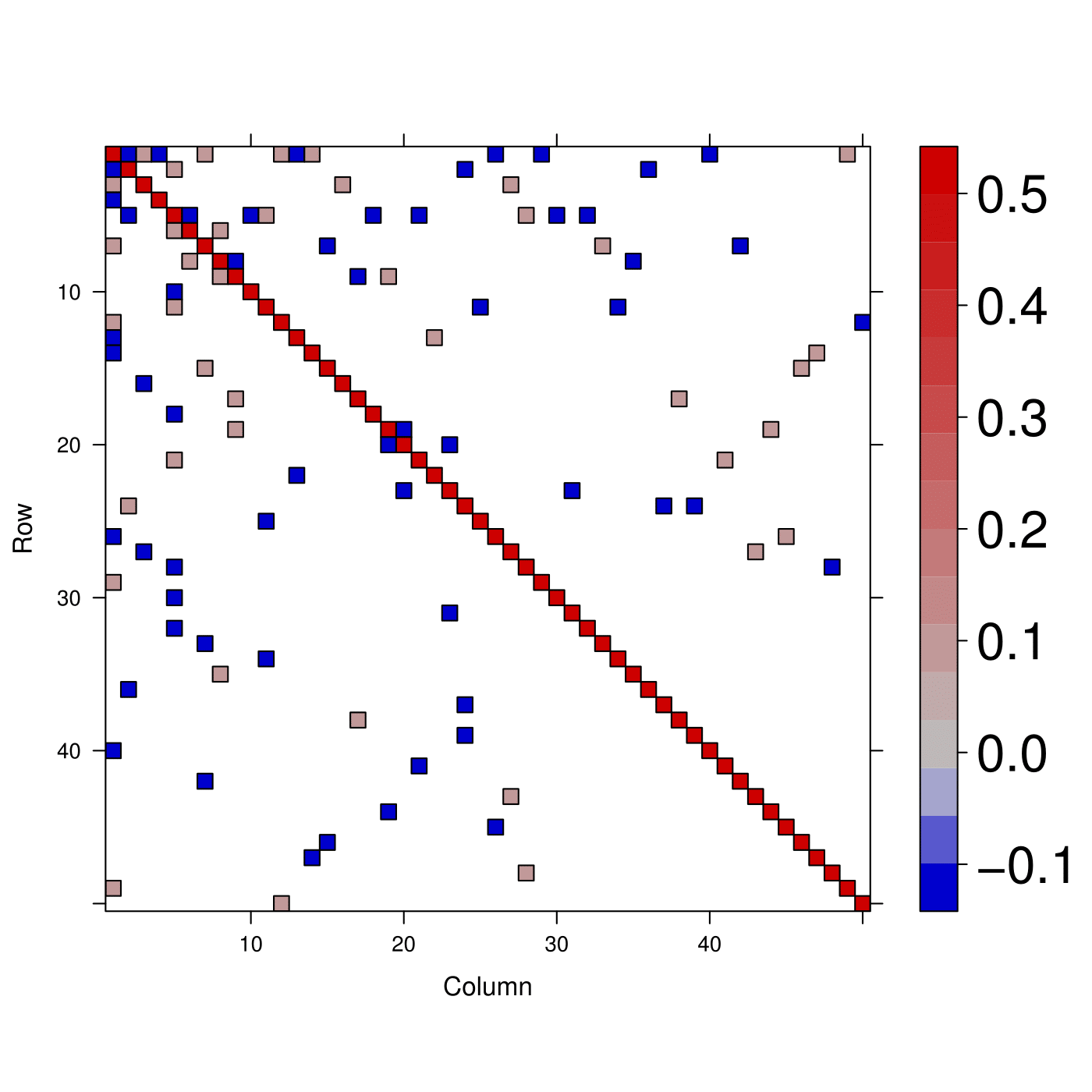}&
		\includegraphics[width=0.4 \textwidth]{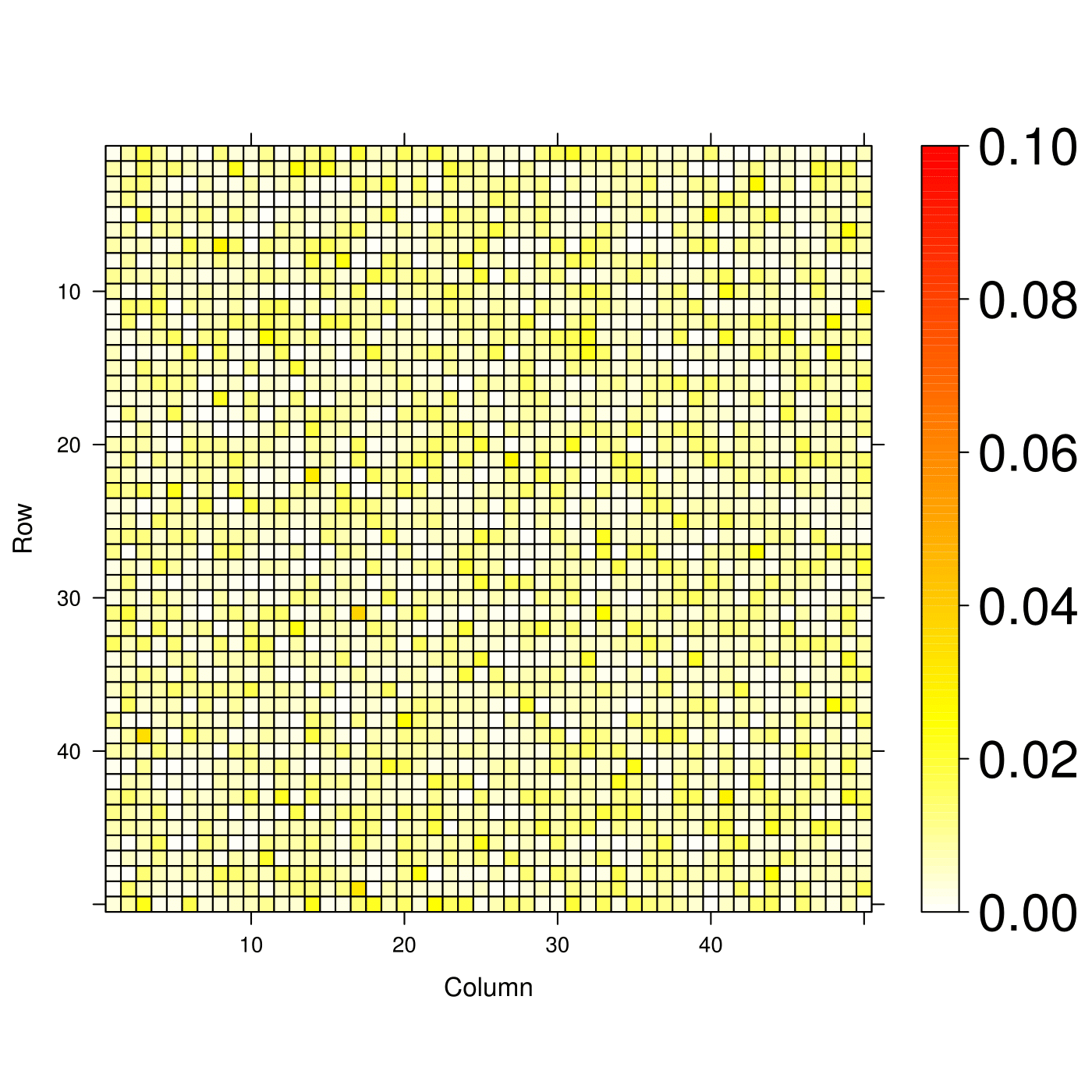} \\
		(a) \enquote{Scale-free} type VAR matrix &(b) Absolute difference of LSE\\
		\includegraphics[width=0.4 \textwidth]{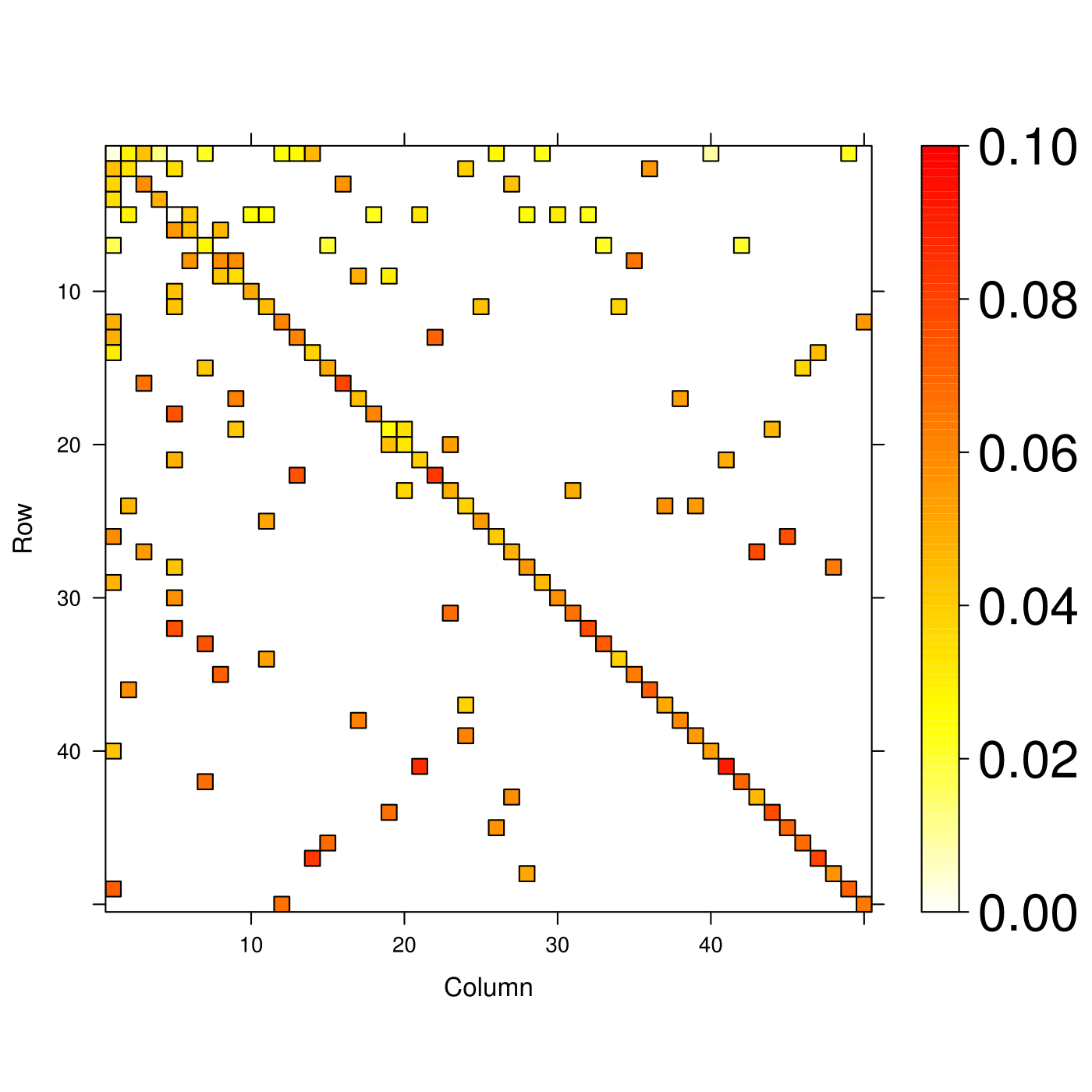} &
		\includegraphics[width=0.4 \textwidth]{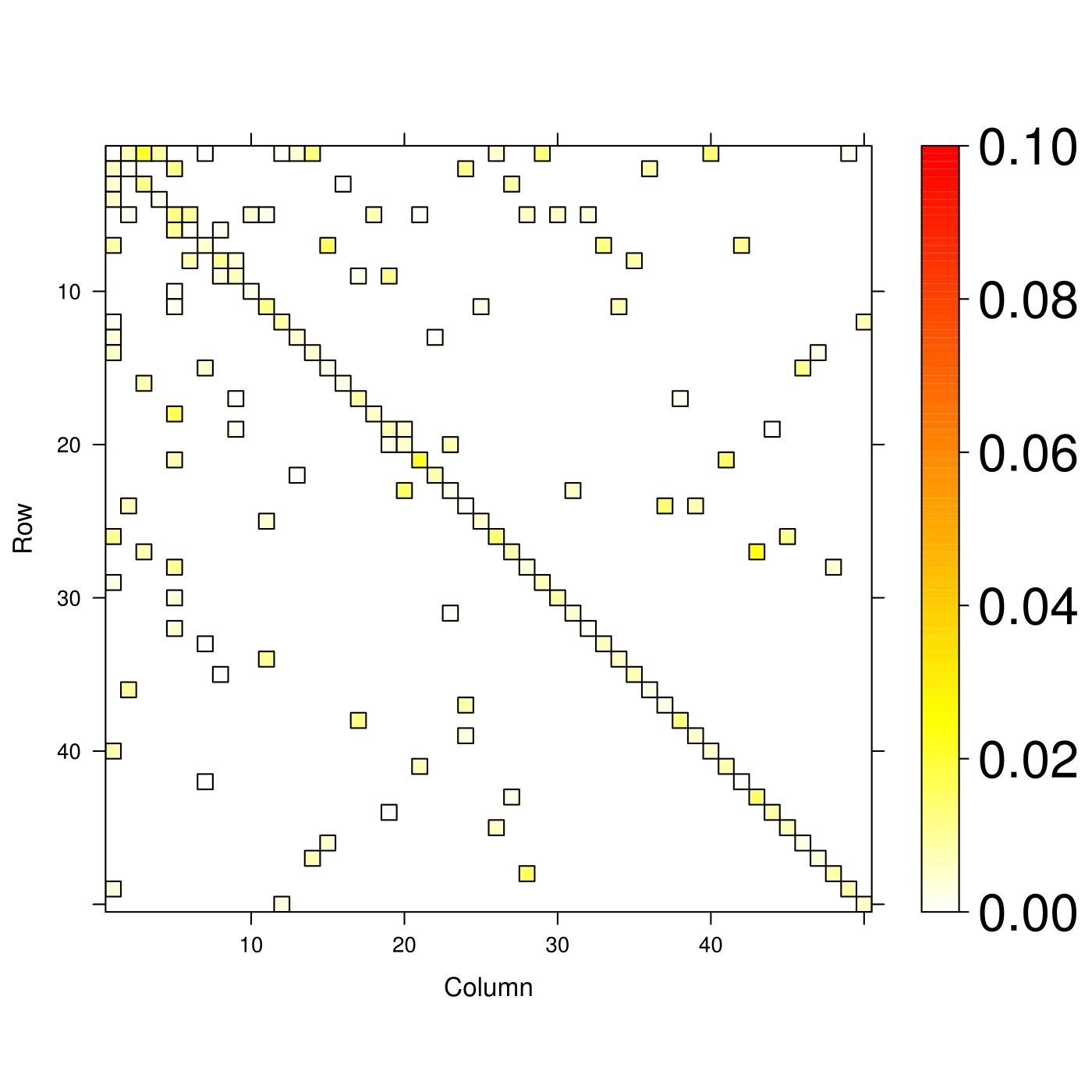}\\
		(c)  Absolute difference of LASSO&(d) Absolute difference of LASSLE
	\end{tabular}
	\caption{Comparison of specificity of true zero on \enquote{Scale-free} type data with Gaussian noise. Figure \ref{scale_compare_normal}(a) demonstrates the true VAR(1) coefficient matrix with $P=50$. Figure \ref{scale_compare_normal}(b),(c),(d) yield the absolute difference between true matrix and estimated matrix by LSE, LASSO and LASSLE method respectively.}
	\label{scale_compare_normal}
\end{figure}

To generate \enquote{Scale-free} type data, we assign 0.5 to all diagonal entries of connectivity matrix, and 0.1 or -0.1 randomly to other non-diagonal entries with small probability (seen in Figure \ref{scale_compare_normal}(a)). Figure \ref{scale_compare_normal}(b),(c),(d) give the visualized estimate results given by LSE, LASSO and LASSLE method separately. MSE comparison can be found in Table 2.

\begin{table}[H]\centering
\small
\begin{tabular}{|c| c| c| c| c| c|}\hline\hline

\multicolumn{3}{|c|}{VAR Parameter Setting} & \multicolumn{3}{c|}{MSE $\times 10^{-3}$} \\
\hline
Number & $P$ & $d$ & LSE & LASSO & LASSLE \\
\hline
100 & 10 & 1 & 7 & 86 & 2 * \\
\hline
500 & 10 & 5 & 53 & 158 & 16 * \\
\hline
1,000 & 10 & 10 & 130 & 472 & 72 * \\
\hline
2,500 & 50 & 1 & 191 & 432 & 9 * \\
\hline
5,000 & 50 & 2 & 480 & 877 & 37 * \\
\hline
10,000 & 100 & 1 & 762 & 921 & 19 * \\
\hline

\end{tabular}
\caption {Comparison of MSE between three methods on \enquote{Scale-free} type data} 
\label{scale}

\end{table}

\subsubsection{Discussion}
From visualized results, we can find that LSE is unable to give specificity for true zero coefficients, since its estimates do not contain blank squares. However, its estimate has general lower bias across all the entries, which is implied by the light color of absolute difference. LASSO is able to identify most true zero entries, but darker color of rectangles indicate that this method has high bias for the estimate. Our method, constrained with LASSO in Step 1, has inherited the specificity of true zero values from LASSO, consequently can capture true zero values as well as LASSO. Thus in the sense of specificity of true zero, the comparison result is: LASSLE = LASSO $>>$ LSE.

For another important criteria MSE, the proposed LASSLE approach has substantial advantage over LSE and LASSO. In 50-channel \enquote{Cluster} setting, the MSE given by LSE and LASSO are approximately 5 times and 10 times the MSE provided by our method. Also, in 50-channel \enquote{Scale-free} settlement, LSE and LASSO provide at least 10 times and 20 times higher MSE compared with LASSLE. With the increase of dimensions, the advantage is also increasing geometrically. Therefore, LASSLE performs better with respect to the MSE criterion compared to both LSE and LASSO.

\subsection{Bootstrap-based inference}
For each VAR setting and its simulation, we follow the bootstrap algorithm in Section 2.7 to generate 1,000 bootstrapped trials and re-estimate the VAR parameters with LASSLE method. We (1.) investigated whether the 95\% bootstrap confidence interval given by the empirical distribution of each VAR parameter captured the true value and (2.) compared the center of the bootstrap distribution to the true value of the quantities of interest (VAR parameters and true PDC values). To answer (1.), we plot the empirical distribution of each VAR parameter and compare its 95\% empirical confidence interval with the true value. To compare (2.), we obtain the median of bootstrapped estimates for each entry, then use these medians to form a matrix and compare its absolute difference with the true connectivity matrix. 

\begin{figure}[H]\centering
	\includegraphics[width=\textwidth, height = 0.8\textwidth]{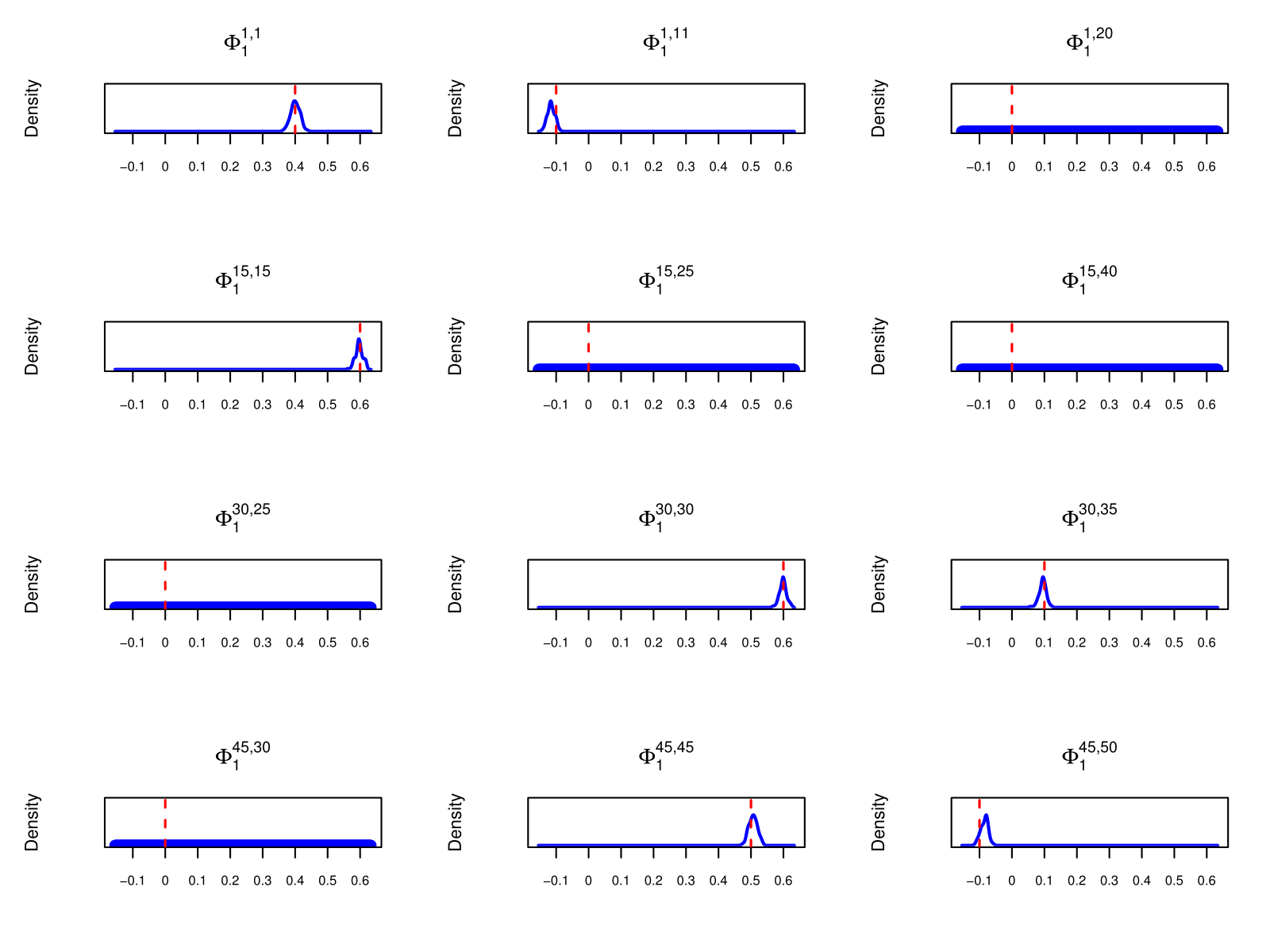}\\
	\caption{Sample empirical distribution of \enquote{Cluster} type bootstrapped estimates. Red dashed line indicates the true value of these example coefficients. Blue curve is the smoothed estimated density curve of each empirical distribution. All 1,000 bootstrap estimates of $\Phi_1^{1,20}, \Phi_1^{15,25}, \Phi_1^{15,40}, \Phi_1^{30,25}$ and $\Phi_1^{45,30}$ are zero, so we are not able to plot the point-mass density for these coefficients.}
	\label{dist_boot_cluster}
\end{figure}

\begin{figure}[H]\centering
	\includegraphics[width=\textwidth, height = 0.8\textwidth]{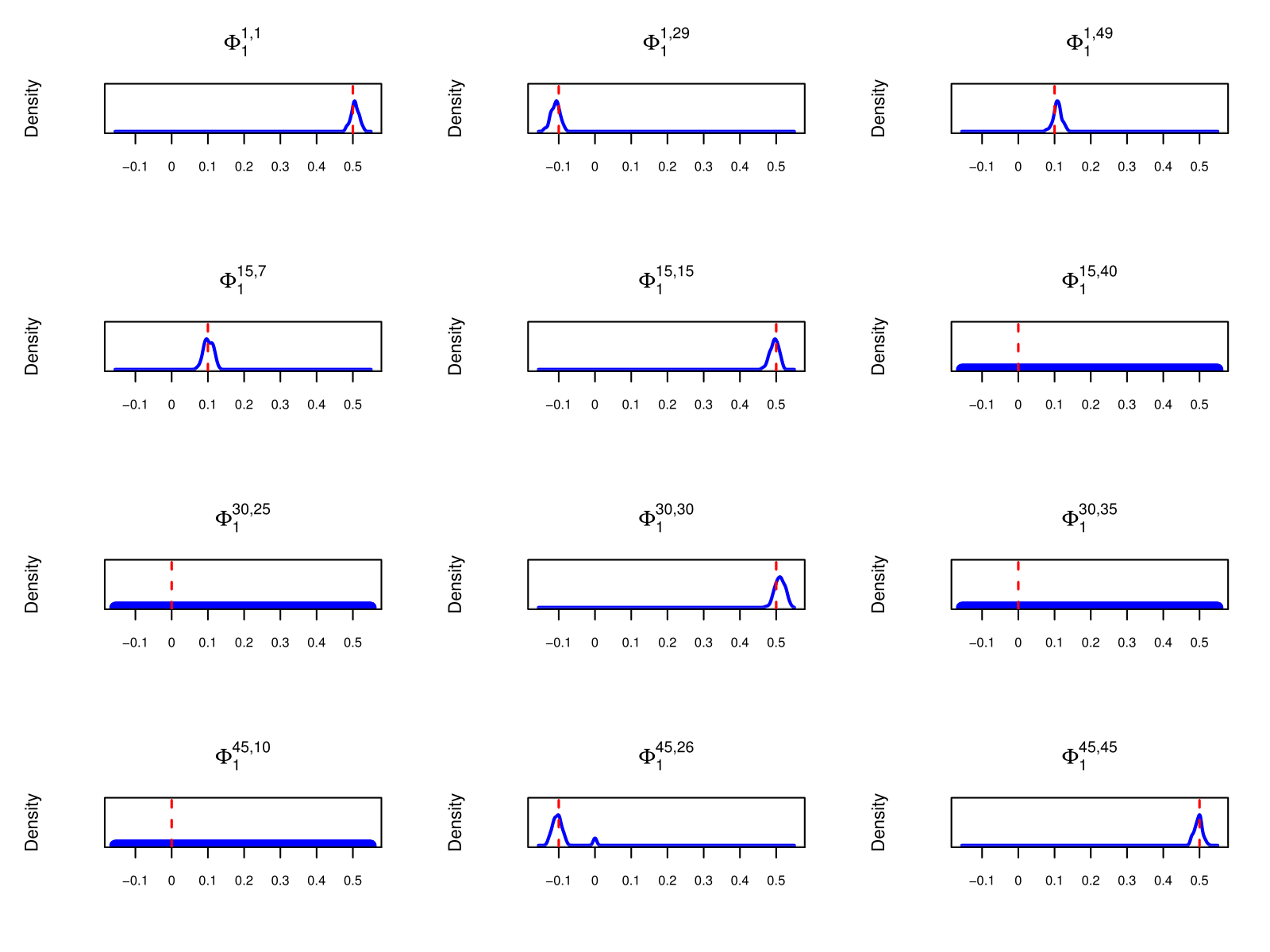}\\
	\caption{Sample empirical distribution of \enquote{Scale-free} type bootstrapped estimates. Red dashed line indicates the true value of these example coefficients. Blue curve is the smoothed estimated density curve of each empirical distribution. All 1,000 bootstrap estimates of $\Phi_1^{15,40}, \Phi_1^{30,25}, \Phi_1^{30,35}$ and $\Phi_1^{45,10}$ are zero, so we are not able to plot the point-mass density for these coefficients.}
	\label{dist_boot_scale}
\end{figure}

Figure \ref{dist_boot_cluster} and Figure \ref{dist_boot_scale} demonstrate examples of empirical distribution derived from 1,000 bootstrap estimates. The red dashed line indicates the true value of these example coefficients. The blue curve is the smoothed estimated density curve of each empirical distribution. For some coefficients, e.g., $\Phi_1^{20}, \Phi_{15}^{25}$ in Figure \ref{dist_boot_cluster} and $\Phi_{15}^{40},\Phi_{30}^{25}$ in Figure \ref{dist_boot_scale}, we are not able to provide a density curve as the empirical distribution is a point-mass density at $x=0$, in other words, all 1,000 bootstrapped LASSLE estimates of these zero coefficients are exactly zero. We can conclude that 95\% confidence interval or set of the empirical distribution can capture the true parameter value in the simulation study.       

\begin{figure}[H]\centering
	\begin{tabular}{cc}
		\includegraphics[width=0.4 \textwidth]{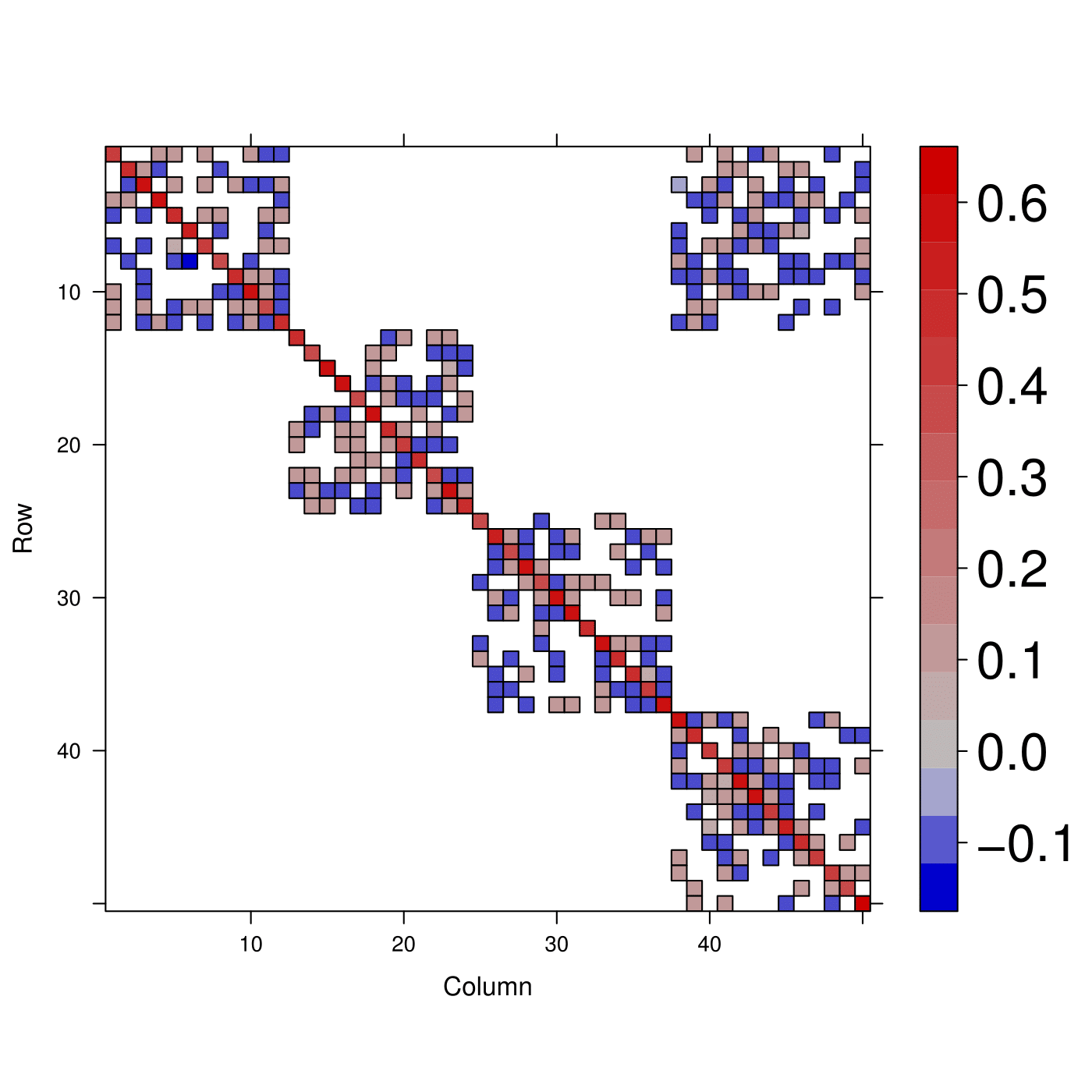}&
		\includegraphics[width=0.4 \textwidth]{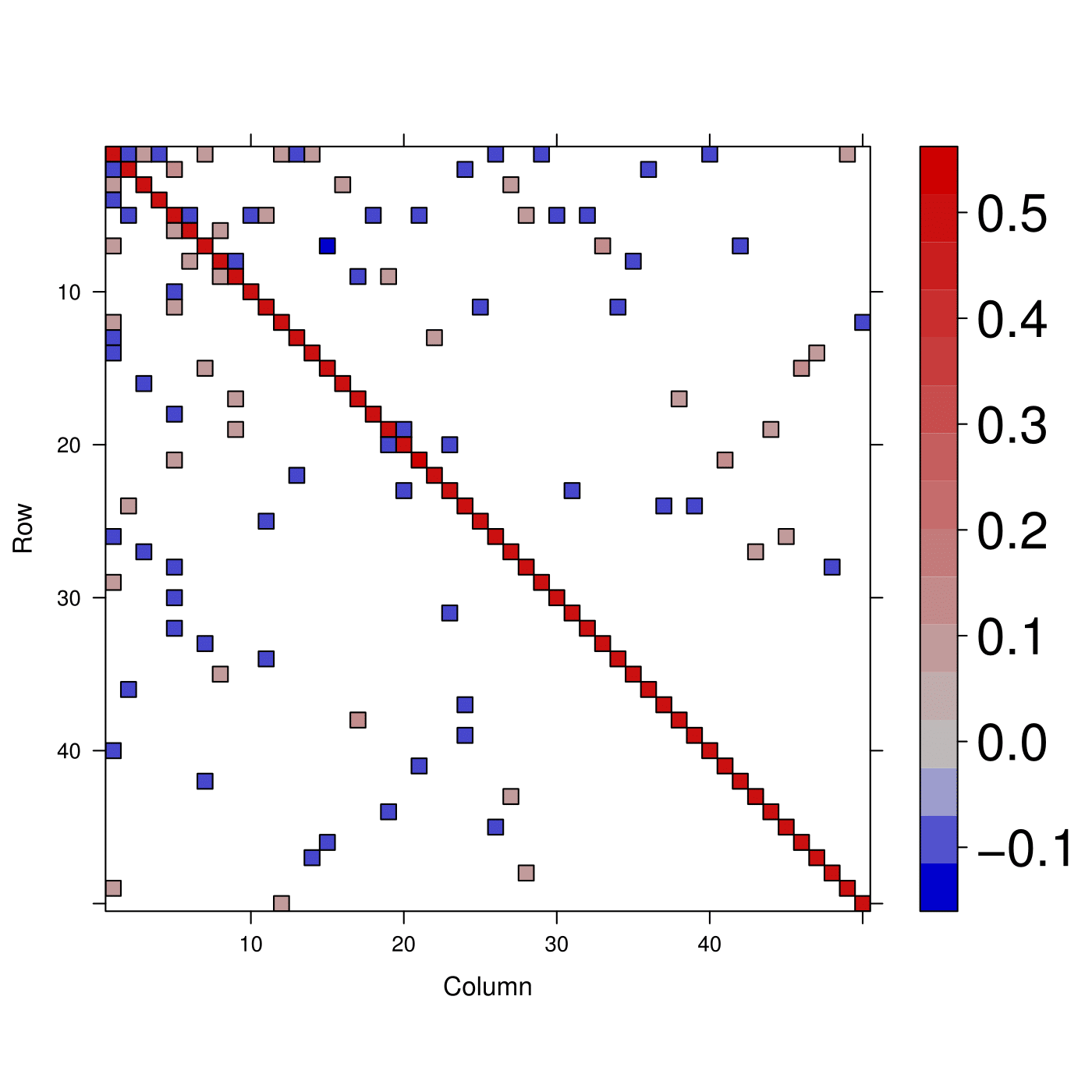} \\
		(a) Bootstrap median of \enquote{Cluster} type &(b) Bootstrap median of \enquote{Scale-free} type\\
		\includegraphics[width=0.4 \textwidth]{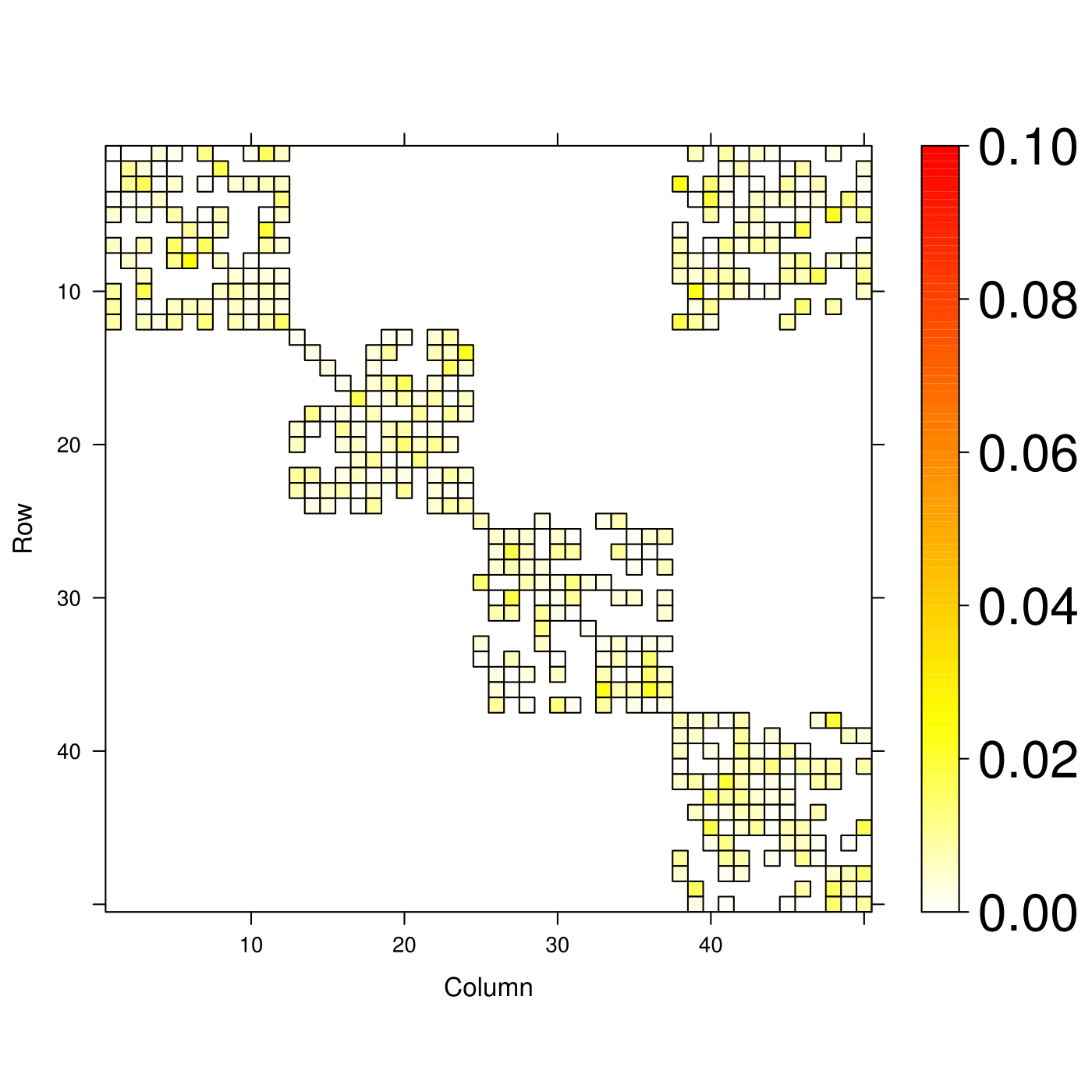}&
		\includegraphics[width=0.4 \textwidth]{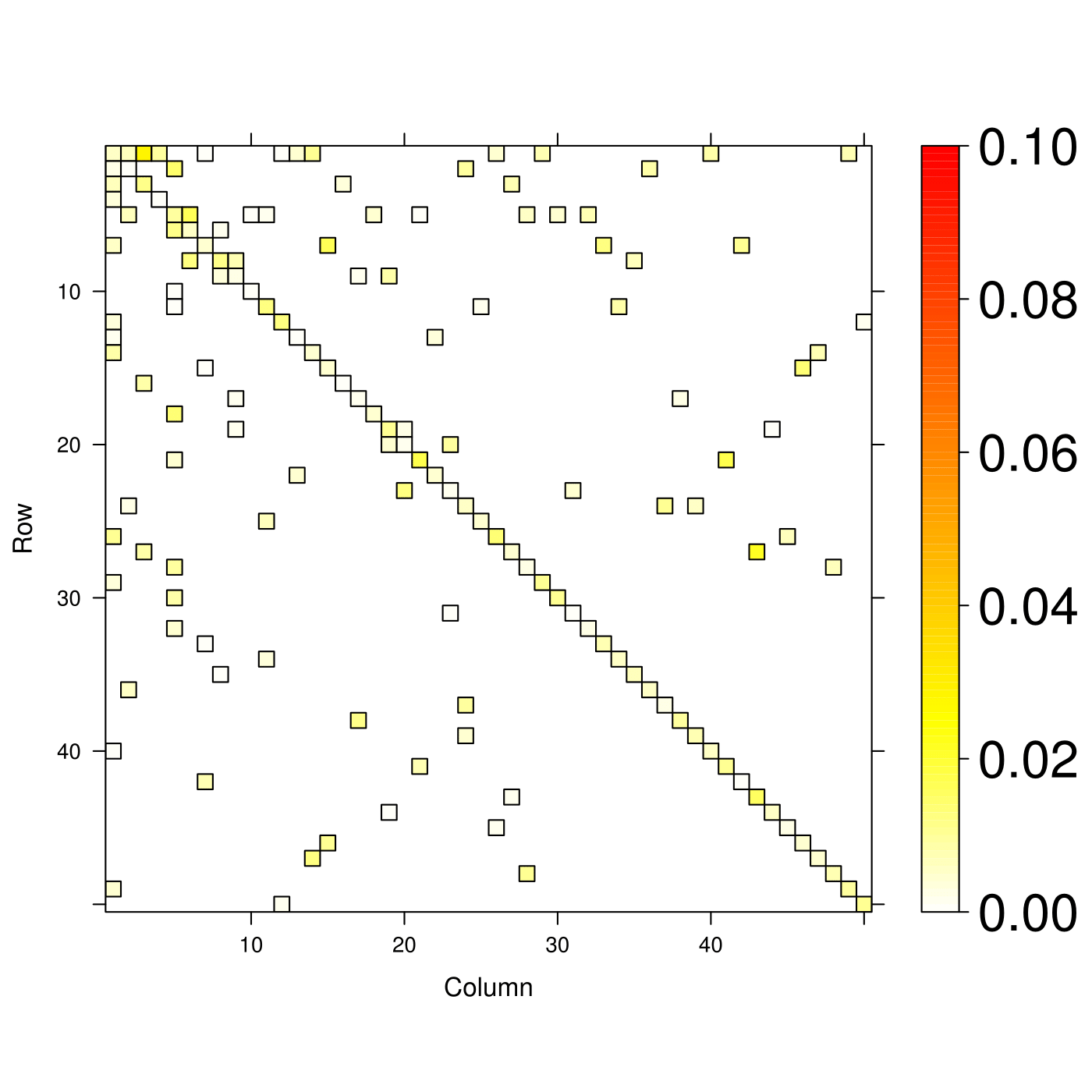} \\
		(c) Absolute difference of \enquote{Cluster} type &(d) Absolute difference of \enquote{Scale-free} type\\
	\end{tabular}
	\caption{Bootstrap median of 1,000 bootstrapped LASSLE estimates. Figure \ref{median_boot}(a),(b) give the median of 1,000 bootstrapped LASSLE estimates for \enquote{Cluster} type and \enquote{Scale-free} type data. Figure \ref{median_boot}(c),(d) demonstrate the absolute difference between the median estimated matrix and true coefficient matrix.}
	\label{median_boot}
\end{figure}

Figure \ref{median_boot}(a),(b) demonstrate the median of 1,000 bootstrapped LASSLE estimates of \enquote{Cluster} and \enquote{Scale-free} type data under previous setting $\{P=50,d=1\}$ respectively. Figure \ref{median_boot}(c),(d) yield their absolute difference with the true connectivity matrix. Given the color of most nonzero median estimates of LASSLE is very light, we can conclude that the empirical distributions generated by bootstrap are well centered around the true coefficient values.

\subsection{Robustness of LASSLE method}
Previous simulation study are conducted under the assumption that $\varepsilon_t$, the noise of VAR($d$) at time $t$, follows a multivariate Gaussian distribution. In addition to Gaussian noise, we are interested to investigate whether the LASSLE method has better specificity, lower bias, lower variance (and thus lower MSE) than the LSE and LASSO methods for other noise distributions, e.g., student's $t$-noise (with low degree of freedom) and shifted $\chi^2$ noise. To explore this, we generated time series datasets using Equation~(\ref{var}) under different VAR setting \{$P,d$\} with $P$ independent $t$-noise and $P$ independent $\chi^2$ noise respectively. Then we apply all three methods to estimate the VAR coefficient matrices under each setting and compare their performance in terms of both the specificity of true zero and the general mean squared error. 

\subsubsection{Results from student's $t$-noise}
\begin{figure}[H]\centering
	\begin{tabular}{cc}
		\includegraphics[width=0.4 \textwidth]{phi_cluster50}&
		\includegraphics[width=0.4 \textwidth]{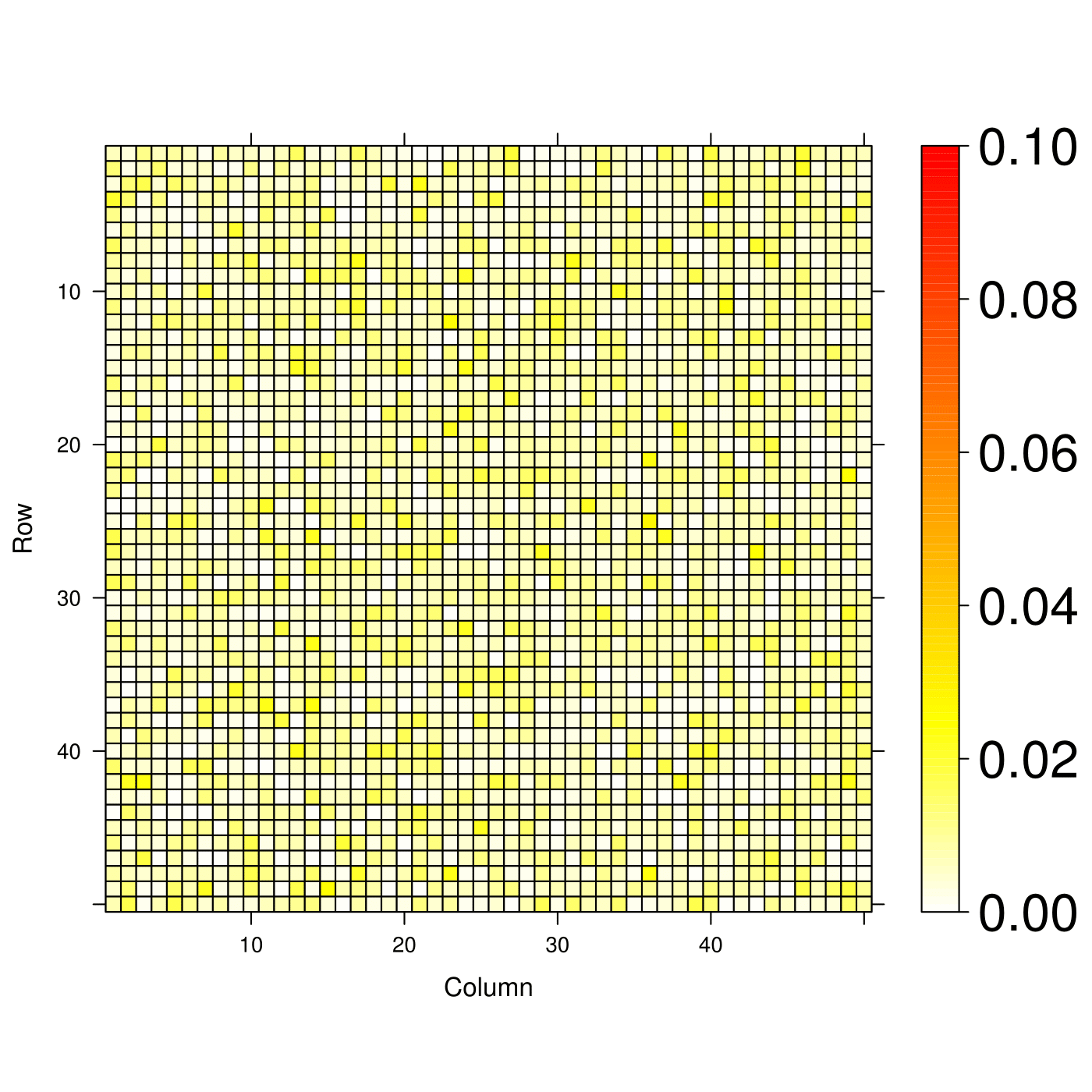} \\
		(a) \enquote{Cluster} type VAR matrix &(b) Absolute difference of LSE\\
		\includegraphics[width=0.4 \textwidth]{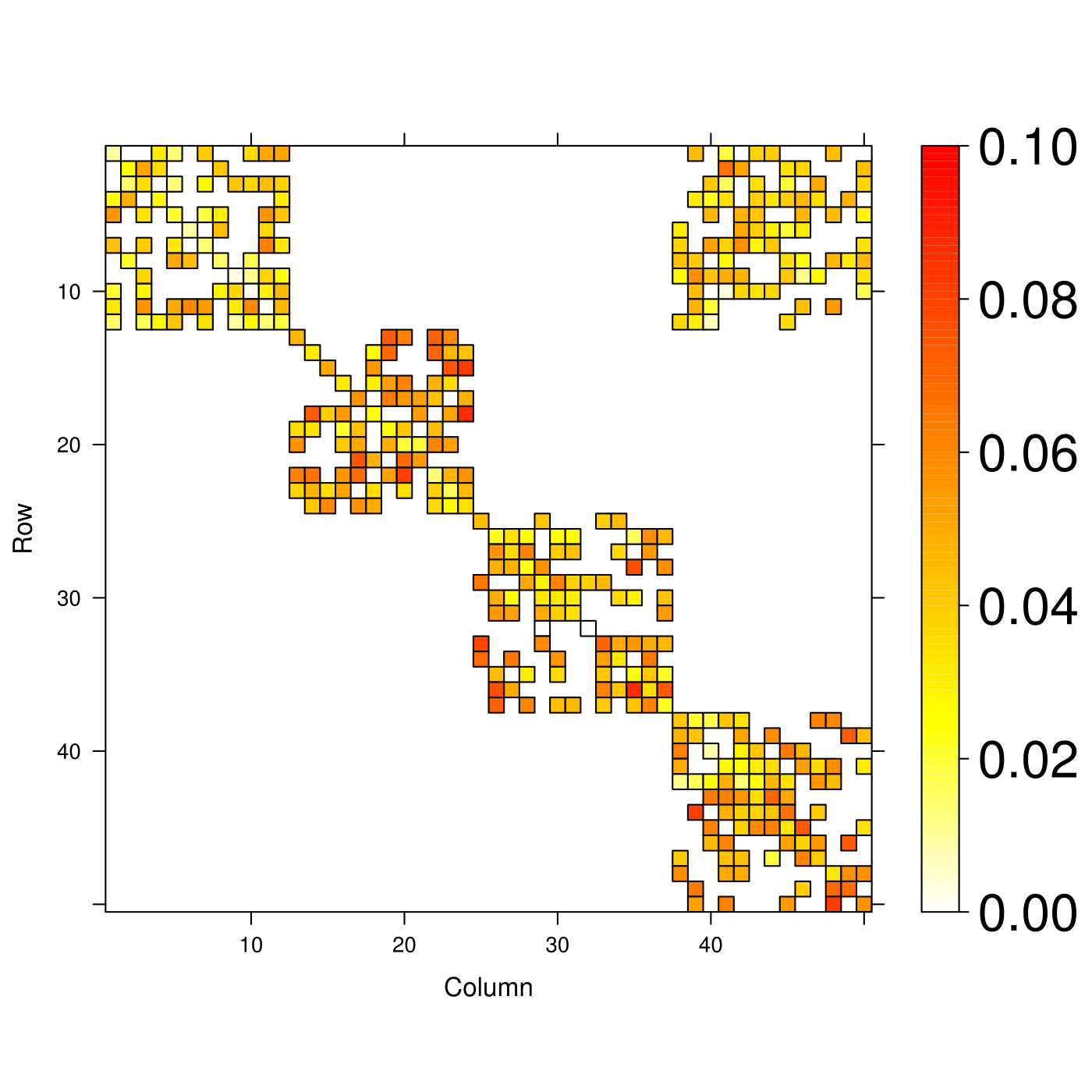} &
		\includegraphics[width=0.4 \textwidth]{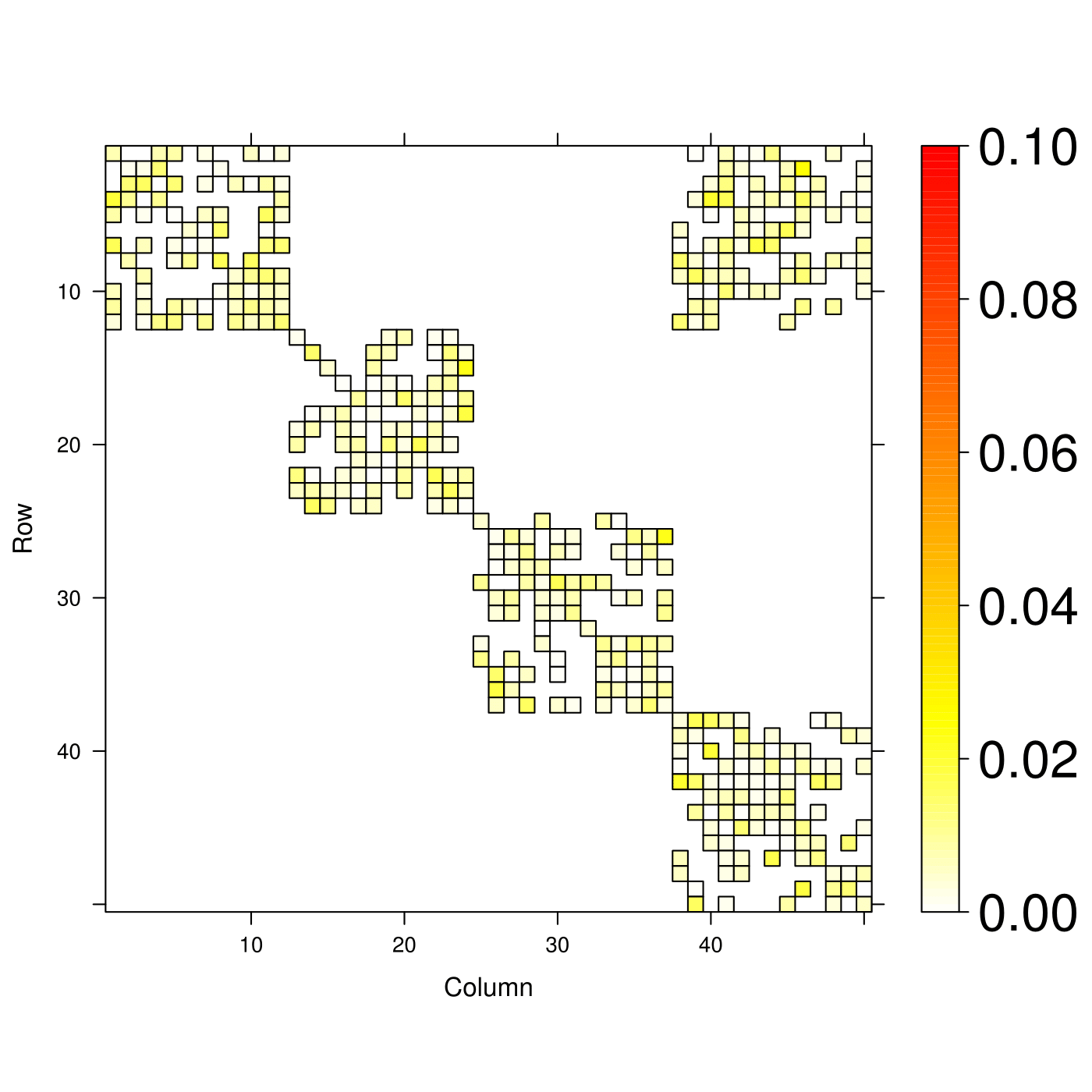} \\
		(c) Absolute difference of LASSO &(d) Absolute difference of LASSLE
	\end{tabular}
	\caption{Comparison of specificity of true zero on \enquote{Cluster} type data with student's $t$-noise.  Figure \ref{fig:specificity_cluster_t_noise}(a) demonstrates the true VAR(1) coefficient matrix with $P=50$. Figure \ref{fig:specificity_cluster_t_noise}(b),(c),(d) give the absolute difference between true matrix and estimated matrix by LSE, LASSO and LASSLE method respectively.}
	\label{fig:specificity_cluster_t_noise}
\end{figure}

\begin{figure}[H]\centering
	\begin{tabular}{cc}
		\includegraphics[width=0.4 \textwidth]{phi_scale50}&
		\includegraphics[width=0.4 \textwidth]{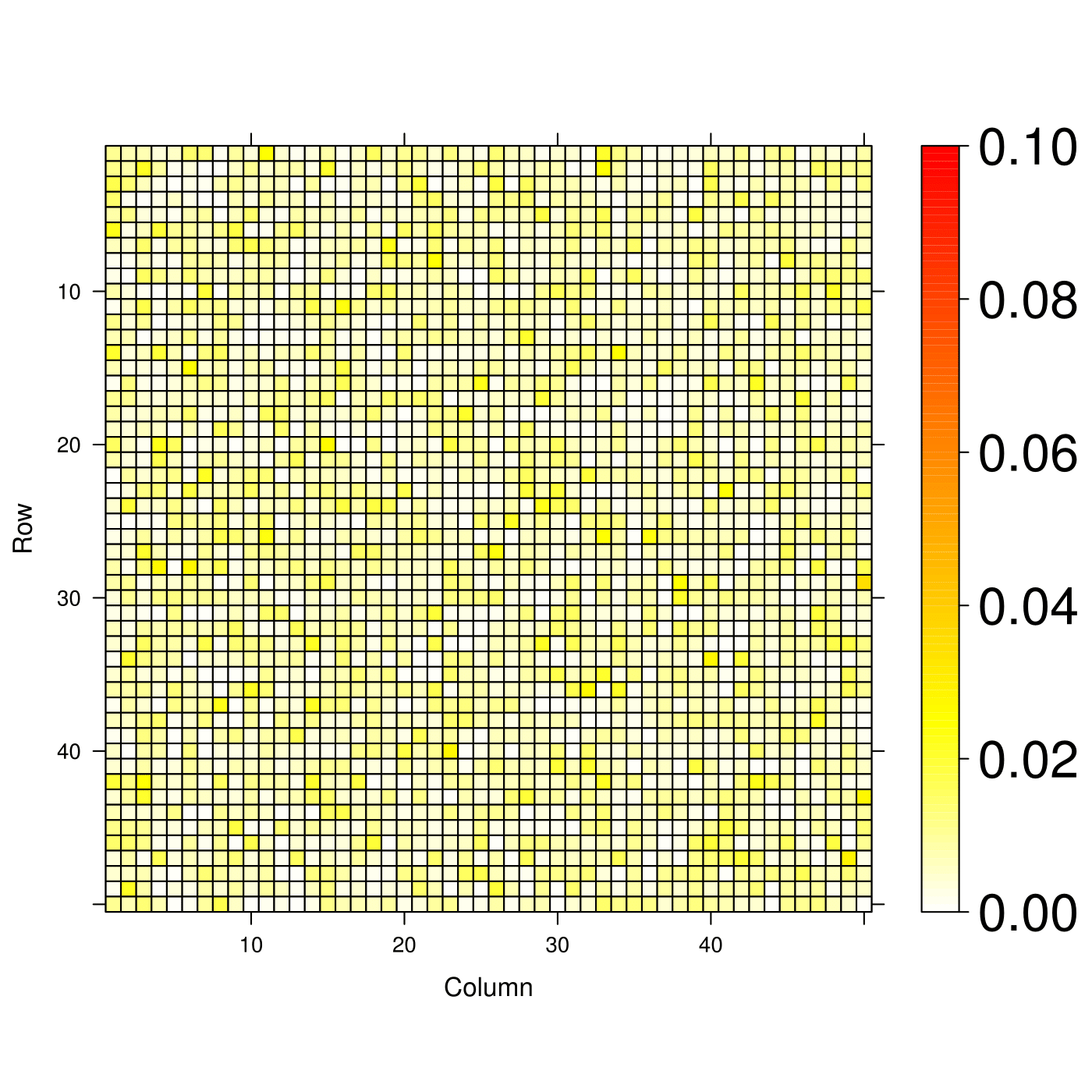} \\
		(a) \enquote{Scale-free} type VAR matrix &(b) Absolute difference of LSE\\
		\includegraphics[width=0.4 \textwidth]{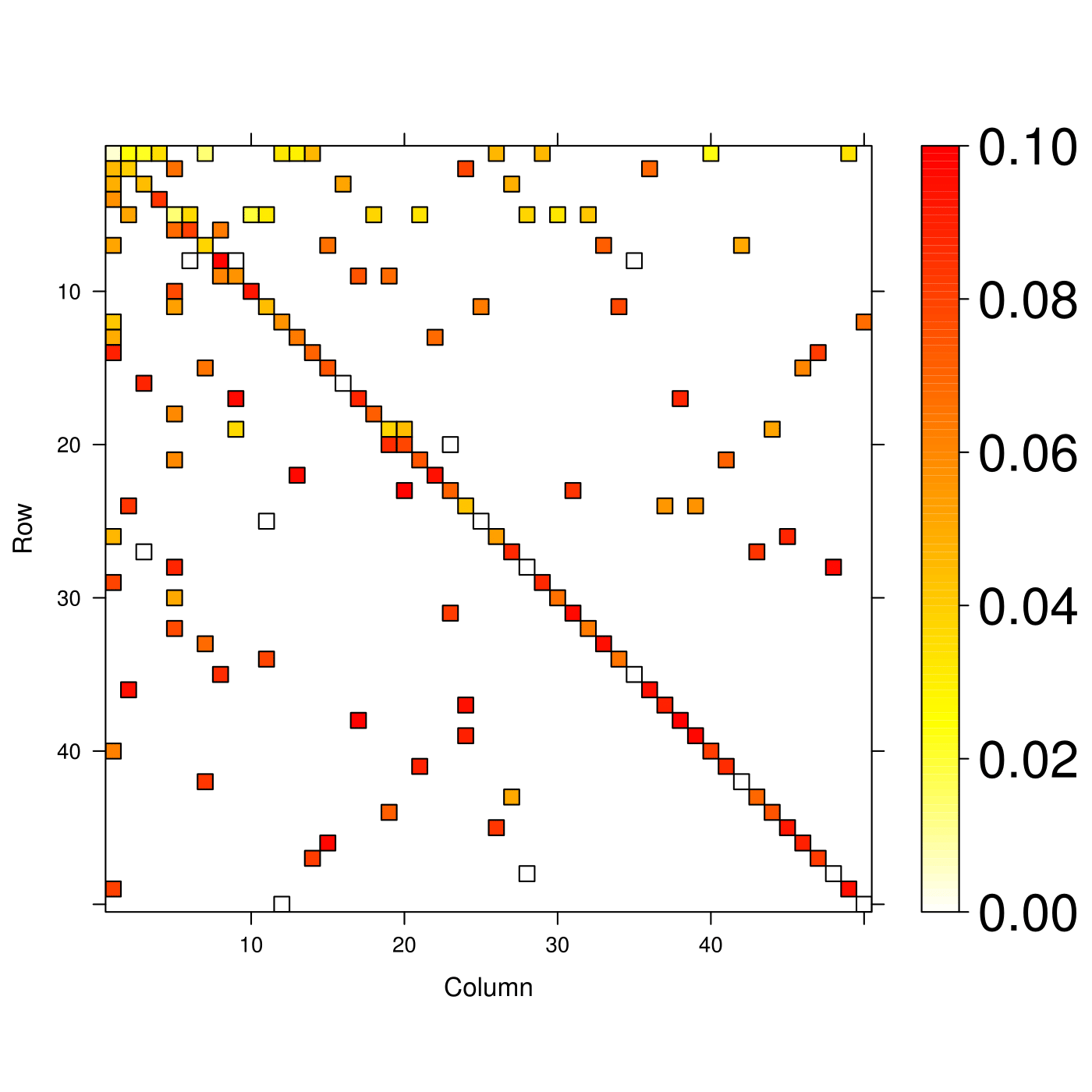} &
		\includegraphics[width=0.4 \textwidth]{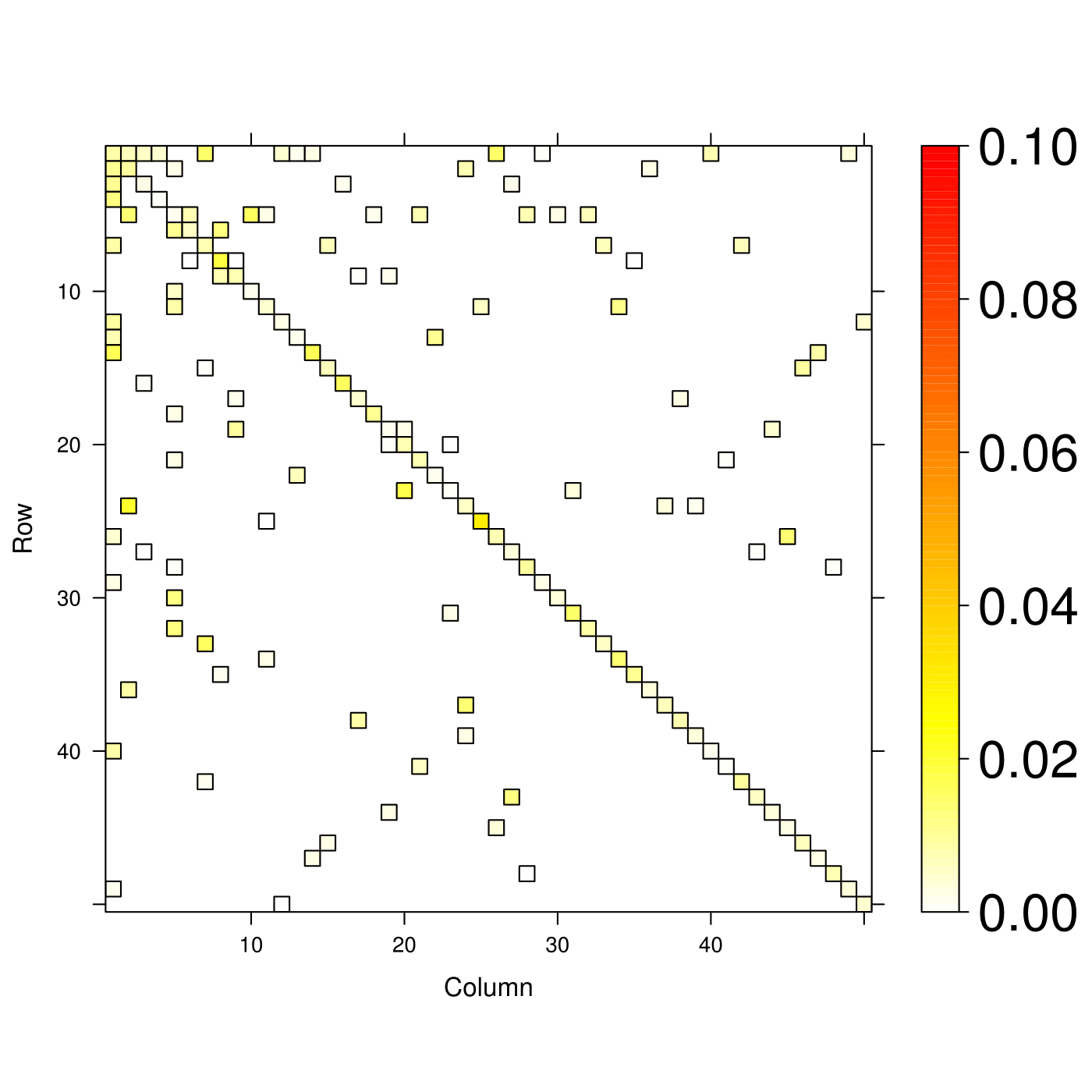} \\
		(c) Absolute difference of LASSO &(d) Absolute difference of LASSLE
	\end{tabular}
	\caption{Comparison of specificity of true zero on \enquote{Scale-free} type data with student's $t$-noise. Figure \ref{fig:specificity_scale_t_noise}(a) demonstrates the true VAR(1) coefficient matrix with $P=50$. Figure \ref{fig:specificity_scale_t_noise}(b),(c),(d) give the absolute difference between true matrix and estimated matrix by LSE, LASSO and LASSLE method respectively.}
	\label{fig:specificity_scale_t_noise}
\end{figure}

We use $P$ independent $\sqrt{0.06}*t(5)$, of which mean equals to 0 and variance equals to 0.1, to generate the student's $t$-noise at time $t$. Figure \ref{fig:specificity_cluster_t_noise} and \ref{fig:specificity_scale_t_noise} demonstrate the performance comparison of three methods on \enquote{Cluster} type and \enquote{Scale-free} type data regarding their specificity of true zero under setting \{$P=50$, $d=1$\}. The visualization results imply that LASSLE can capture the true zero coefficient substantially better than LSE. In addition, LASSLE has much lower absolute difference on non-zero coefficients compared to LASSO.

Table \ref{t_compare_cluster} and \ref{t_compare_scale} list the MSE results of all three methods on both \enquote{Cluster} type and \enquote{Scale-free} type data with student's $t$-noise under different VAR parameter setting. We can see that LASSLE method still has overwhelming advantage over LSE and LASSO when the number of parameter is large enough ($\geq 2,500$). On the other hand, LSE has slightly lower MSE than LASSLE under low-dimensional parameter setting.

\begin{table}[H]\centering
	\small
	\begin{tabular}{|c| c| c| c| c| c|}\hline\hline
		\multicolumn{3}{|c|}{VAR Parameter Setting} & \multicolumn{3}{c|}{MSE $\times 10^{-3}$} \\
		\hline
		Number & $P$ & $d$ & LSE & LASSO & LASSLE \\
		\hline
		100 & 10 & 1 & 7 * & 184 & 28 \\
		\hline
		500 & 10 & 5 & 57 * & 603 & 73 \\
		\hline
		1,000 & 10 & 10 & 147 * & 741 & 216 \\
		\hline
		2,500 & 50 & 1 & 175 & 836 & 37 * \\
		\hline
		5,000 & 50 & 2 & 455 & 1242 & 98 * \\
		\hline
		10,000& 100 & 1 & 686 & 1726 & 88 * \\
		\hline
		\end{tabular}
	\caption {Comparison of MSE for \enquote{Cluster} type data with student's $t$-noise} 
	\label{t_compare_cluster}
\end{table}

\begin{table}[H]\centering
	\small
	\begin{tabular}{|c| c| c| c| c| c|}\hline\hline
		\multicolumn{3}{|c|}{VAR Parameter Setting} & \multicolumn{3}{c|}{MSE $\times 10^{-3}$} \\
		\hline
		Number & $P$ & $d$ & LSE & LASSO & LASSLE \\
		\hline
		100 & 10 & 1 & 6 * & 134 & 11 \\
		\hline
		500 & 10 & 5 & 44 & 278 & 12 * \\
		\hline
		1,000 & 10 & 10 & 124 * & 728 & 217 \\
		\hline
		2,500 & 50 & 1 & 181 & 778 & 50 * \\
		\hline
		5,000 & 50 & 2 & 450 & 1056 & 24 * \\
		\hline
		10,000& 100 & 1 & 747 & 1641 & 160 * \\
		\hline
	\end{tabular}
	\caption {Comparison of MSE for \enquote{Scale-free} type data with student's $t$-noise} 
	\label{t_compare_scale}
	
\end{table}

\subsubsection{Results from shifted zero-mean $\chi^2$ noise}
\begin{figure}[H]\centering
	\begin{tabular}{cc}
		\includegraphics[width=0.4 \textwidth]{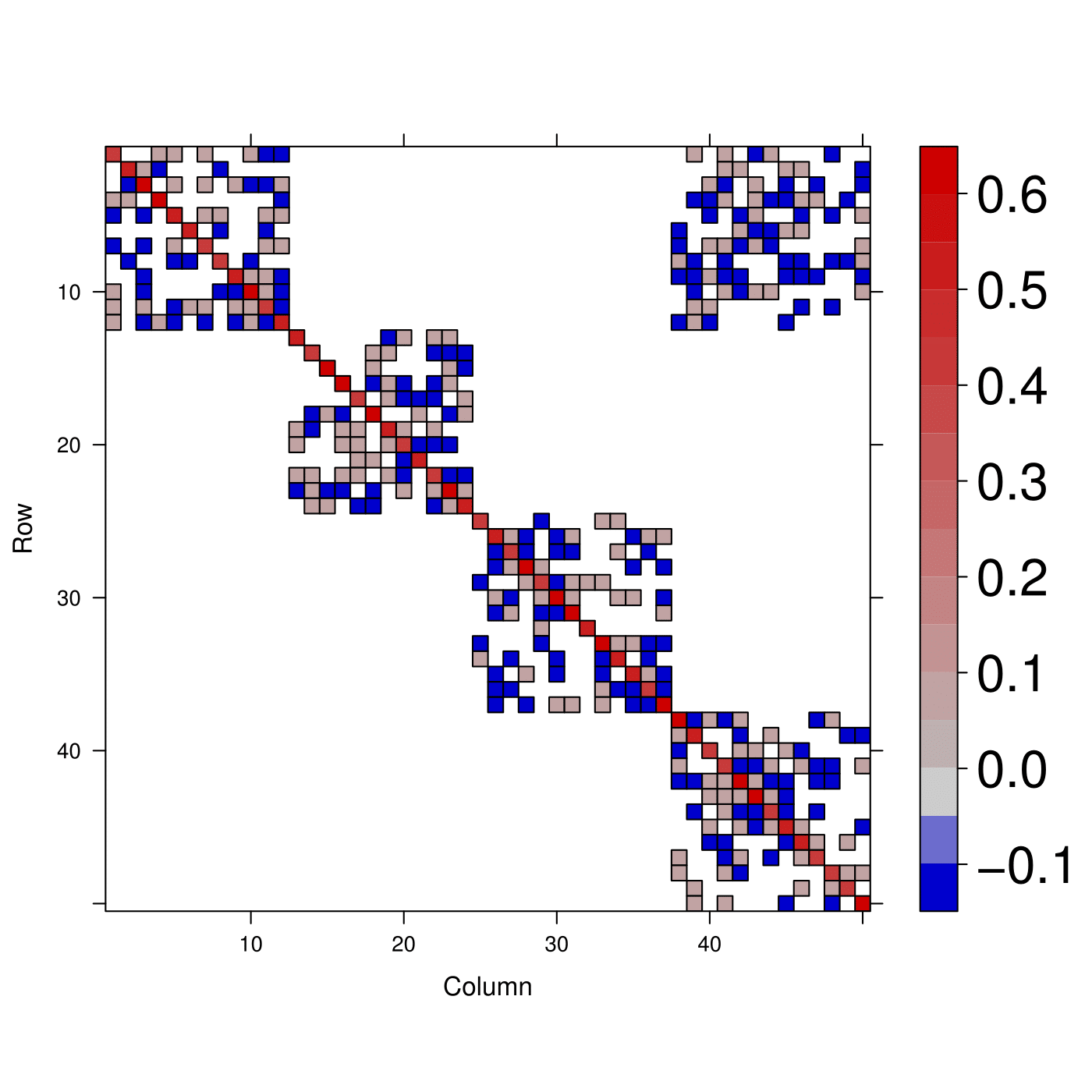}&
		\includegraphics[width=0.4 \textwidth]{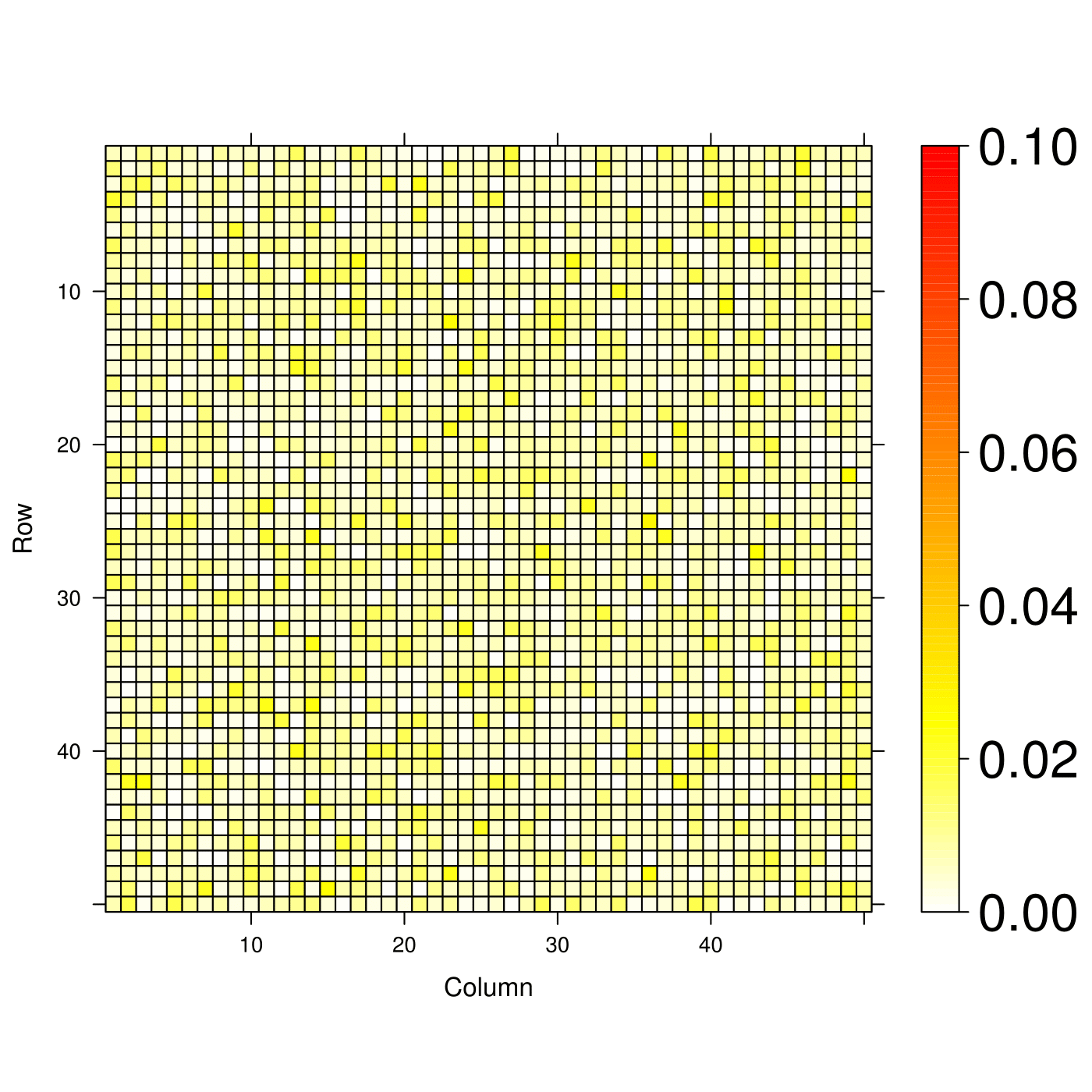} \\
		(a) \enquote{Cluster} type VAR matrix &(b) Absolute difference of LSE\\
		\includegraphics[width=0.4 \textwidth]{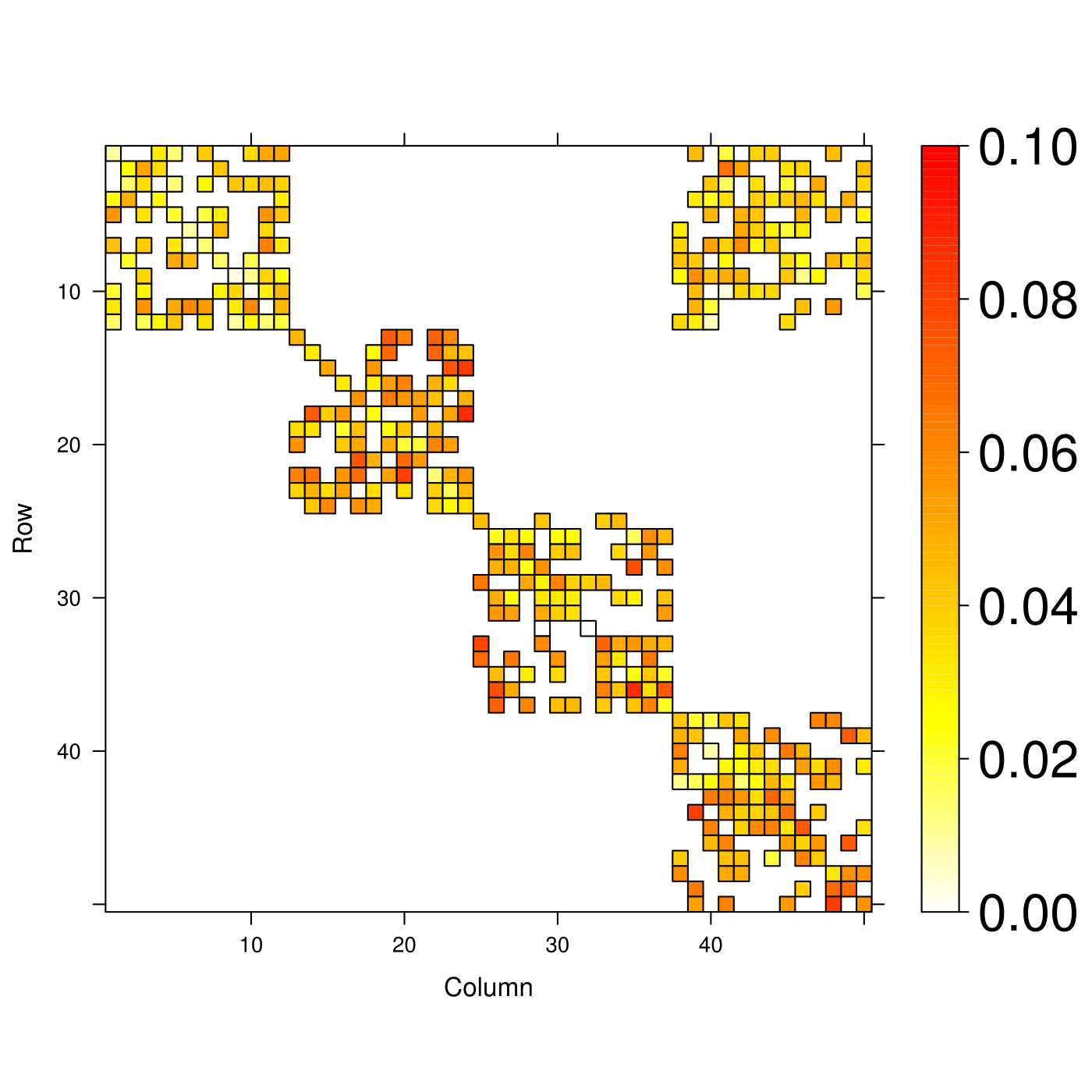} &
		\includegraphics[width=0.4 \textwidth]{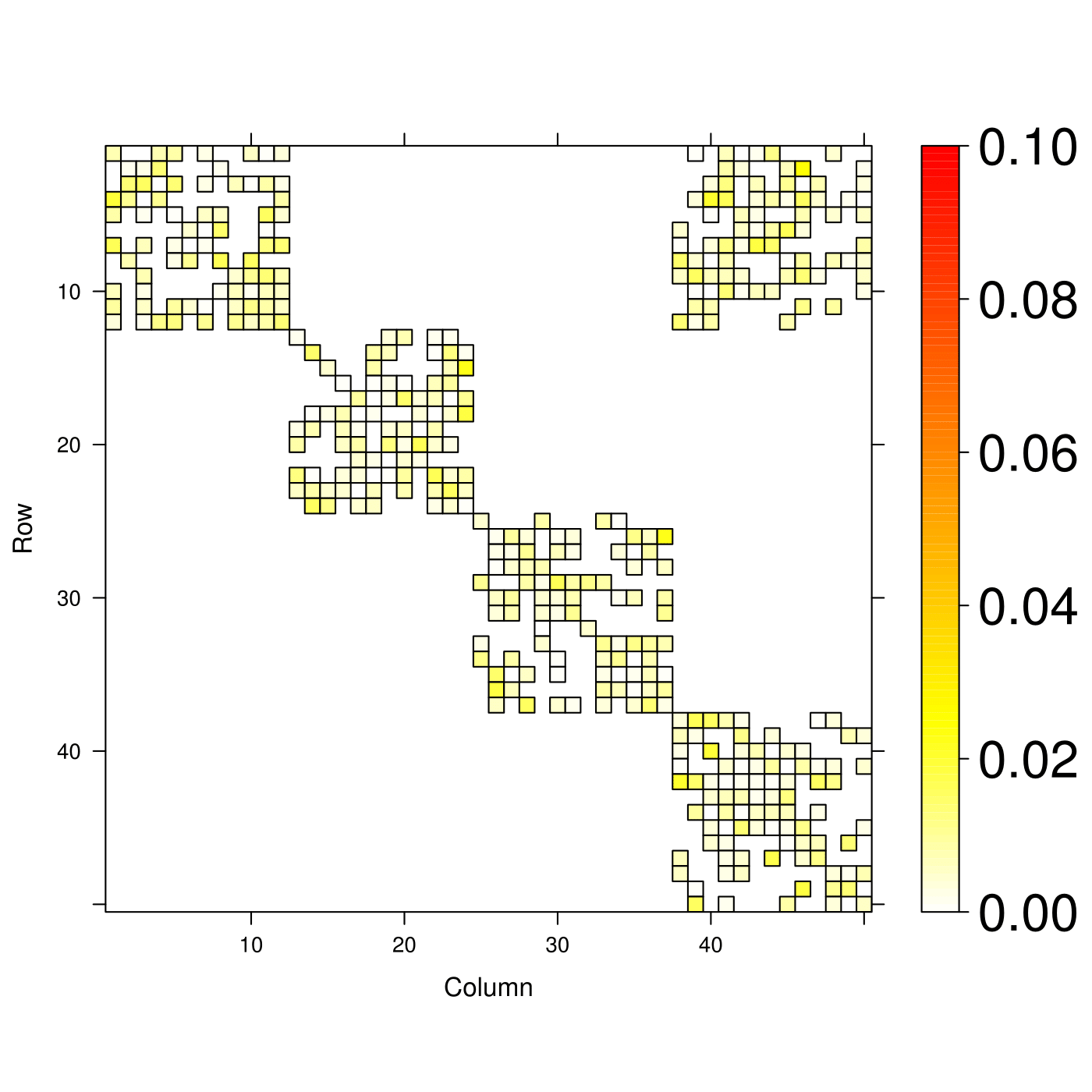} \\
		(c) Absolute difference of LASSO &(d) Absolute difference of LASSLE
	\end{tabular}
	\caption{Comparison of specificity of true zero on \enquote{Cluster} type data with shifted zero-mean $\chi^2$ noise. Figure \ref{fig:specificity_cluster_chi_noise}(a) demonstrates the true VAR(1) coefficient matrix with $P=50$. Figure \ref{fig:specificity_cluster_chi_noise}(b),(c),(d) give the absolute difference between true matrix and estimated matrix by LSE, LASSO and LASSLE method respectively.}
	\label{fig:specificity_cluster_chi_noise}
\end{figure}

\begin{figure}[H]\centering
	\begin{tabular}{cc}
		\includegraphics[width=0.4 \textwidth]{phi_scale50}&
		\includegraphics[width=0.4 \textwidth]{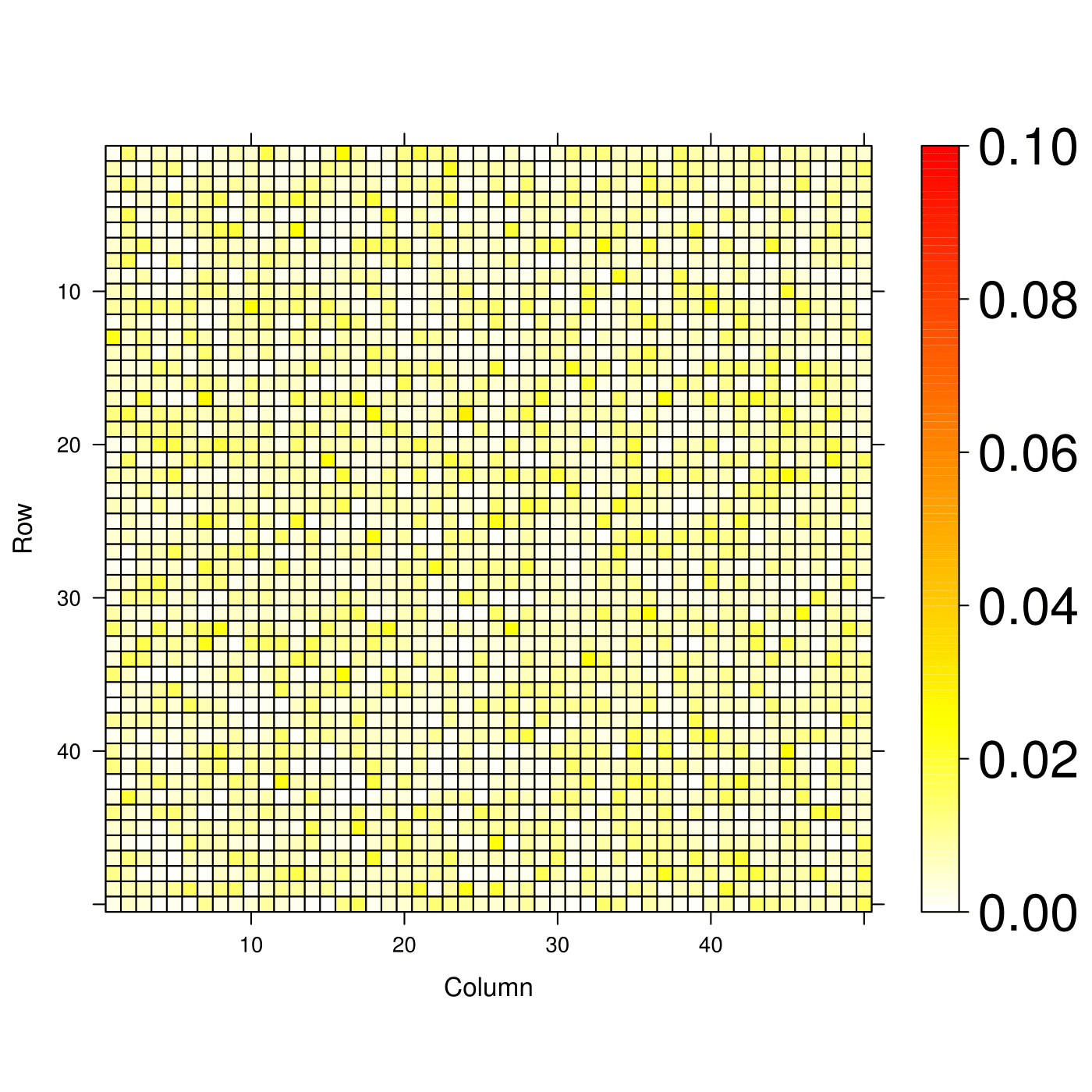} \\
		(a) \enquote{Scale-free} type VAR matrix &(b) Absolute difference of LSE\\
		\includegraphics[width=0.4 \textwidth]{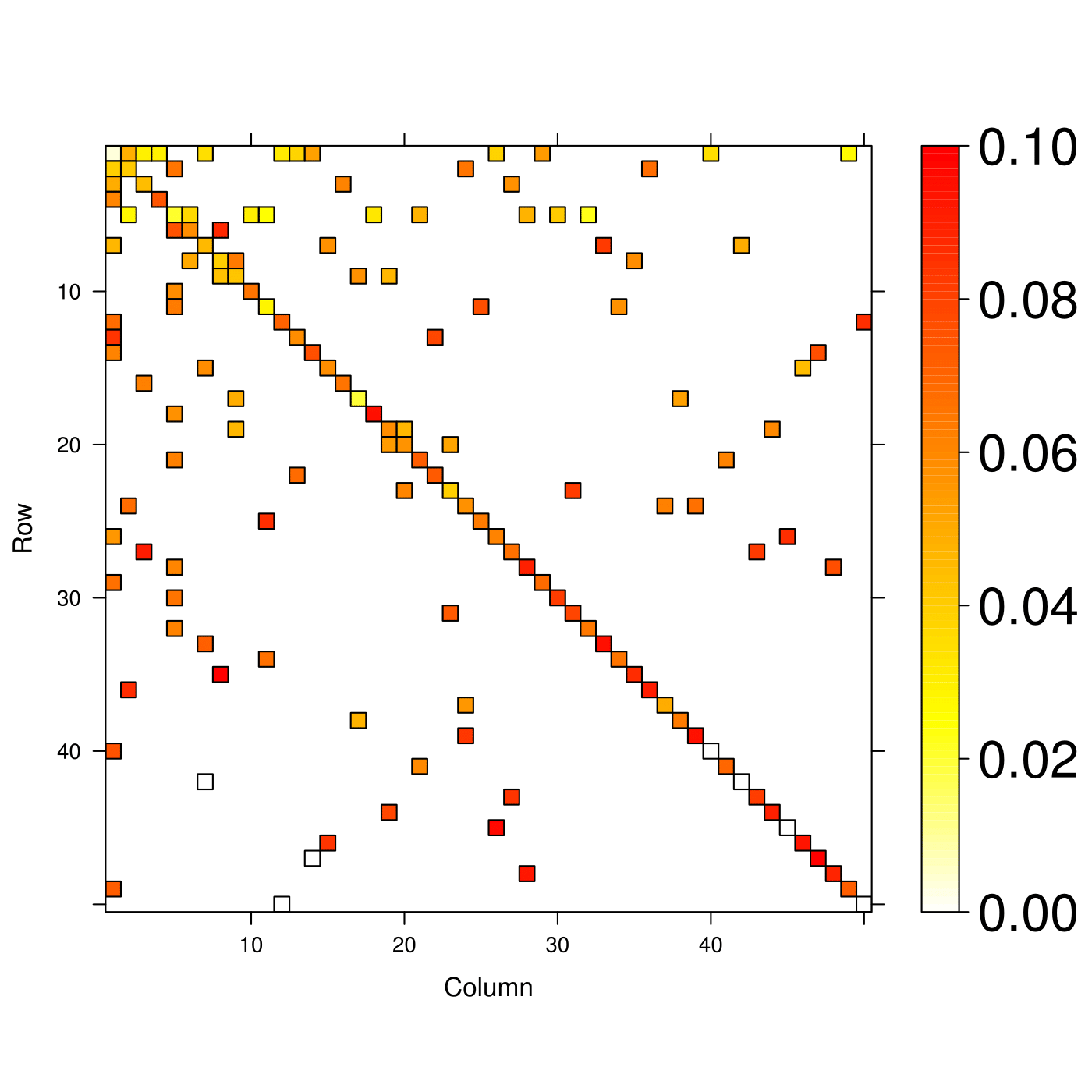} &
		\includegraphics[width=0.4 \textwidth]{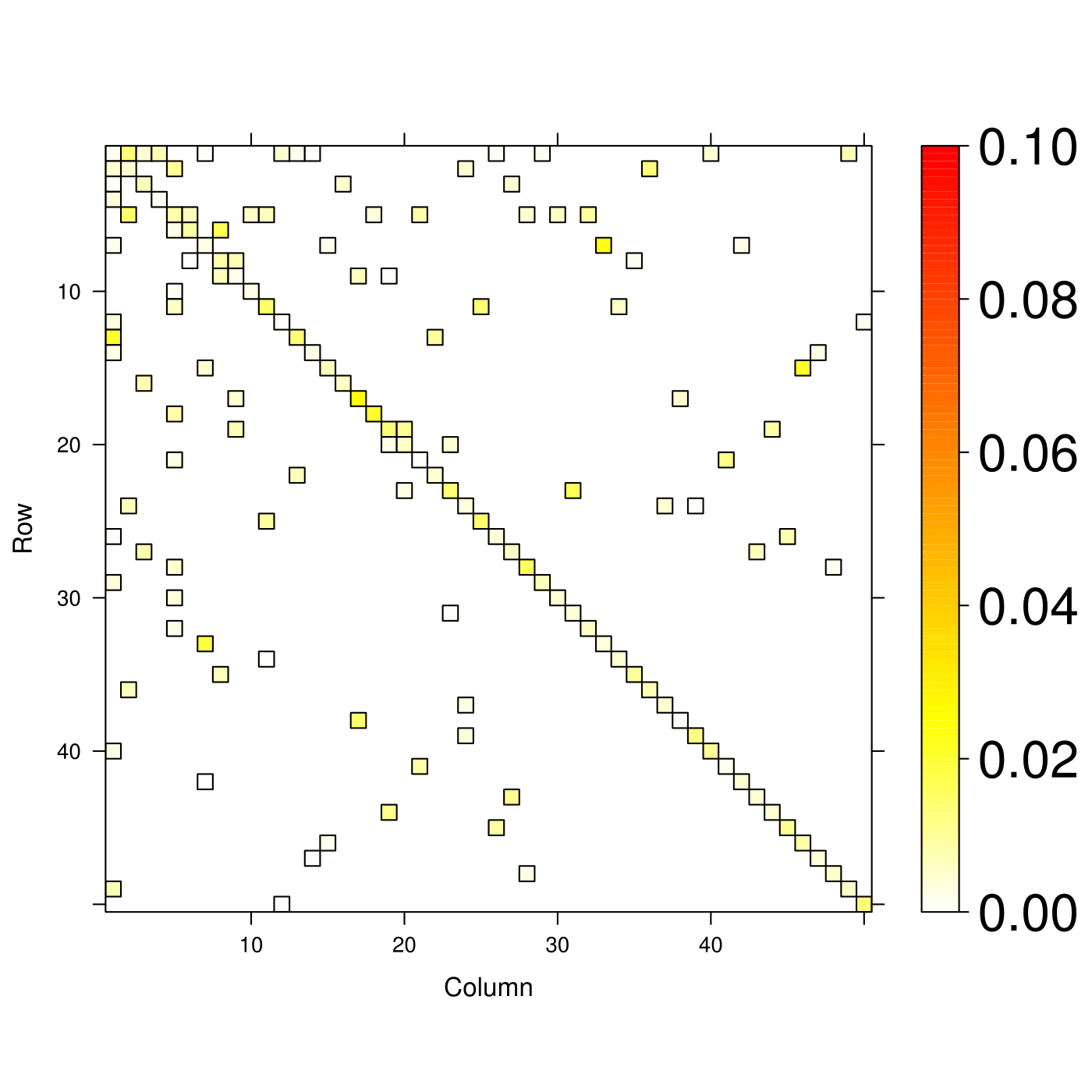} \\
		(c) Absolute difference of LASSO &(d) Absolute difference of LASSLE
	\end{tabular}
	\caption{Comparison of specificity of true zero on \enquote{Scale-free} type data with shifted zero-mean $\chi^2$ noise. Figure \ref{fig:specificity_scale_chi_noise}(a) demonstrates the true VAR(1) coefficient matrix with $P=50$. Figure \ref{fig:specificity_scale_chi_noise}(b),(c),(d) give the absolute difference between true matrix and estimated matrix by LSE, LASSO and LASSLE method respectively.}
	\label{fig:specificity_scale_chi_noise}
\end{figure}

To generate $P$-dimensional shifted $\chi^2$ noise at time $t$, we employ $P$ independent $\sqrt{0.0125}*\chi^2_4-\sqrt{0.2}$, with mean of 0 and variance of 0.1. Figure \ref{fig:specificity_cluster_chi_noise} and \ref{fig:specificity_scale_chi_noise} demonstrate the comparison results of three methods on \enquote{Cluster} type and \enquote{Scale-free} type data in terms of their specificity of true zero under setting \{$P=50$, $d=1$\}. Based on the visualization results, we can see that LASSLE estimate has very good specificity of true zero coefficients regardless of $\chi^2$ noise.

Table \ref{chi_compare_cluster} and \ref{chi_compare_scale} list the MSE results of all three methods on both \enquote{Cluster} type and \enquote{Scale-free} type data with shifted $\chi^2$ noise under different VAR parameter setting. It implies that LASSLE method has significantly lower MSE under most high dimension parameter setting than LSE and LASSO.

\begin{table}[H]\centering
	\small
	\begin{tabular}{|c| c| c| c| c| c|}\hline\hline
		\multicolumn{3}{|c|}{VAR Parameter Setting} & \multicolumn{3}{c|}{MSE $\times 10^{-3}$} \\
		\hline
		Number & $P$ & $d$ & LSE & LASSO & LASSLE \\
		\hline
		100 & 10 & 1 & 7 * & 159 & 14 \\
		\hline
		500 & 10 & 5 & 56 & 465 & 41 * \\
		\hline
		1,000 & 10 & 10 & 139 & 561 & 136 * \\
		\hline
		2,500 & 50 & 1 & 182 & 666 & 31 * \\
		\hline
		5,000 & 50 & 2 & 453 & 1039 & 75 * \\
		\hline
		10,000& 100 & 1 & 696 & 1572 & 87 * \\
		\hline
	\end{tabular}
\caption {Comparison of MSE for \enquote{Cluster} type data with shifted zero-mean $\chi^2$ noise} 
\label{chi_compare_cluster}
\end{table}		
		
\begin{table}[H]\centering
	\small
	\begin{tabular}{|c| c| c| c| c| c|}\hline\hline
		\multicolumn{3}{|c|}{VAR Parameter Setting} & \multicolumn{3}{c|}{MSE $\times 10^{-3}$} \\
		\hline
		Number & $P$ & $d$ & LSE & LASSO & LASSLE \\
		\hline
		100 & 10 & 1 & 8 & 143 & 3 * \\
		\hline
		500 & 10 & 5 & 55 & 224 & 16 * \\
		\hline
		1,000 & 10 & 10 & 129 * & 654 & 153 \\
		\hline
		2,500 & 50 & 1 & 194 & 663 & 22 * \\
		\hline
		5,000 & 50 & 2 & 451 & 967 & 23 * \\
		\hline
		10,000& 100 & 1 & 747 & 1291 & 40 * \\
		\hline
	\end{tabular}
	\caption {Comparison of MSE for \enquote{Scale-free} type data with shifted zero-mean $\chi^2$ noise} 
	\label{chi_compare_scale}
\end{table}

%% file: Application_revise.tex
\section{Application to effective connectivity in multichannel LFPs}
In this section, we will fit a VAR model to LFP data recorded from multiple electrodes as rats perform a non-spatial sequence memory task (\cite{allen2016nonspatial}) and apply the LASSLE method to estimate the VAR parameters and consequently partial directed coherence. Our objective is to examine and quantify potential connectivity (i,.e., effective) among electrodes located in hippocampal region CA1, which is clinically meaningful as this form of sequence memory shows strong behavioral parallels in rats and humans (\cite{allen2014sequence}), and depends on the hippocampus for both species (\cite{fortin2016distinct}, \cite{boucquey2015cross}), and is impaired in normal aging (\cite{allen2015memory}) .
\subsection{Data description}
\begin{figure}[H]\centering
\includegraphics[width = 0.5\textwidth]{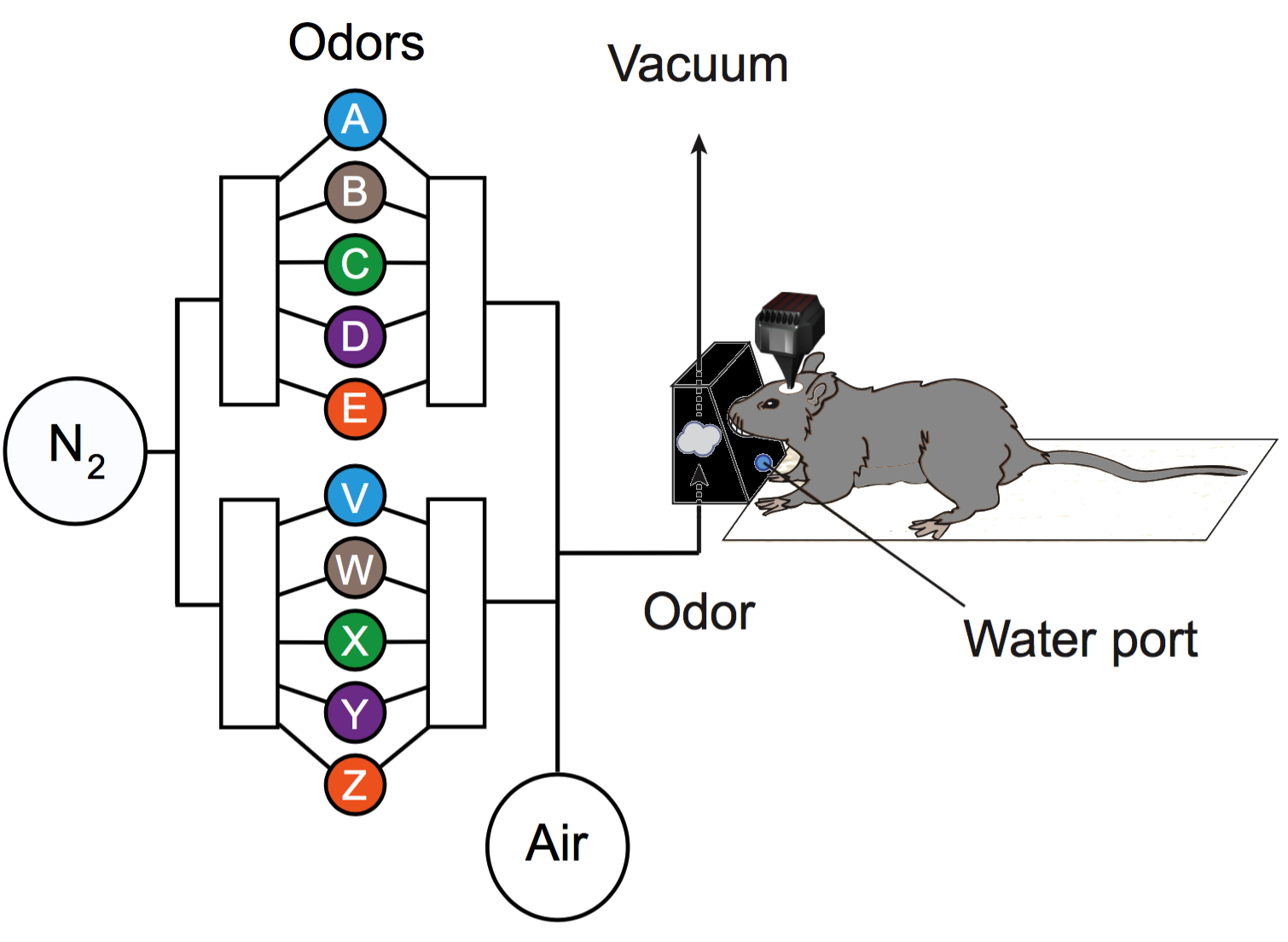}\\
\caption{A non-spatial sequence memory experiment in rats. Rats were presented with repeated sequences of five odors (A,B,C,D and E) in a single odor port. Each odor presentation was initiated by a nose poke and rats were required to correctly identify the odor as either InSeq (ABCDE) by holding their nose poke until the signal or OutSeq (e.g.,ABDDE) by withdrawing their nose poke before the signal.}
\label{rat_experiment}
\end{figure}

In the experiment (shown in Figure \ref{rat_experiment}), rats were presented with repeated sequences of five odors in a single odor port. They were trained to identify whether each odor was presented "in sequence" (by holding their nose poke until the signal) or "out of sequence" (by withdrawing their nose poke before the signal) to receive a water reward. The LFP data included here was recorded from CA1 electrodes during a session in which a well-trained rat performed the task at a high level (\cite{allen2016nonspatial}).

\begin{figure}[H]\centering
	\includegraphics[width = 0.8\textwidth, height = 0.6\textwidth]{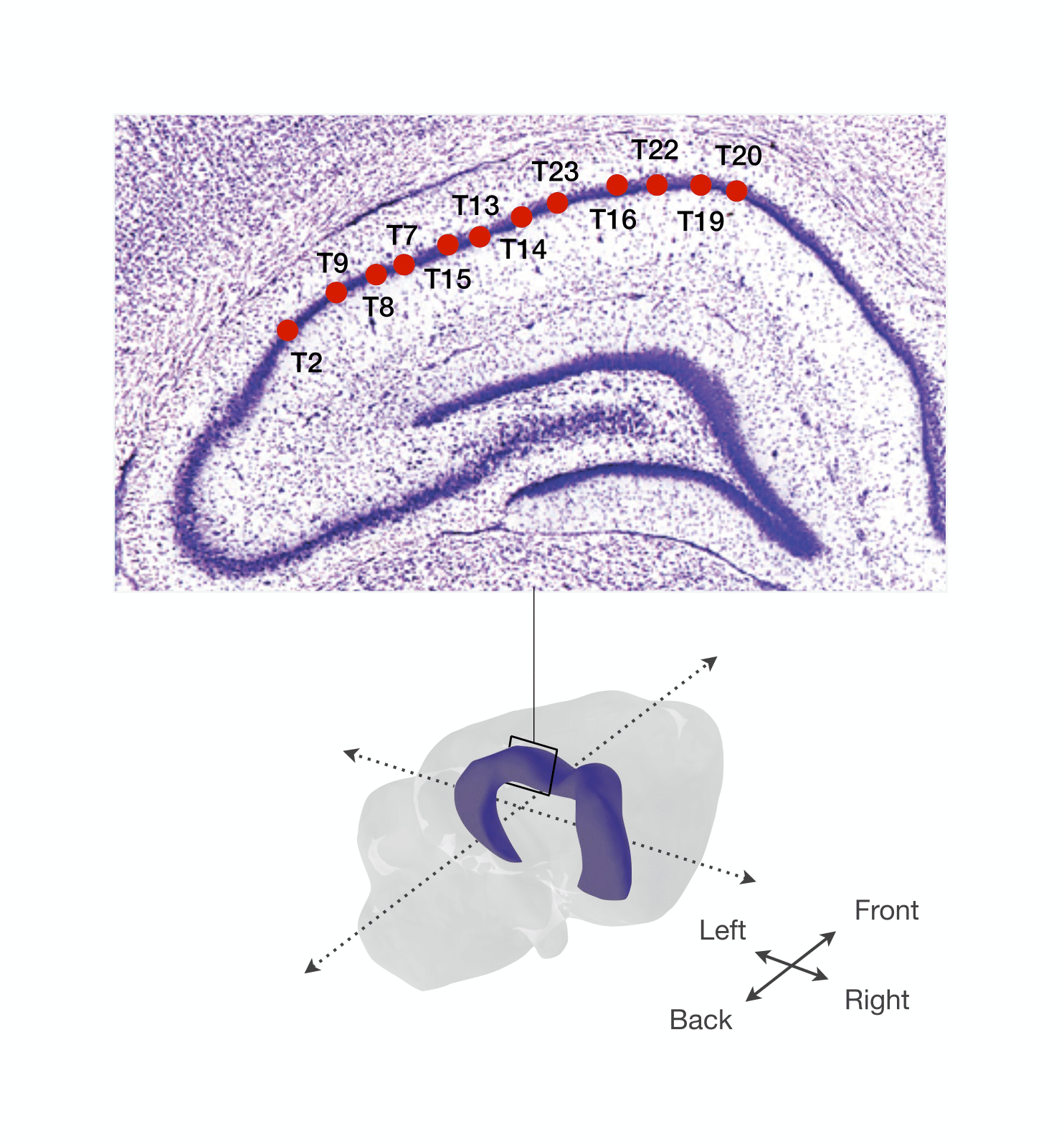}\\
	\caption{Estimated location within the hippocampus (dorsal CA1 region) of subset of 12 tetrodes included in the analyses.}
	\label{tetrode_location}
\end{figure}

\begin{figure}[H]\centering
	\includegraphics[width = 0.8\textwidth, height = 0.6\textwidth]{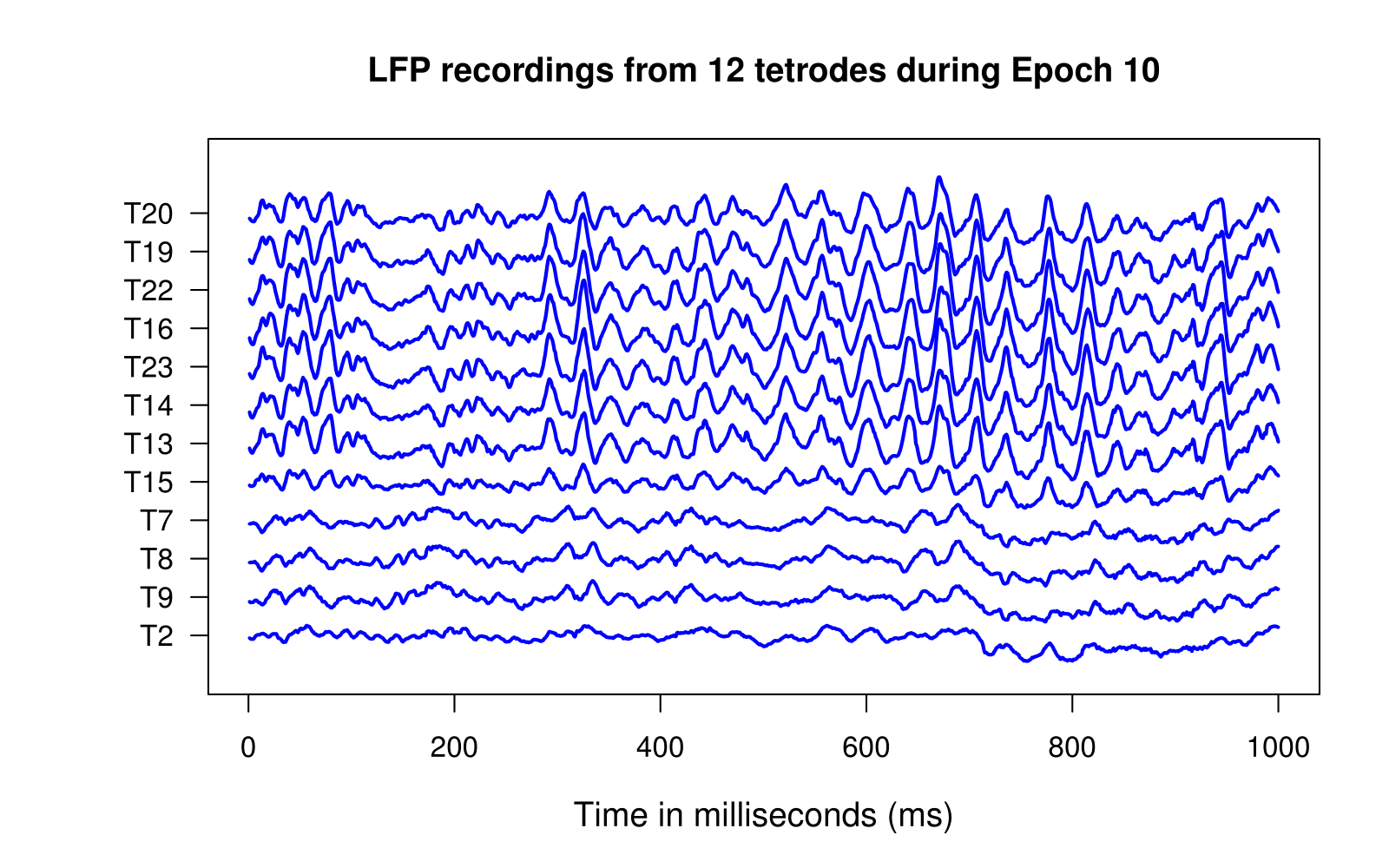}\\
	\caption{LFPs from 12 tetrodes studied in this paper during Epoch 10. These LFPs have temporal patterns that can be separated into two clusters. The first consists of T7, T8, T9 and T2 which are all on the posterior (back) portion of the dorsal CA1 region. The second consists of the remaining channels which are all on the anterior portion (front).}
	\label{lfp_timeseries}
\end{figure}

The full dataset includes LFPs from 23 tetrodes located in the hippocampus and $n = 247$ epochs, where $n = 219$ epochs are "in sequence" (InSeq) and $n = 28$ epochs are "out of sequence" (OutSeq). Each epoch is recorded roughly 1 second with sampling frequency of 1000 Hz and thus has $T = 1000$ time points. We specifically focused our analyses on LFPs from $P=12$ tetrodes, a subset of electrodes that also recorded clear single-cell spiking activity and were confirmed to be located in the pyramidal layer of CA1 (see estimated tetrode locations in Figure \ref{tetrode_location}). In addition, LFPs of Epoch 10 can be found in Figure \ref{lfp_timeseries}. We observe that time series of LFPs from tetrode T13 to tetrode T23 have highly similar temporal pattern, while time series of the remaining tetrodes are highly similar. This is because tetrodes near each other are likely to behave more similarly than those that are far apart. 

\begin{figure}[H]\centering
\begin{tabular}{cc}
\includegraphics[width=0.5\textwidth,height=0.4\textwidth]{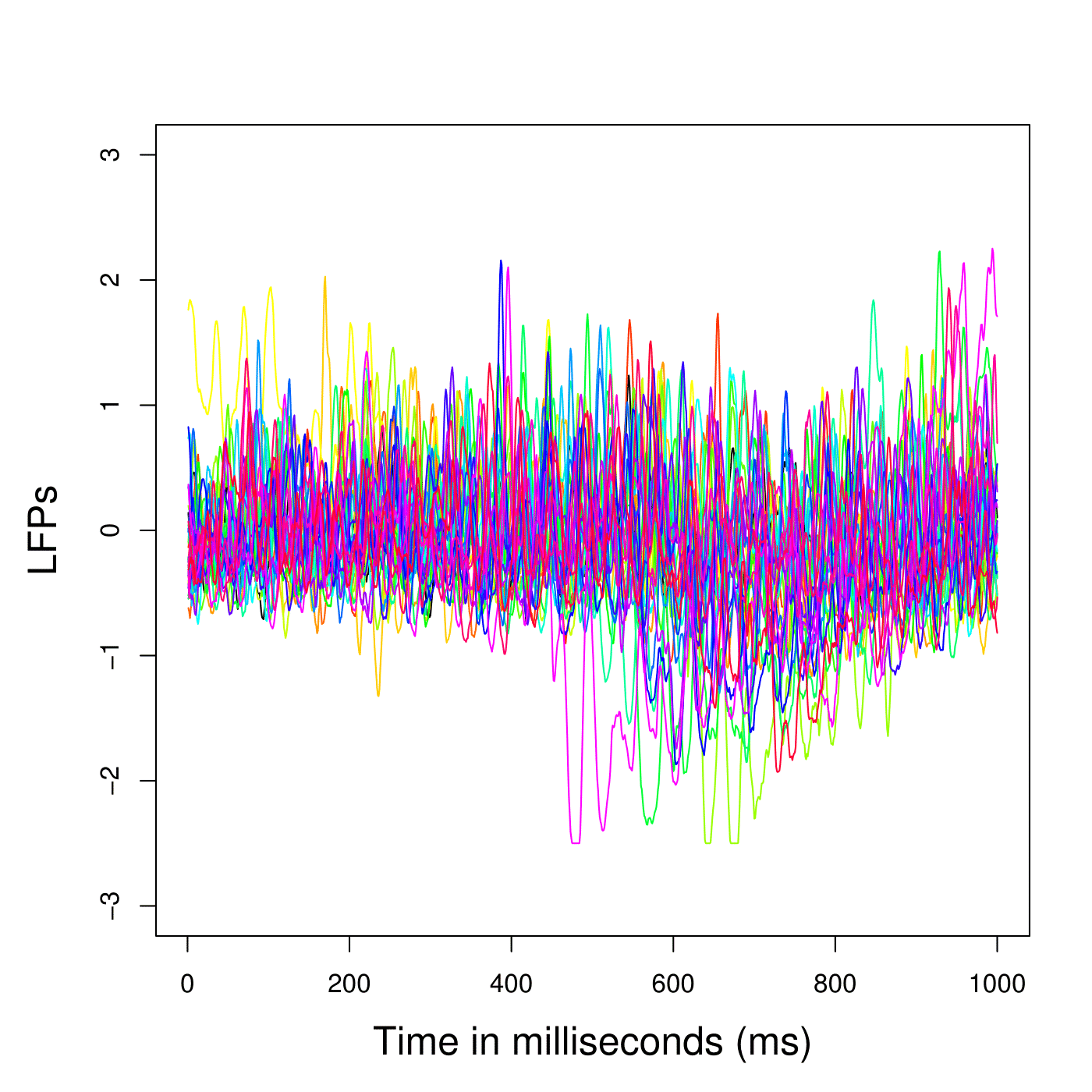} &
\includegraphics[width=0.5\textwidth,height=0.4\textwidth]{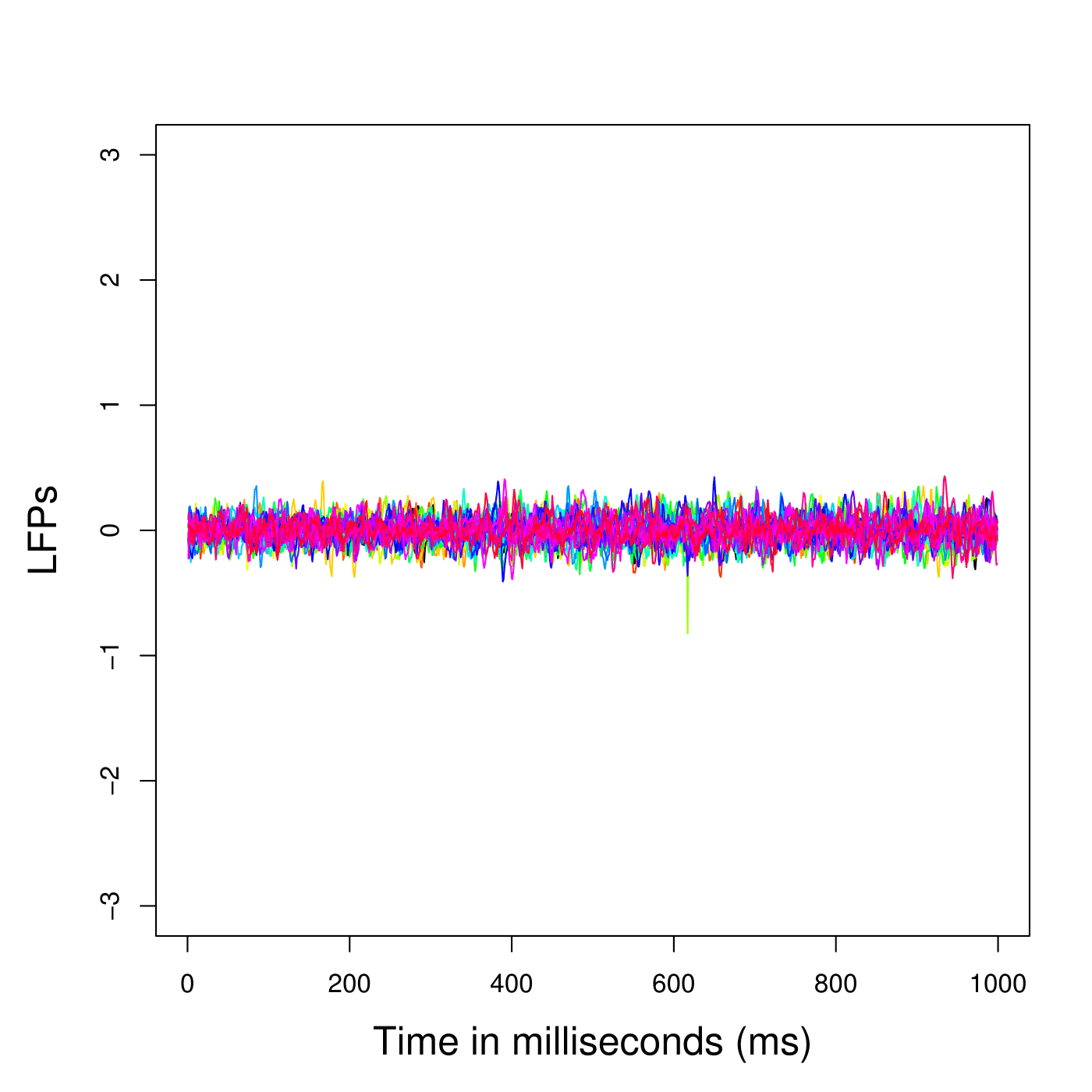}\\
\includegraphics[width=0.5\textwidth,height=0.4\textwidth]{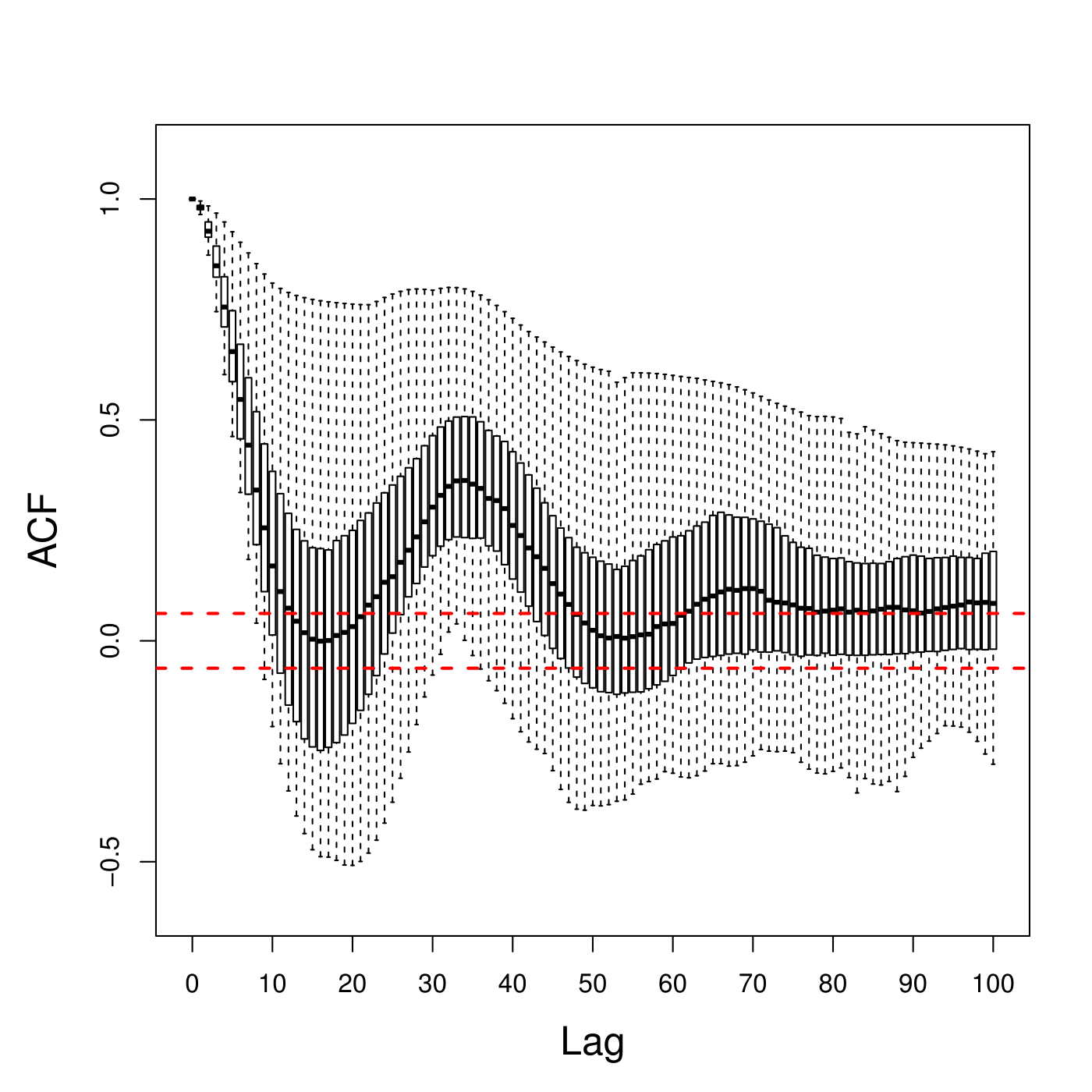}&
\includegraphics[width=0.5\textwidth,height=0.4\textwidth]{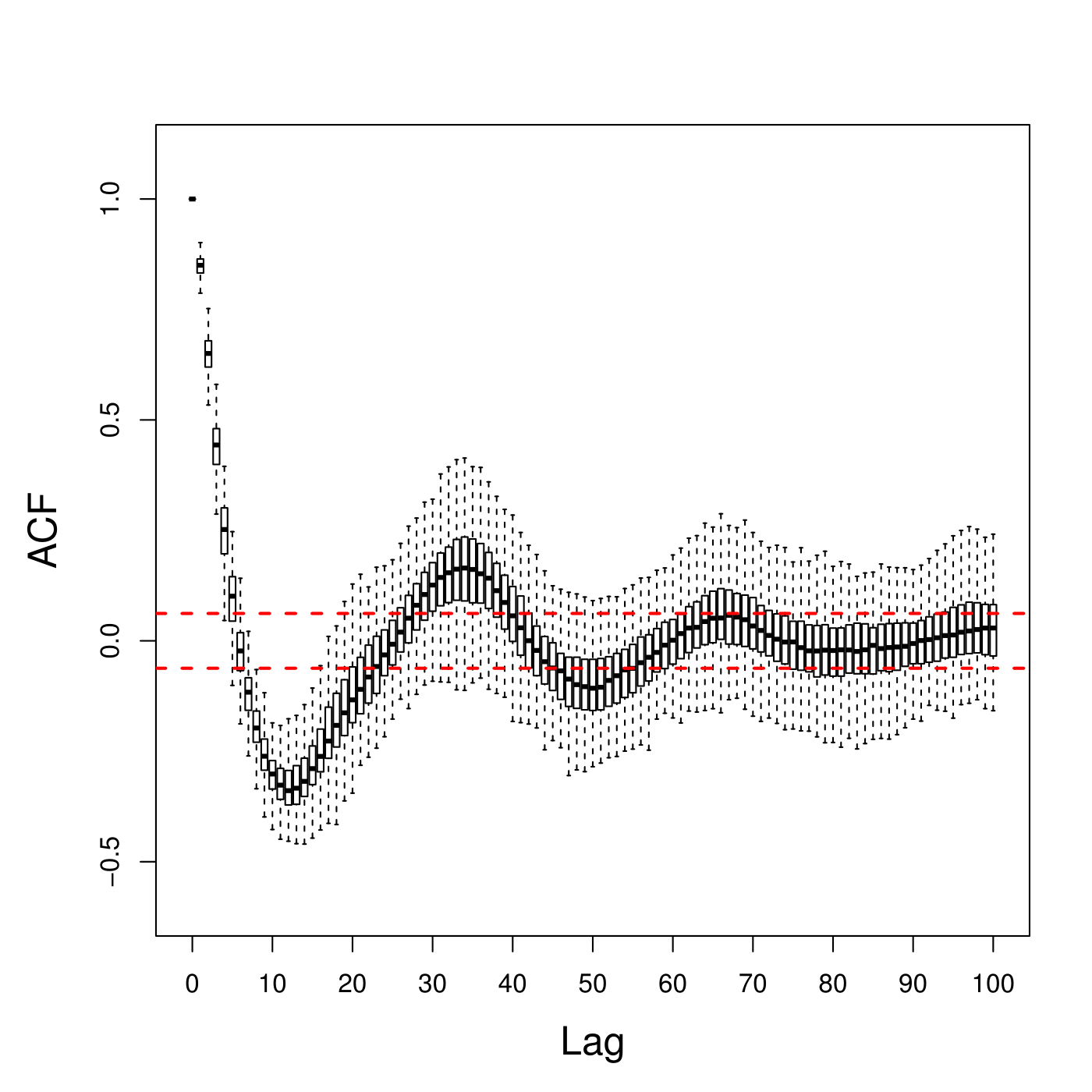}\\
\includegraphics[width=0.5\textwidth,height=0.4\textwidth]{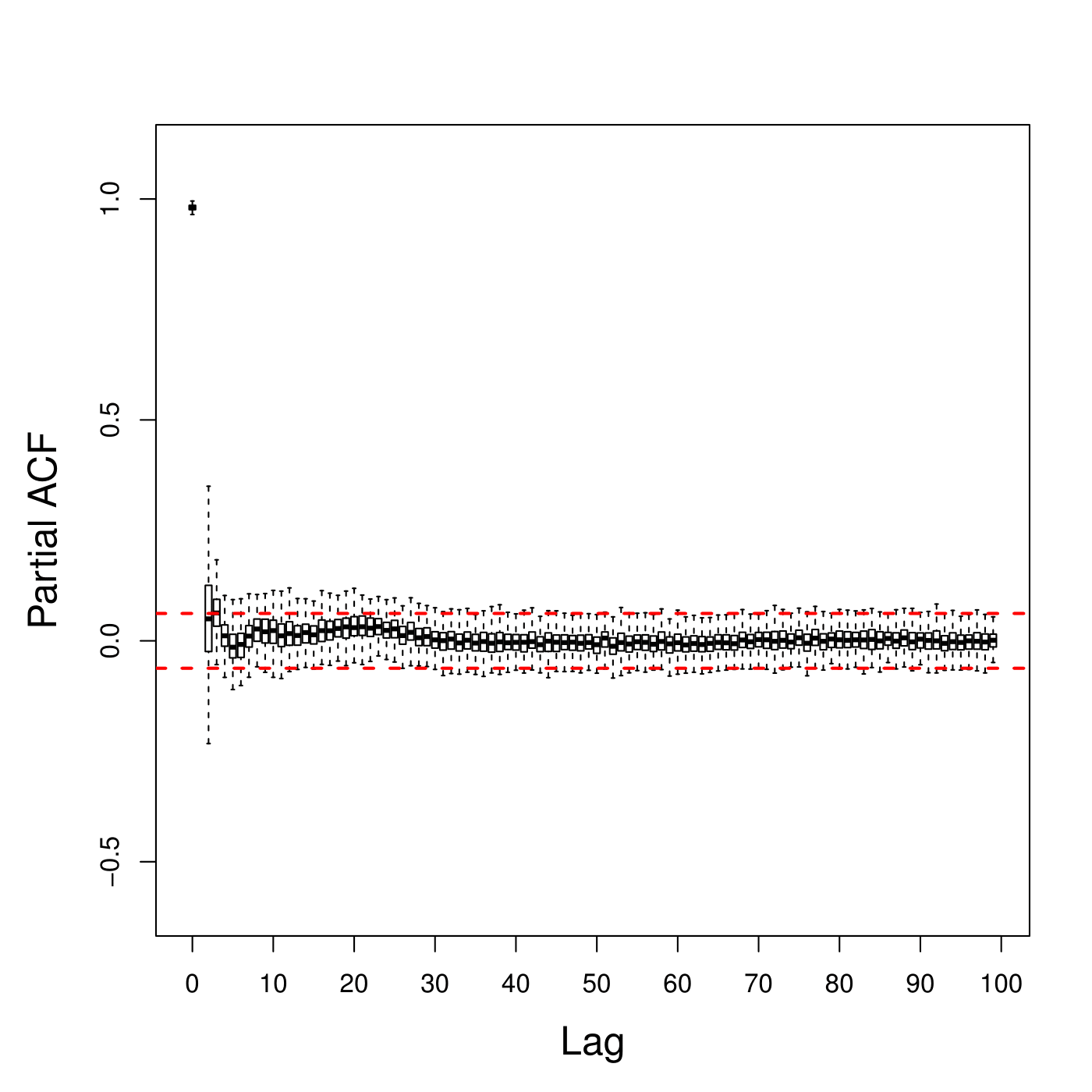}&
\includegraphics[width=0.5\textwidth,height=0.4\textwidth]{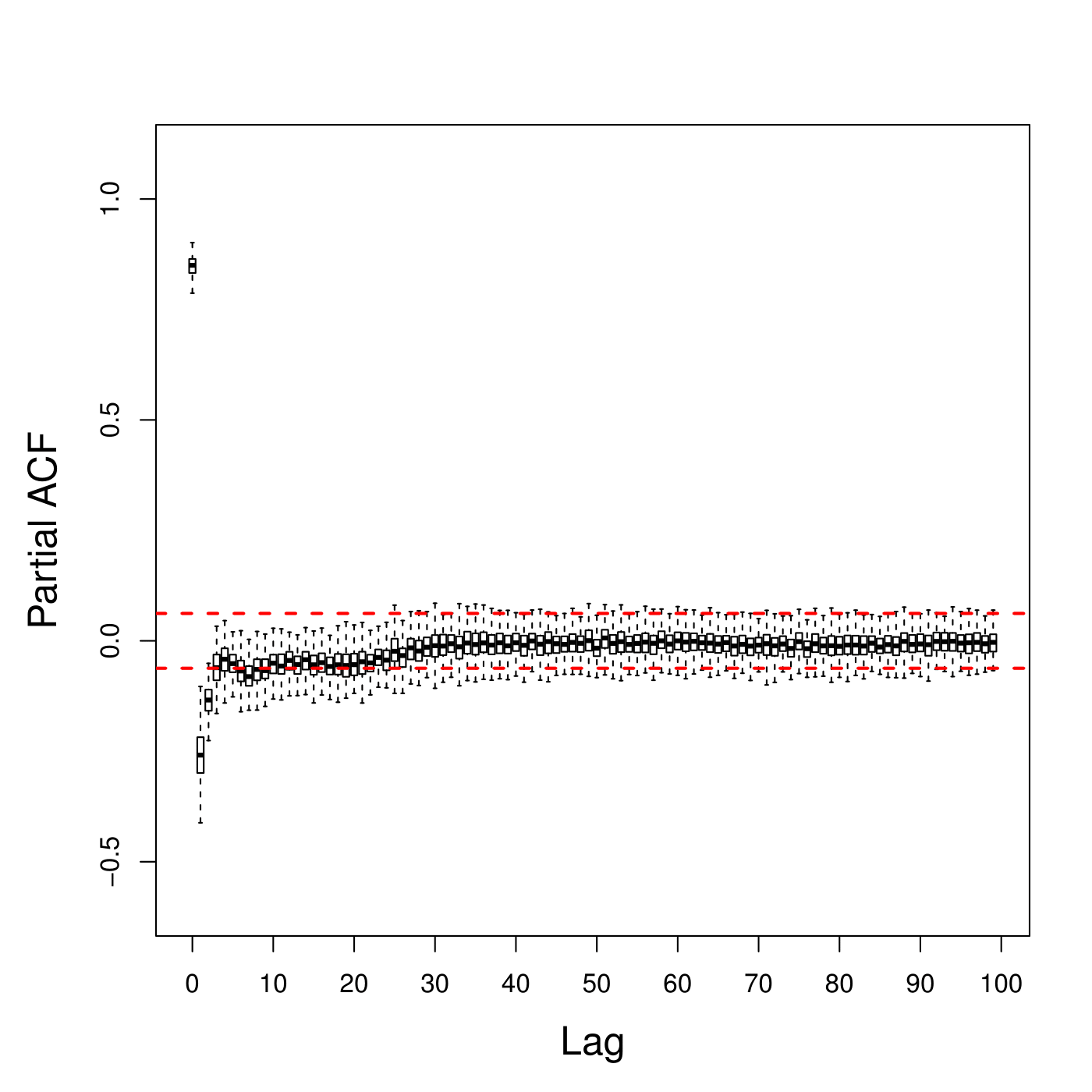}\\
\end{tabular}
\caption{Top: The LFPs time series plots of the first 30 epochs at tetrode T22 before (left) and after (right) processing. Middle: The boxplots of auto-correlation function (ACF) from tertrode T22 before (left) and after (right) processing across all epochs. Bottom: The boxplots of partial auto-correlation function (PACF) from tetrode T22 before (left) and after (right) processing across all epochs.}
\label{pre-and-post}
\end{figure}

Figure \ref{pre-and-post} (left) displays the LFPs from tetrode T22 during the first 30 epochs; the boxplots of its auto-correlation function (ACF) across all 247 epochs; and the boxplots of the partial auto-correlation function (PACF) across all epochs. We observe that the boxplots of ACF fail to drop to zero even after very long lags 
and there is a cyclical behavior in the pattern. Both of these could be evidence of non-stationarity (or long-memory). 
These suggest pre-processing the data by taking a first order difference. The results of LFPs after differencing are shown on the right side of Figure \ref{pre-and-post}. Compared to the previous plots, the ACF boxplots eventually decay to zero with smaller interquartile range, which means that the pre-processed data looks more stationary and the correlation drops to zero faster than the original LFPs. Therefore we will fit the VAR model to the first order differenced LFPs.

\subsection{Preliminary analysis of a single epoch}
We first demonstrate fitting the VAR model to to a single epoch (Epoch 10 in this example). To select the best lag order $\widehat{d}$, we fit a VAR($d_j$) model with candidate order $d_j \in \{1,2,...,12\}$ and use LSE to estimate the coefficient matrices. Then we apply Equation~(\ref{SSE}), (\ref{sigmahat}), and (\ref{aic}) to compute AIC for each candidate order $d_j$. For epoch 10, the best order (or the minimizer of AIC) was $\widehat{d} = 3$. Consequently there were 3 coefficient matrices (each of dimension 12 x 12) to estimate. 

\begin{figure}[H] \centering
	\begin{tabular}{ccc}
		\includegraphics[width=0.33\textwidth]{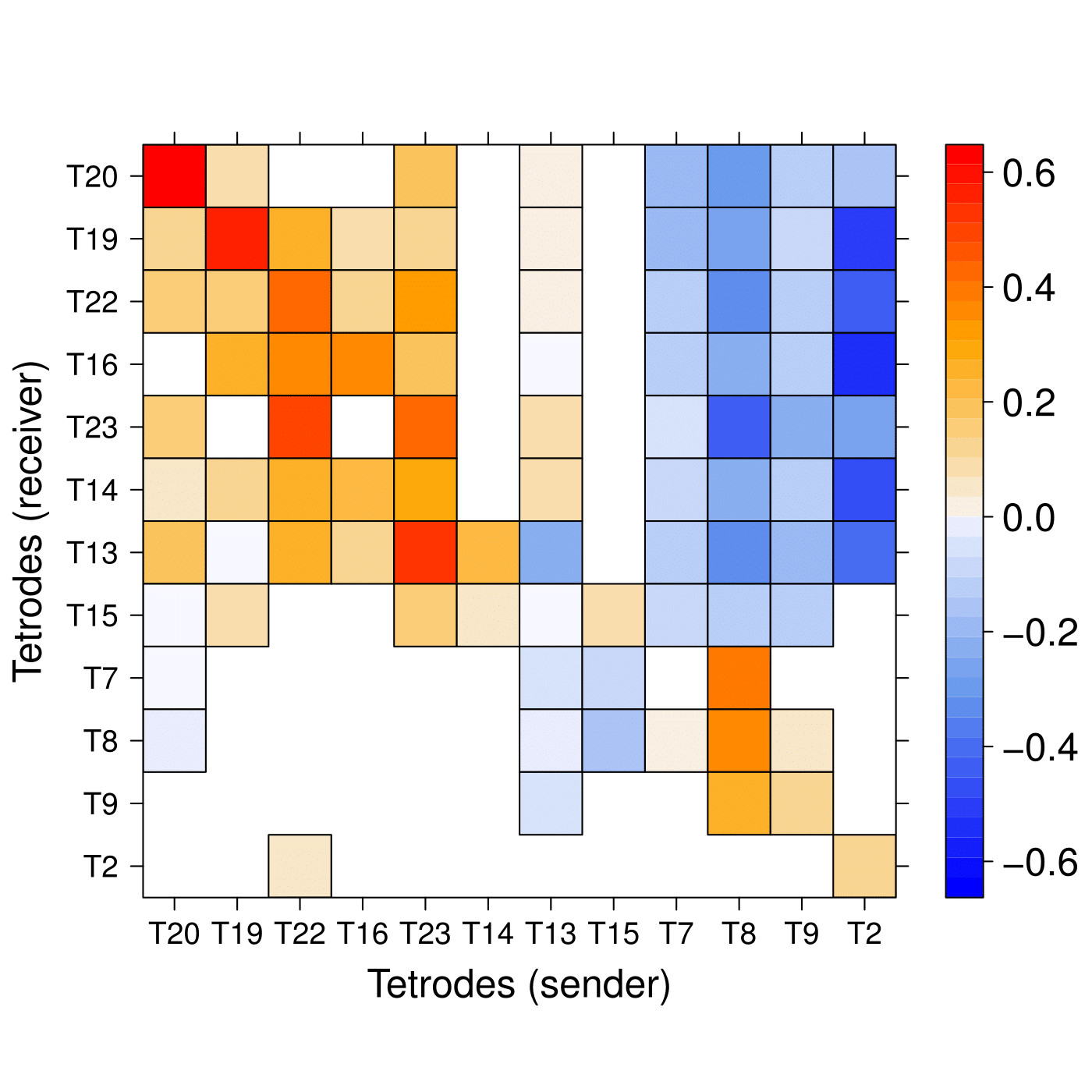} &
		\includegraphics[width=0.33\textwidth]{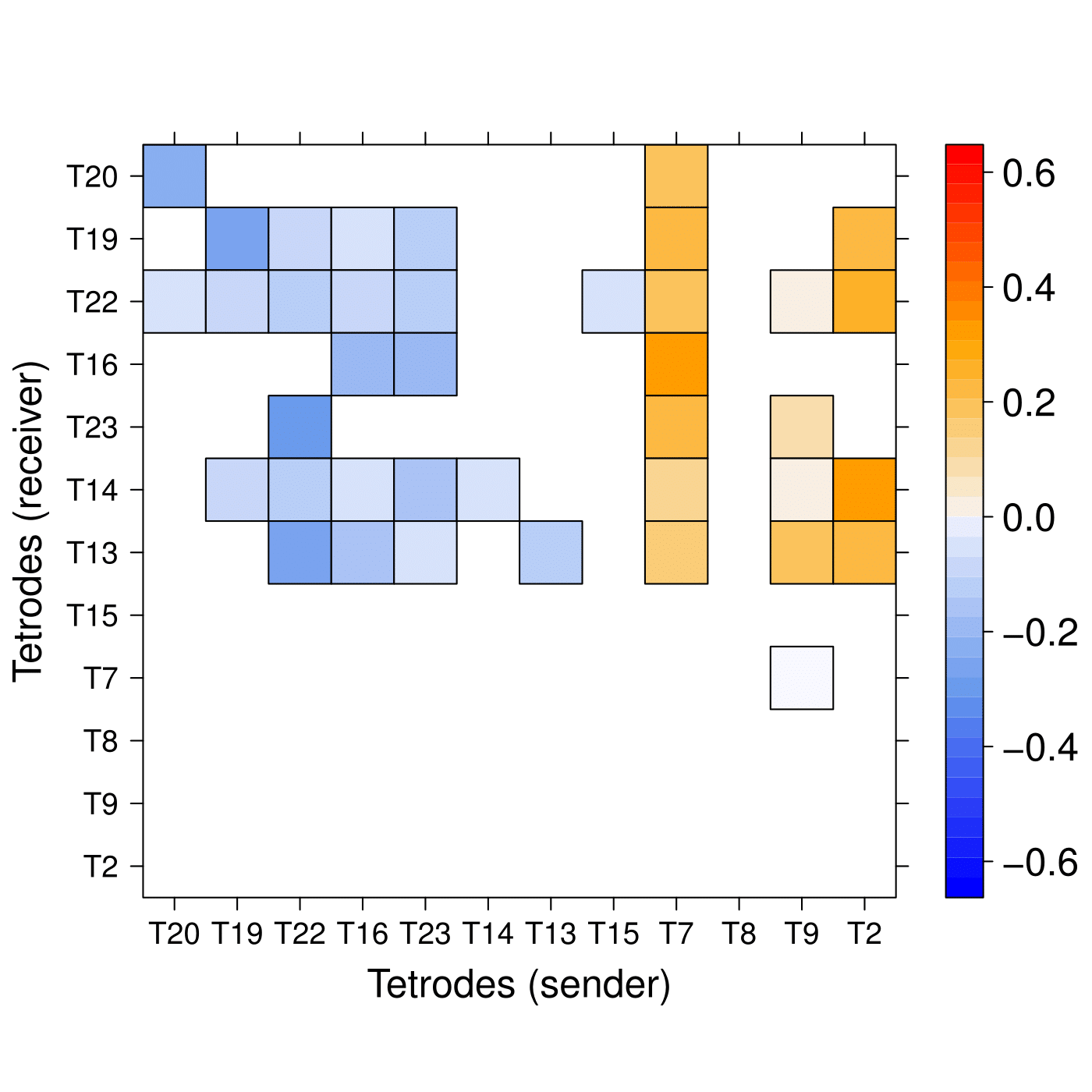} &
		\includegraphics[width=0.33\textwidth]{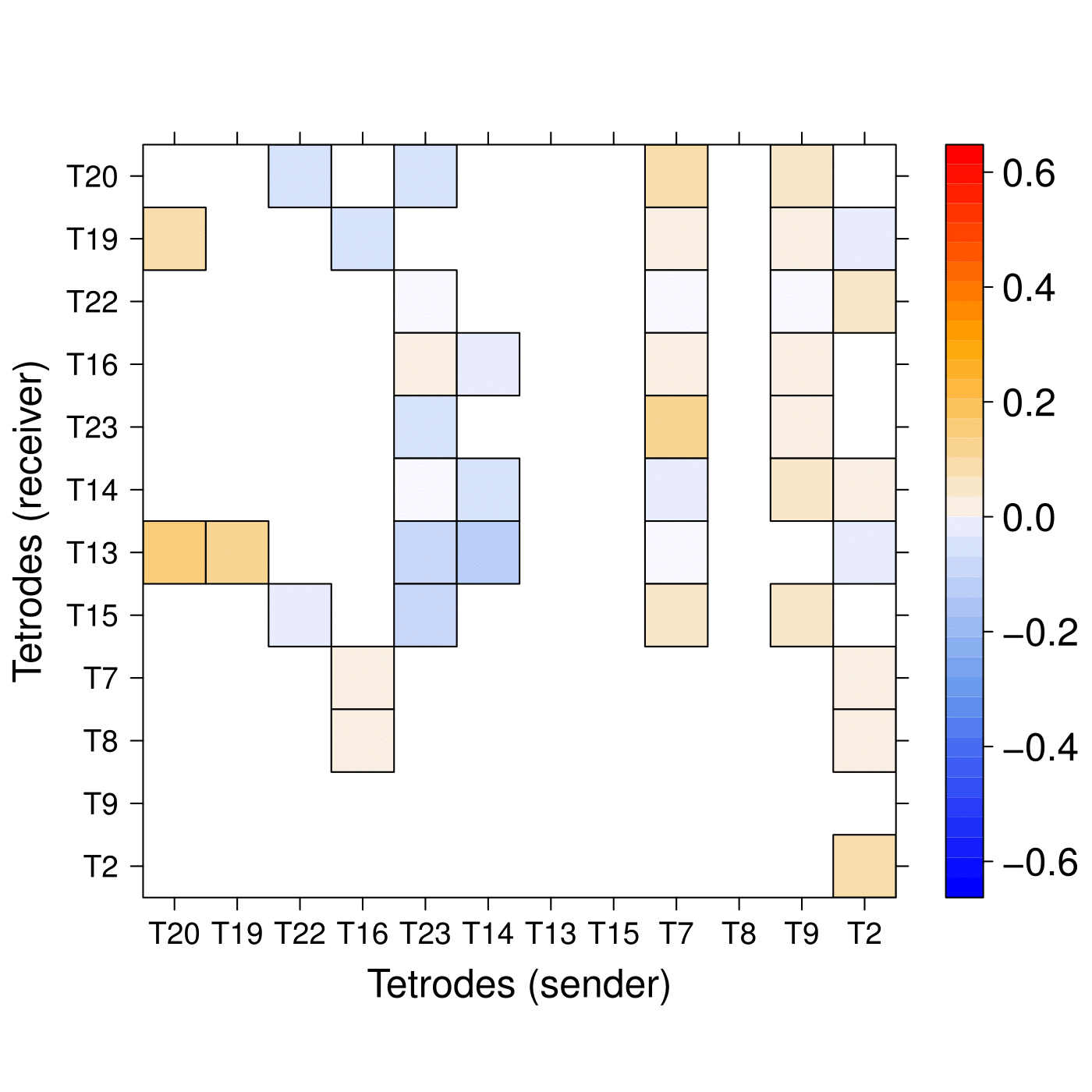}\\
		(a) Estimated $\Phi_1$ by LASSLE &(b) Estimated $\Phi_2$ by LASSLE &(c) Estimated $\Phi_3$ by LASSLE
	\end{tabular}
	\caption{Estimated coefficient matrices $\Phi_1$, $\Phi_2$, $\Phi_3$ in Epoch 10 using the LASSLE method.}
	\label{LASSLE_connectivity_matrices}
\end{figure}

Figure \ref{LASSLE_connectivity_matrices} shows the LASSLE estimates of $\Phi_1$, $\Phi_2$ and $\Phi_3$ for Epoch 10. Blanks are assigned to entries whose value is zero, so non-dependence between tetrodes is easy to tell. For entries whose value is not zero, we assign them with colors of red for positive value and blue for negative value, and the strength of dependence is implied by the color-key.  
As we can see, most diagonal entries of $\Phi_1$ are either red or orange, which implies that signals have strong positive auto-dependence. In addition, upper off-diagonal entries in column 9 to 12 of $\Phi_1$ are mostly blue, which could be evidence that signal of tetrode T7, T8, T9 and T2 at time $t-1$ has significant negative dependence with signals from other tetrodes location at time $t$. Compared to $\Phi_1$, more than half entries in $\Phi_2$ are blank, suggesting there is no auto- and cross-dependence between those tetrodes at time $t-2$ and at time $t$. Column 1 to 8 in $\Phi_2$ are light blue, which implies weak negative dependence between signals from tetrode T20, T19,..., T15 at time $t-2$ and signals from these tetrodes location at time $t$. Also, we believe there is positive dependence between signals from tetrode T7, T9, T2 at time $t-2$ and signals from tetrode T20, T19, ..., T13 at time $t$ as the color of column 9, 11 and 12 in $\Phi_2$ is orange. However, most entries of $\Phi_3$ are blank and limited non-zero estimates are close to zero, which implies that the dependence between LFPs at time $t-3$ and time $t$ is very weak.

\begin{figure}[H]\centering 
\begin{tabular}{ccc}
\includegraphics[width=0.33\textwidth]{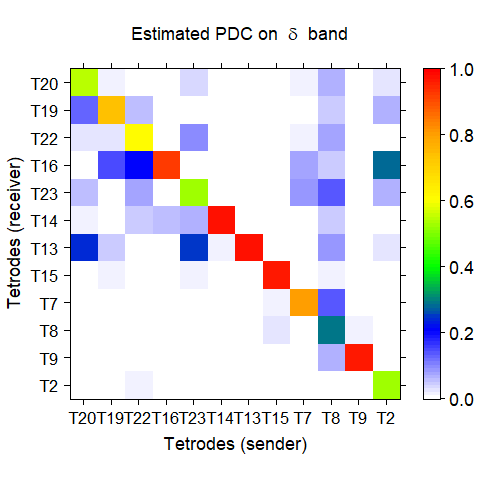} &
\includegraphics[width=0.33\textwidth]{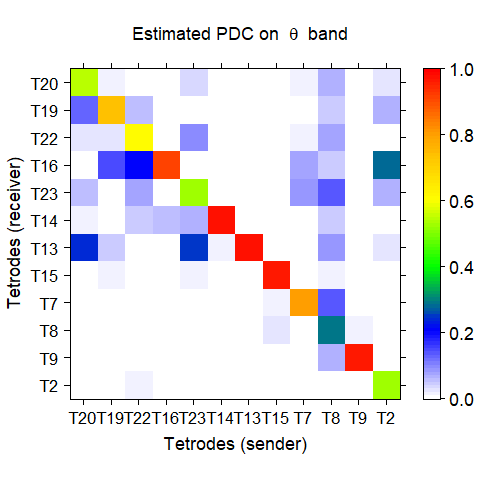} &
\includegraphics[width=0.33\textwidth]{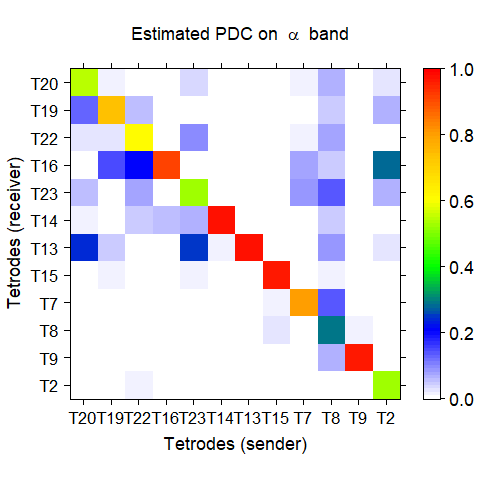}\\
\includegraphics[width=0.33\textwidth]{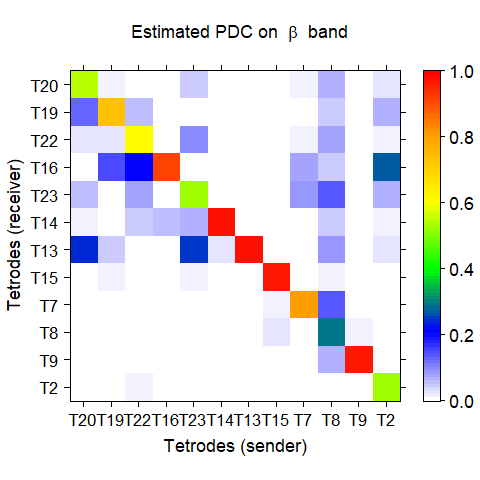} &
\includegraphics[width=0.33\textwidth]{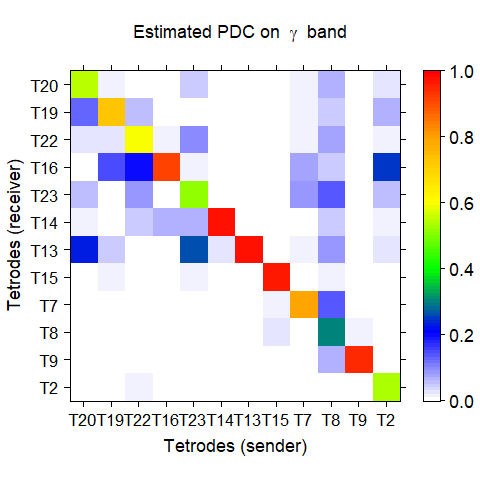}\\
\end{tabular}
\caption{Estimated PDC by LASSLE of Epoch 10}
\label{pdc_epoch1}
\end{figure}

Next we applied Equation~(\ref{fourier}) and (\ref{pdc}) to the LASSLE estimates to calculate partial directed coherence. PDC was computed at the following frequency bands in the study: $\delta$ band (0-4 Hertz), $\theta$ band (4-8 Hertz), $\alpha$ band (8-12 Hertz), $\beta$ band (12-32 Hertz) and $\gamma$ band (32-50 Hertz), which are standard in brain signals analysis. To estimate PDC at specific frequency band, we calculate the average of estimates of PDC over all singleton frequencies in that band. Figure~\ref{pdc_epoch1} demonstrates the estimated PDC results of these frequency bands in Epoch 10. Since there is only slight change on the estimated PDC across different frequency bands, we use the results of the $\gamma$ band (shown in Table~\ref{pdc_value_epoch10}) as representative to explain the PDC. For tetrodes T16, T14, T13, T15, T7 and T9, over 75\% of their information can be explained by their own past while most of their information flowing to other tetrodes are very close to 0. More specifically, tetrode T14 has 2.4\% information that flows to tetrode T13, and 6.1\% information of tetrode T16 flows to tetrode T14. This implies that they tend to have communication with specific tetrodes instead of the entirety. Unlike these tetrodes, tetrodes T20, T19, T22, T23, T8 and T2 have significant amount of information flowing to other tetrodes. For example, the proportion of current tetrode T8 that is explained by its own past is only about 30.0\%. This could be evidence that these tetrodes play an important role of passing information to other tetrodes while the rat was engaged in a non-spatial memory task. Estimated PDCs from tetrodes T20, T19, T22, T16, T23, T14, T13, T15 (sender) to tetrodes T7, T8, T9, T2 (receiver) are almost none (the blank on the bottom left of PDC), which suggest that previous oscillatory activity at the $\gamma$ band of first 8 tetrodes can hardly explain future oscillatory activity at the $\gamma$ band of last 4 tetrodes as they are far apart in spatial distance. 

\begin{table}[H]\centering
\begin{tabular}{|c|c|c|c|c|c|c|c|c|c|c|c|c|}\hline\hline
\multirow{2}{*}{Tetrode (reciever)} & \multicolumn{12}{c|}{Tetrode (sender)} \\
	\cline{2-13}
 & T20 & T19 & T22 & T16 & T23 & T14 & T13 & T15 & T7 & T8 & T9 & T2 \\
\hline
T20 &  0.548 & 0.019 & 0.005 & 0.000 & 0.041 & 0.000 & 0.002 & 0.000 & 0.018 & 0.060 & 0.004 & 0.021\\
\hline
T19 &  0.126 & 0.734 & 0.054 & 0.005 & 0.003 & 0.000 & 0.001 & 0.000 & 0.011 & 0.048 & 0.002 & 0.062\\
\hline
T22 &  0.026 & 0.026 & 0.593 & 0.010 & 0.091 & 0.000 & 0.001 & 0.003 & 0.017 & 0.078 & 0.006 & 0.019\\
\hline
T16 &  0.000 & 0.147 & 0.200 & 0.909 & 0.010 & 0.000 & 0.000 & 0.000 & 0.073 & 0.041 & 0.009 & 0.252\\
\hline
T23 &  0.056 & 0.000 & 0.085 & 0.000 & 0.513 & 0.000 & 0.007 & 0.000 & 0.082 & 0.139 & 0.007 & 0.058\\
\hline
T14 &  0.010 & 0.007 & 0.043 & 0.061 & 0.068 & 0.971 & 0.006 & 0.000 & 0.001 & 0.042 & 0.004 & 0.019\\
\hline
T13 &  0.232 & 0.048 & 0.008 & 0.003 & 0.258 & 0.024 & 0.979 & 0.000 & 0.009 & 0.081 & 0.004 & 0.029\\
\hline
T15 &  0.000 & 0.019 & 0.000 & 0.000 & 0.016 & 0.005 & 0.000 & 0.962 & 0.002 & 0.014 & 0.004 & 0.000\\
\hline
T7  &  0.000 & 0.000 & 0.000 & 0.006 & 0.000 & 0.000 & 0.001 & 0.010 & 0.786 & 0.133 & 0.002 & 0.002\\
\hline
T8  &  0.001 & 0.000 & 0.000 & 0.006 & 0.000 & 0.000 & 0.001 & 0.025 & 0.001 & 0.300 & 0.012 & 0.002\\
\hline
T9  &  0.000 & 0.000 & 0.000 & 0.000 & 0.000 & 0.000 & 0.002 & 0.000 & 0.000 & 0.063 & 0.946 & 0.000\\
\hline
T2  &  0.000 & 0.000 & 0.012 & 0.000 & 0.000 & 0.000 & 0.000 & 0.000 & 0.000 & 0.000 & 0.000 & 0.536\\
\hline
\end{tabular}
\caption {Estimated PDC value at the $\gamma$ band in Epoch 10. The estimated PDC from tetrode T16 to tetrode T22 is 0.010. The estimated PDC from tetrode T22 to T16 is 0.200. }
\label{pdc_value_epoch10}
\end{table}

\subsection{Change of brain connectivity across epochs}
\begin{figure}[H]\centering
\includegraphics[width = 0.66\textwidth, height = 0.5\textwidth]{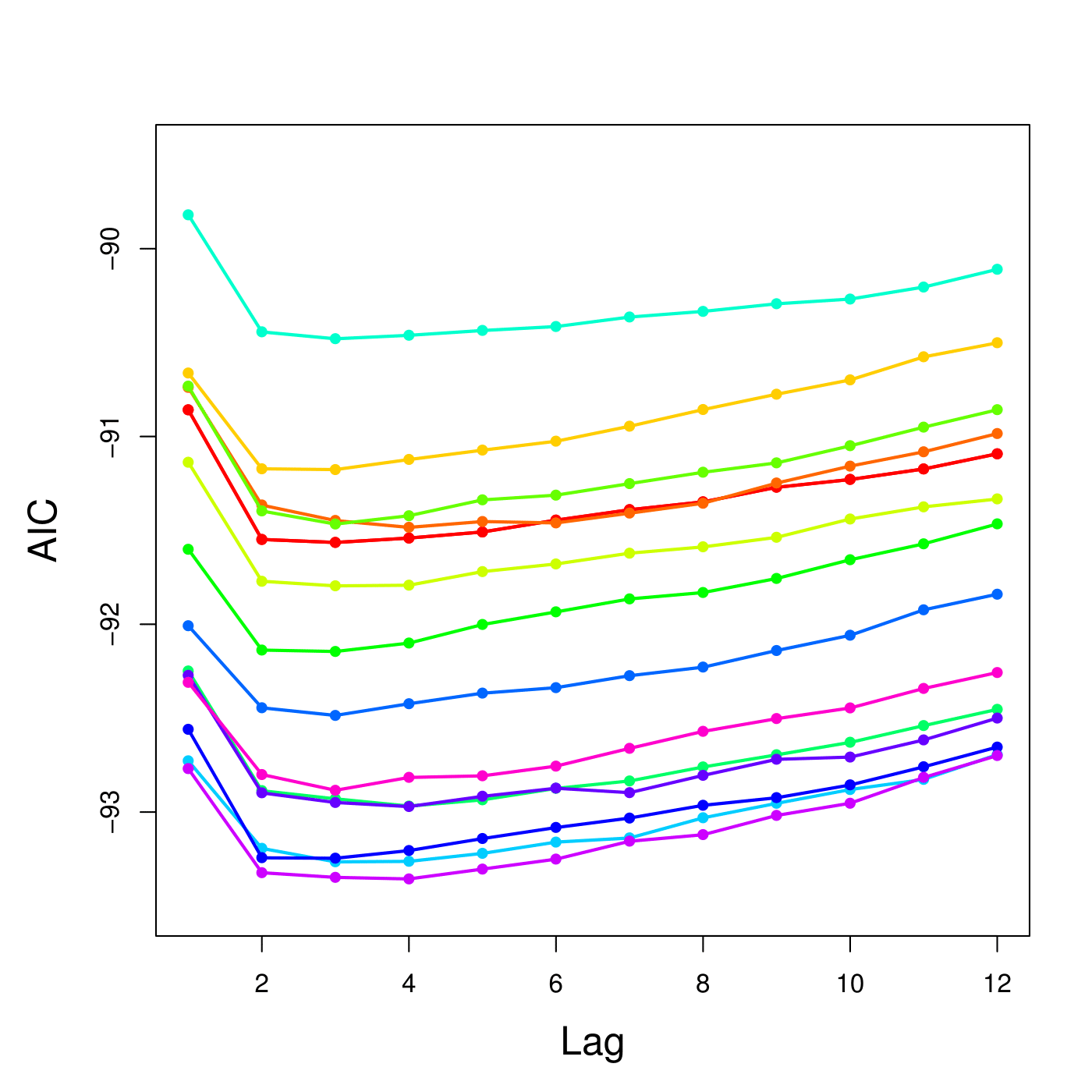}\\
\caption{AIC of fitted VAR on first 15 epochs, lag order range: 1,2,...,12.}
\label{aic_epochs}
\end{figure}

\begin{figure}[H]\centering
\begin{tabular}{cc}
\includegraphics[width=0.5\textwidth,height=0.4\textwidth]{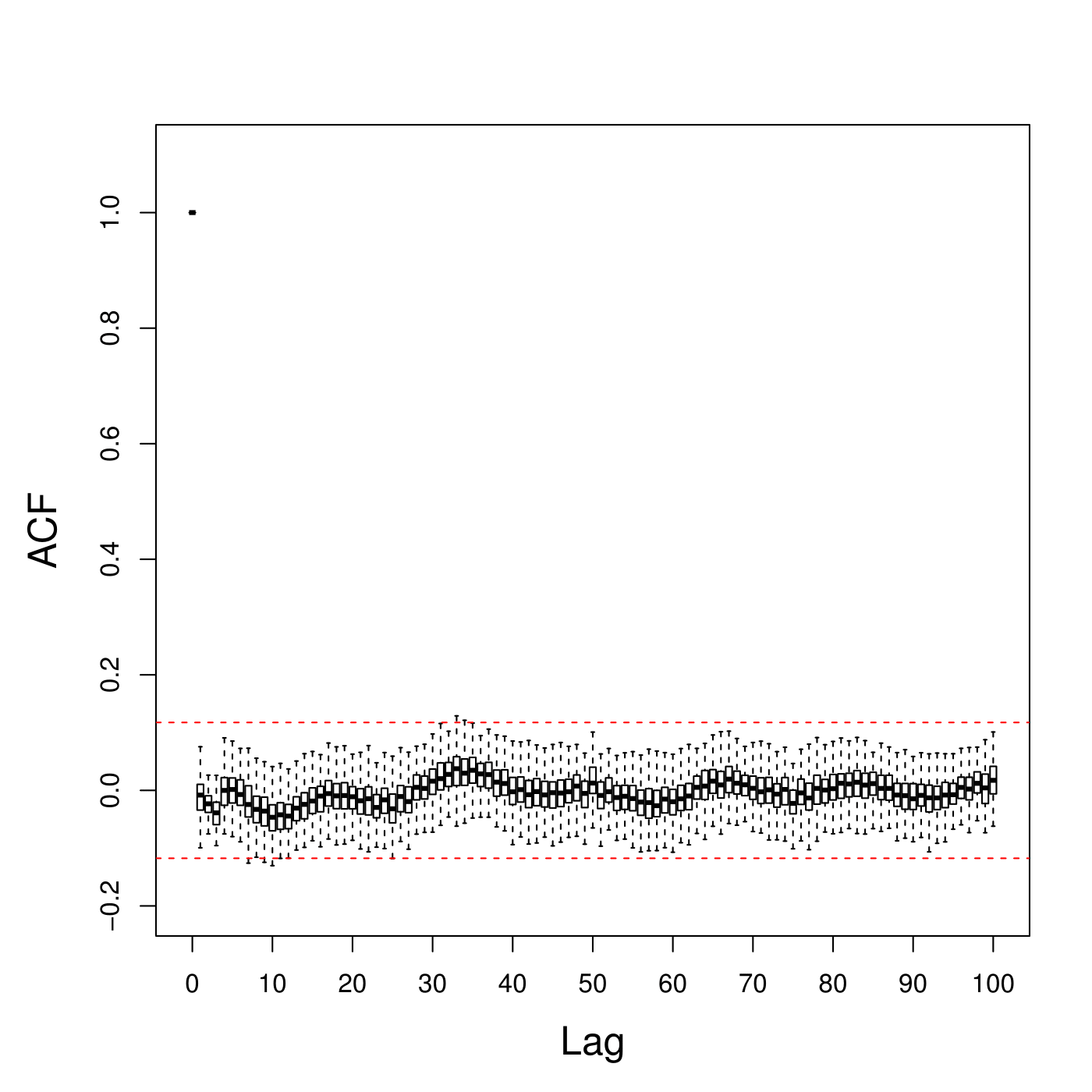} &
\includegraphics[width=0.5\textwidth,height=0.4\textwidth]{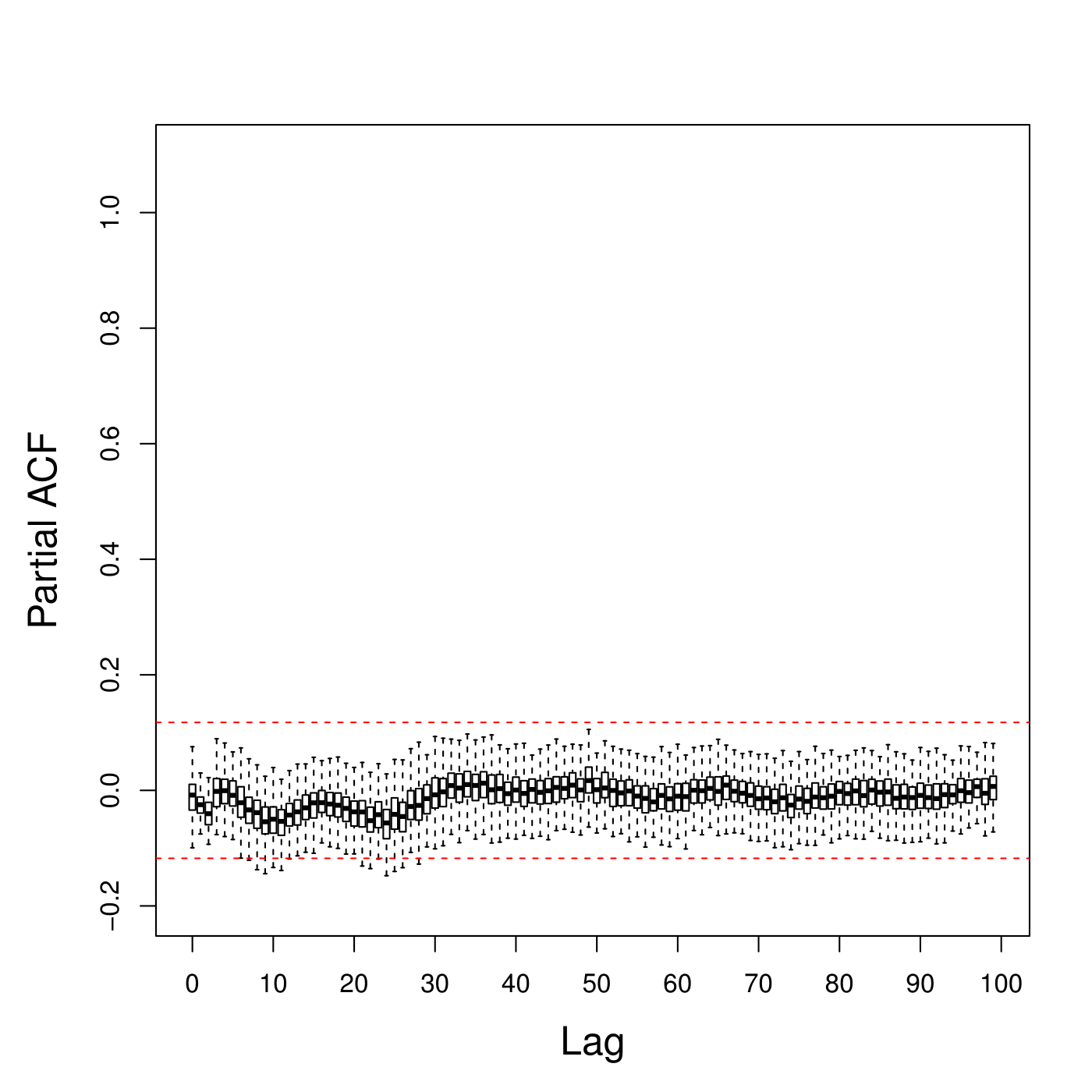}\\
\end{tabular}
\caption{The boxplots of auto-correlation function (ACF) of residuals fitted from tetrode T22 across all epochs (left). The boxplots of partial auto-correlation function (PACF) of residuals fitted from tetrode T22 across all epochs (right).}
\label{acf_pacf_residual}
\end{figure}

We repeat the same procedure for all epochs and select the best VAR order separately. Figure~\ref{aic_epochs} demonstrates the AIC curves of the first 15 epochs, from which we can see some epochs reach the lowest AIC at $\widehat{d} = 3$ and some of them are $\widehat{d} = 4$. Table~\ref{bestorder} shows the distribution of $\widehat d$ across all 247 epochs. We fit VAR($\widehat d$) to each epoch and estimate the corresponding coefficient matrices by LASSLE method. Figure \ref{acf_pacf_residual} shows the boxplots of ACF and PACF of residuals fitted from tetrode T22 across all 247 epochs, which is strong evidence that the residuals from tetrode T22 are white noise. The same phenomenon is observed for residuals fitted from other tetrodes.

\begin{table}[H]\centering
\begin{tabular}{|c|c|c|c|c|}\hline\hline
Selected lag order&  2 & 3 & 4 & 5 \\
\hline
Number of epochs & 64 & 158 & 23 & 2 \\
\hline
Proportion (\%) & 25.9 & 64.0 & 9.3 & 0.8 \\
\hline
\end{tabular}
\caption {Distribution of selected VAR lag order} 
\label{bestorder}
\end{table}

\begin{figure}[H]\centering
	\includegraphics[width = 0.9\textwidth, height = 0.6\textwidth]{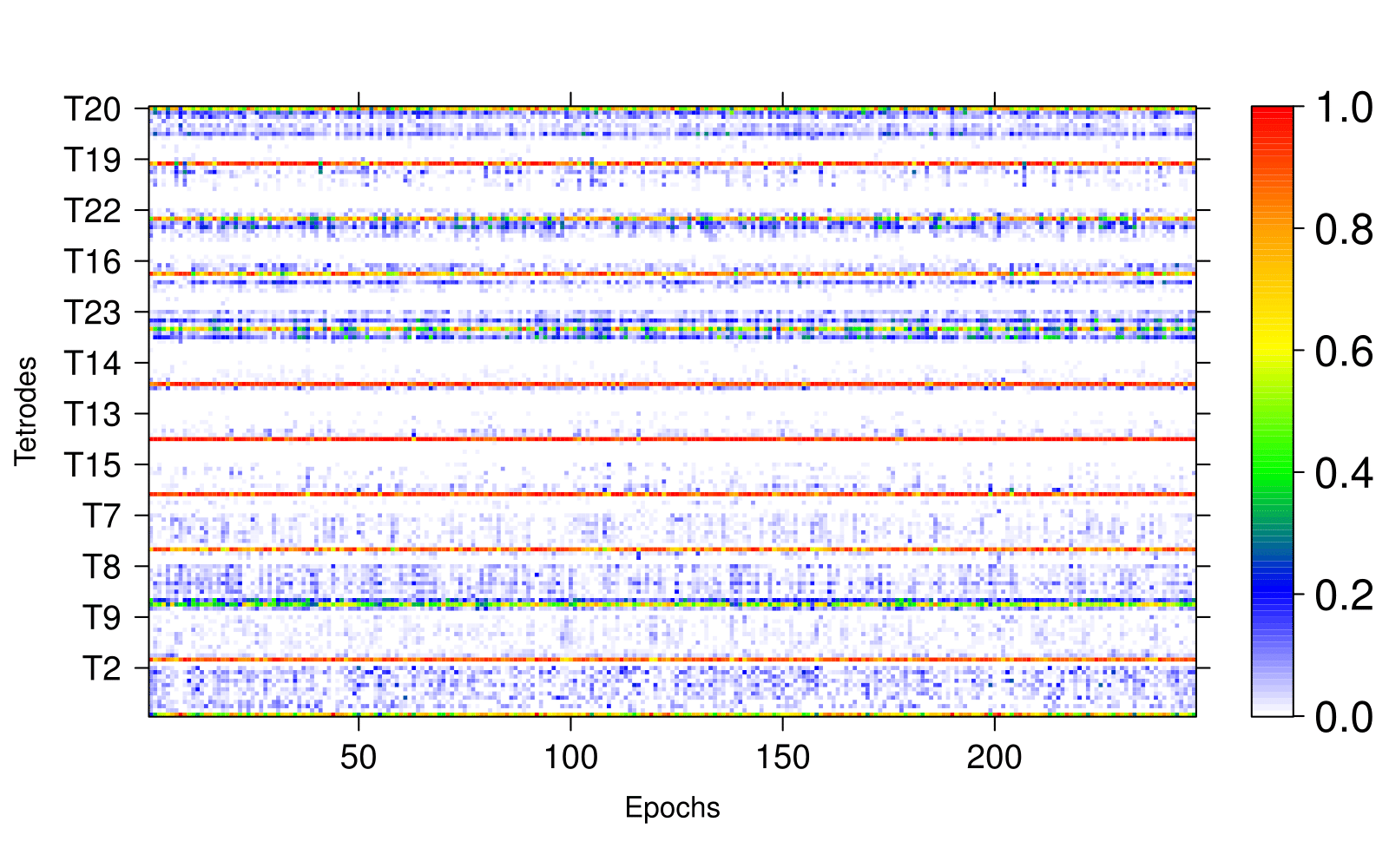}\\
	\caption{PDC on $\gamma$ band across all epochs. X-axis is the index of epochs. At each epoch, PDC matrix ($12\times12$) is converted to a column vector of $12\times12=144$ elements, where every 12 elements are the PDC values of one tetrode to all 12 tetrodes.}
	\label{pdc_gamma_allepochs}
\end{figure}

\begin{figure}[H]\centering
	\includegraphics[width = 0.9\textwidth, height = 0.5\textwidth]{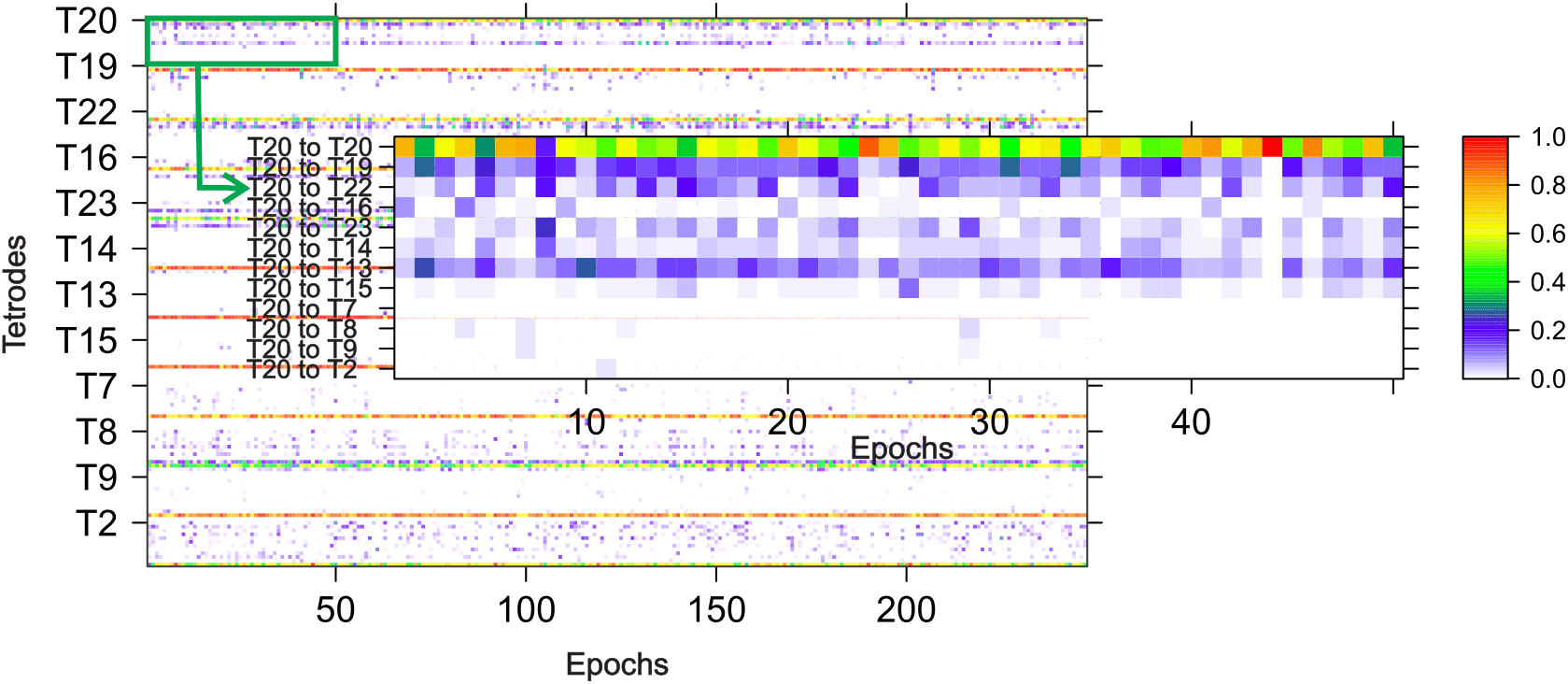}\\
	\caption{Illustration of Figure \ref{pdc_gamma_allepochs}. Every 12 rows in Figure \ref{pdc_gamma_allepochs} indicate the PDCs from one tetrode to all 12 trodes across 247 epochs. For example, the first 12 rows demonstrate the PDCs from T20 to T20, T19, ..., T2 at all epochs.}
	\label{pdc_gamma_allepochs_explain}
\end{figure}

After computing all the PDCs, we obtain the 95\% confidence interval of PDC by summarizing from the empirical distribution if we assume all epochs carry the same connectivity information. However, this assumption may not be true and we are more interested in the variation of PDCs across all epochs, which can help us understand the dynamics of rat's brain connectivity in this memory experiment. To visualize the evolution, we develop Figure~\ref{pdc_gamma_allepochs} and Figure~\ref{pdc_gamma_allepochs_explain}, where PDC matrix ($12\times12$) is converted to a column vector of $12\times12=144$ elements at each epoch and x-axis indicates the index of epochs. We can clearly see that estimated PDCs at the $\gamma$ band are quite stable on some tetrode pairs, e.g., tetrode T13 to tetrode T13 (always red color), while PDC estimates of other tetrode pairs are varying with epochs.

To compare the variation of estimated PDCs at the $\gamma$ band between InSeq epochs and OutSeq epochs, the Kolmogorov-Smirnov (KS) test is used (\cite{kolmogorov1933sulla}). The null hypothesis of KS test is that the empirical distribution of PDCs from InSeq epochs and that of OutSeq epochs are identical. 
Here we use permutation to obtain the empirical distribution of KS test statistics. Since it is necessary to preserve the inherent correlation across different epochs, the entire 247 epochs were partitioned into 50 groups where 5 consecutive epochs are within the same group (for the last group, we replicate Epoch 246 and Epoch 247 to make 5 epochs). This idea is inspired by the block bootstrap procedure for time series. Then, we randomly selected 5 groups (containing 25 epochs) from 50 groups as experimental OutSeq epochs, using the rest as experimental InSeq epochs, and compute the KS-statistic for this new Inseq and OutSeq grouping. This procedure is repeated 10,000 times to obtain the empirical distribution of KS-statistics. Finally, the proportion of permuted KS-statistics with larger values than the real KS-statistic is used as the p-value. 

\begin{figure}[H]\centering
\includegraphics[width = \textwidth, height = 0.8\textwidth]{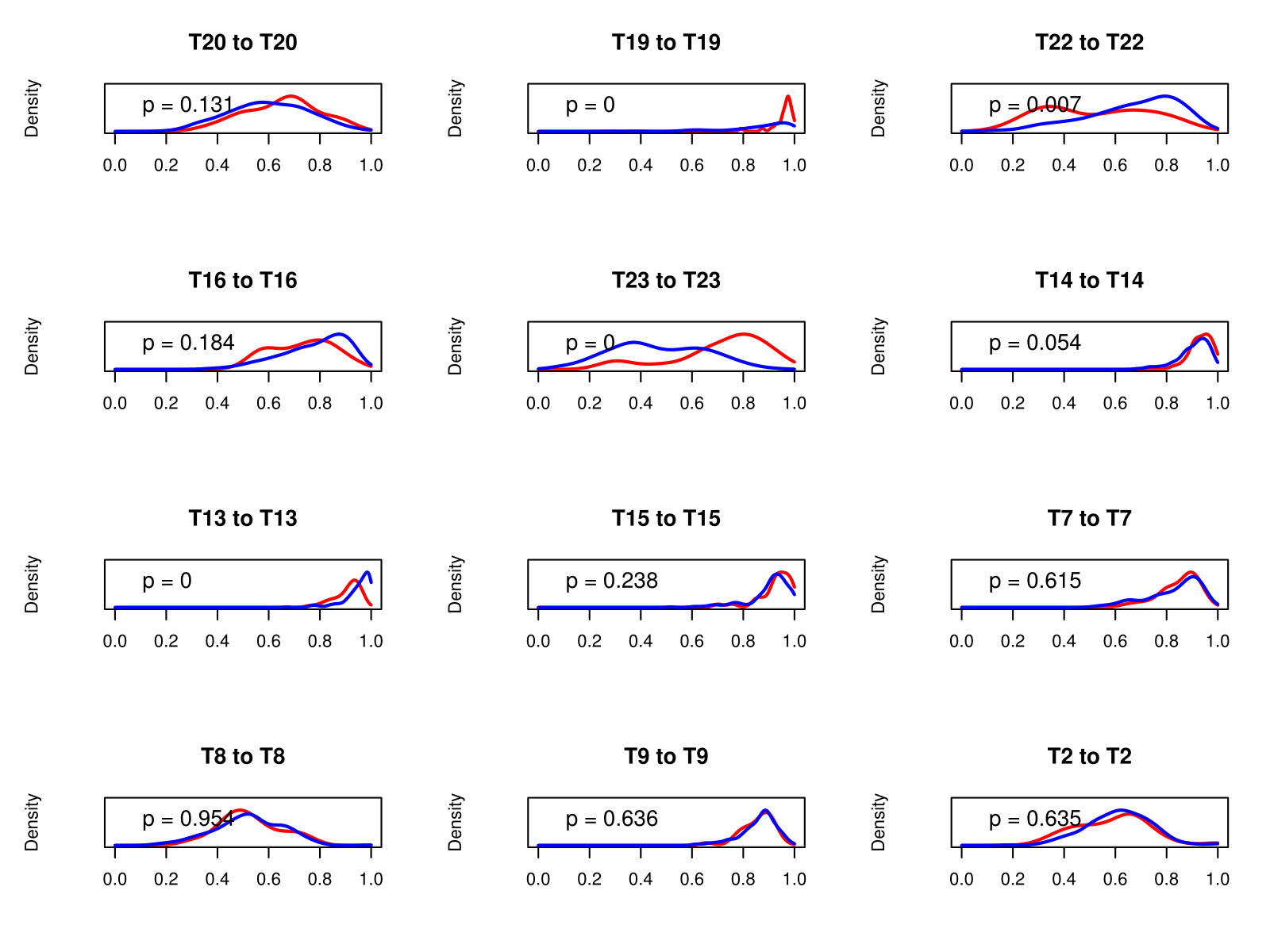}\\
\caption{Density curves of auto-PDCs across all 247 epochs. Blue one is the density curve of InSeq epochs only. Red one is the density curve of OutSeq epochs only. Kolmogorov-Smirnov test is used, where the null hypothesis is that two empirical distributions are the same. P-value is obtained from permutation and we reject the null hypothesis when $p<=0.05$.}
\label{ks_compare1}
\end{figure}

\begin{figure}[H]\centering
\includegraphics[width = \textwidth, height = 0.8\textwidth]{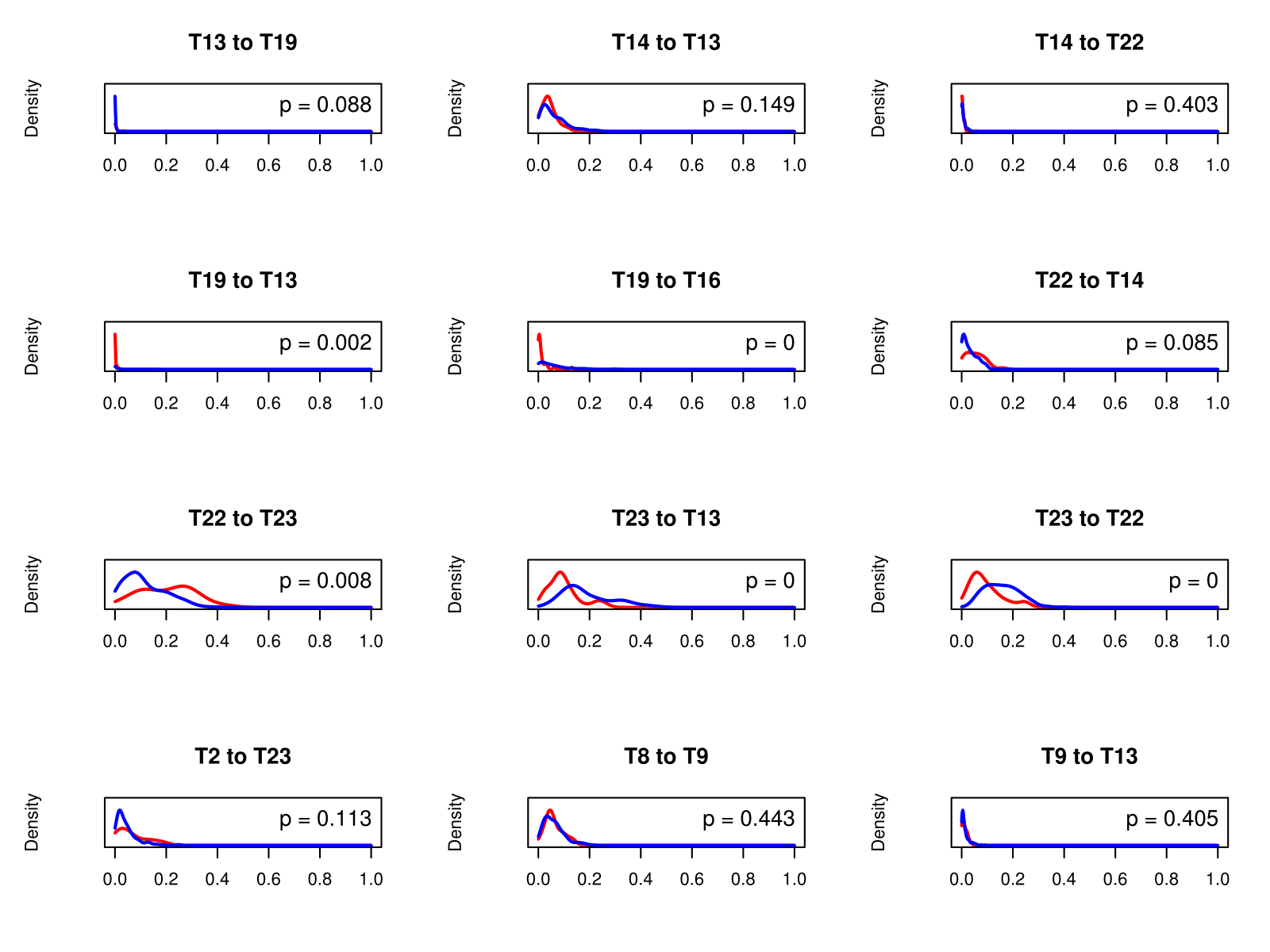}\\
\caption{Density curves of some cross-PDCs across all 247 epochs. Blue one is the density curve of InSeq epochs only. Red one is the density curve of OutSeq epochs only. Kolmogorov-Smirnov test is used, where the null hypothesis is that two empirical distributions are the same. P-value is obtained from permutation and we reject the null hypothesis when $p<=0.05$.}
\label{ks_compare2}
\end{figure}

Figure \ref{ks_compare1} and Figure \ref{ks_compare2} demonstrate the empirical distributions of estimated PDCs for all tetrodes given by Inseq epochs (blue curve) and Outseq epochs (red curve). Based on the p-values of KS test, there is strong evidence showing that the variation of auto-PDC of tetrode T19, T22, T23 and T13 are different between Inseq epochs and Outseq epochs (Figure \ref{ks_compare1}). For these tetrodes, the proportion of their current oscillatory activity that can be explained by their own past activity is influenced by whether odors are presented in the correct or incorrect sequence position (Inseq or Outseq, respectively). However, for the remaining tetrodes the variation in their estimated auto-PDC is quite stable across Inseq epochs and Outseq epochs. As shown in Figure \ref{ks_compare2}, p-values also indicate that the variation of estimated PDC from tetrode T19, T22 and T23 (sender) to some other tetrodes (receiver) are significantly different between Inseq epochs and Outseq epochs, which suggests that the information flowing from these tetrode locations to others is also influenced by the Inseq/Outseq status of the presented odor.

\subsection{Comparison of three methods on PDC across all epochs}
\begin{figure}[H]\centering
	\includegraphics[width = \textwidth, height = 0.8\textwidth]{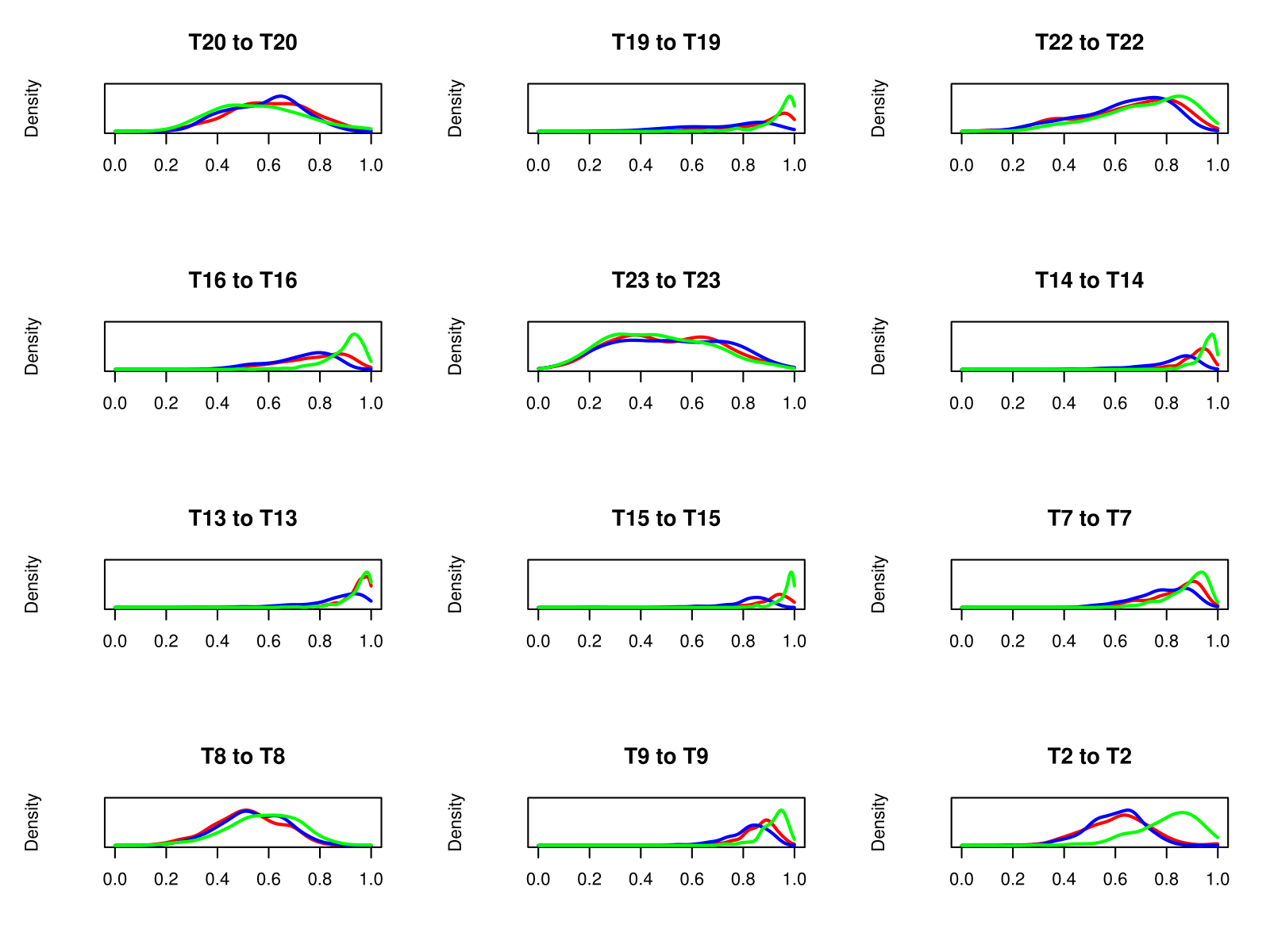}\\
	\caption{Density curves of auto-PDCs across all 247 epochs given by three methods. The red curve is LASSLE, the blue one is LSE, and the green one is LASSO.}
	\label{application_compare1}
\end{figure}

\begin{figure}[H]\centering
	\includegraphics[width = \textwidth, height = 0.8\textwidth]{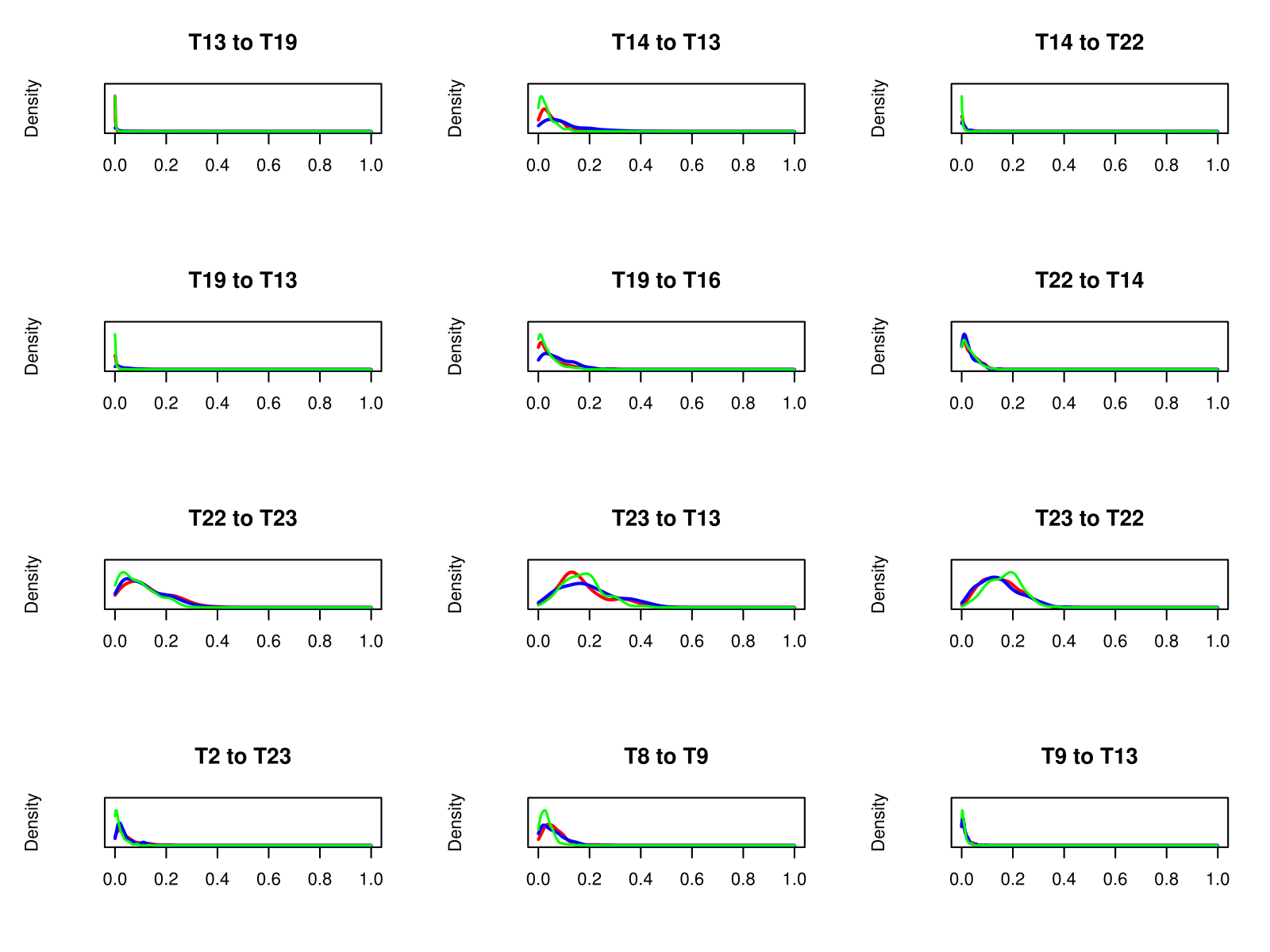}\\
	\caption{Density curves of some cross-PDCs across all 247 epochs given by three methods. The red curve is LASSLE, the blue one is LSE, and the green one is LASSO.}
	\label{application_compare2}
\end{figure}

We also apply traditional methods (LSE only and LASSO only) to estimate VAR coefficients and then compute the PDC for each epoch separately. Figure~\ref{application_compare1} and Figure~\ref{application_compare2} demonstrate the density curve of auto-PDCs and some cross-PDCs estimated by three methods across all epochs. The red curve is given by LASSLE method, the blue one is via LSE, and the green one is achieved by LASSO. As we can see, the red curve is close to the blue one for most PDCs. This is because each estimated PDC is mostly influenced by some dominant non-zero VAR coefficients, of which the estimates are close to each other by LASSLE and LSE separately. Noted that LASSO method has shrinked many VAR coefficients to zero and its non-zero estimates are very different from those by LSE. Cosequently the green curve is dissimilar to the blue one for most PDCs. 

%% file: Conclusion.tex
\section{Conclusion}
In this paper, we proposed a hybrid LASSLE (LASSO+LSE) method to estimate the coefficients of vector auto-regressive models characterizing the effective and directional connectivity for multichannel brain physiological signals. This method uses regularization to control for sparsity on the first stage and then use least squares to improve bias and mean-squared error of the estimator on the second stage. Compared to the separate LASSO and LSE, the advantage of our method is that it is able to both indicate the most important effective connectivity and give a more accurate estimate of the connectivity strength. Note that sparse VAR coefficient estimates can still capture complex dependency structures in a multivariate time series. In addition, we employ partial directed coherence to measure the directional connectivity between the channels. PDC is a directed frequency-specific measure that explains the extent to which the present oscillatory activity in a sender channel influences the future oscillatory activity in a specific receiver channel relative to all possible receivers in the network. The proposed modeling approach provided key insights into potential functional relationships among simultaneously recorded sites during performance of a complex memory task. Specifically, this novel method was successful in revealing patterns of effective connectivity across tetrode locations, by quantifying how present oscillatory activity in each tetrode is influenced by past oscillatory activity in other tetrodes. This approach was also successful in capturing how this effective connectivity varied across trial epochs and trial types (Inseq or Outseq).

%% file: Acknowledgement.tex
\section*{Acknowledgement}
N.J. Fortin's research was supported in part by the National Science Foundation (awards IOS-1150292 and BCS-1439267), the Whitehall Foundation (award 2010-05-84), and the University of California, Irvine. \\H. Ombao's work was supported in part by grants from the US NSF Division of Mathematical Sciences (DMS 15-09023) and the Division of Social and Economic Sciences (SES 14-61534).

%% file: modeling_high_dimensional_multichannel_brain_signals_revise.bbl
\begin{thebibliography}{28}
\providecommand{\natexlab}[1]{#1}
\providecommand{\url}[1]{\texttt{#1}}
\expandafter\ifx\csname urlstyle\endcsname\relax
  \providecommand{\doi}[1]{doi: #1}\else
  \providecommand{\doi}{doi: \begingroup \urlstyle{rm}\Url}\fi

\bibitem[Allen et~al.(2014)Allen, Morris, Mattfeld, Stark, and
  Fortin]{allen2014sequence}
Timothy~A Allen, Andrea~M Morris, Aaron~T Mattfeld, Craig~EL Stark, and
  Norbert~J Fortin.
\newblock A sequence of events model of episodic memory shows parallels in rats
  and humans.
\newblock \emph{Hippocampus}, 24\penalty0 (10):\penalty0 1178--1188, 2014.

\bibitem[Allen et~al.(2015)Allen, Morris, Stark, Fortin, and
  Stark]{allen2015memory}
Timothy~A Allen, Andrea~M Morris, Shauna~M Stark, Norbert~J Fortin, and
  Craig~EL Stark.
\newblock Memory for sequences of events impaired in typical aging.
\newblock \emph{Learning \& Memory}, 22\penalty0 (3):\penalty0 138--148, 2015.

\bibitem[Allen et~al.(2016)Allen, Salz, McKenzie, and
  Fortin]{allen2016nonspatial}
Timothy~A Allen, Daniel~M Salz, Sam McKenzie, and Norbert~J Fortin.
\newblock Nonspatial sequence coding in ca1 neurons.
\newblock \emph{Journal of Neuroscience}, 36\penalty0 (5):\penalty0 1547--1563,
  2016.

\bibitem[Baccal{\'a} and Sameshima(2001)]{baccala2001partial}
Luiz~A Baccal{\'a} and Koichi Sameshima.
\newblock Partial directed coherence: a new concept in neural structure
  determination.
\newblock \emph{Biological cybernetics}, 84\penalty0 (6):\penalty0 463--474,
  2001.

\bibitem[Baccal{\'a} and Sameshima(2014)]{baccala2014partial}
Luiz~A Baccal{\'a} and Koichi Sameshima.
\newblock Partial directed coherence.
\newblock In \emph{Methods in Brain Connectivity Inference through Multivariate
  Time Series Analysis}, pages 57--73. CRC Press, 2014.

\bibitem[Boucquey et~al.(2015, submitted)Boucquey, Allen, Huffman, Fortin, and
  Stark]{boucquey2015cross}
Veronique~K Boucquey, Timothy~A Allen, Derek~J Huffman, Norbert~J Fortin, and
  Craig~EL Stark.
\newblock A cross-species sequence memory task reveals hippocampal and medial
  prefrontal cortex activity and interactions in humans.
\newblock \emph{Hippocampus}, 2015, submitted.

\bibitem[Fiecas and Ombao(2011)]{fiecas2011generalized}
Mark Fiecas and Hernando Ombao.
\newblock The generalized shrinkage estimator for the analysis of functional
  connectivity of brain signals.
\newblock \emph{The Annals of Applied Statistics}, pages 1102--1125, 2011.

\bibitem[Fiecas et~al.(2010)Fiecas, Ombao, Linkletter, Thompson, and
  Sanes]{fiecas2010functional}
Mark Fiecas, Hernando Ombao, Crystal Linkletter, Wesley Thompson, and Jerome
  Sanes.
\newblock Functional connectivity: Shrinkage estimation and randomization test.
\newblock \emph{NeuroImage}, 49\penalty0 (4):\penalty0 3005--3014, 2010.

\bibitem[Fortin et~al.(2016)Fortin, Asem, Ng, Quirk, Allen, and
  Elias]{fortin2016distinct}
NJ~Fortin, JSA Asem, CW~Ng, CR~Quirk, TA~Allen, and GA~Elias.
\newblock Distinct contributions of hippocampal, prefrontal, perirhinal and
  nucleus reuniens regions to the memory for sequences of events.
\newblock \emph{Society for Neuroscience Abstracts (San Diego, CA)}, 2016.

\bibitem[Friedman et~al.(2010)Friedman, Hastie, and
  Tibshirani]{friedman2010regularization}
Jerome Friedman, Trevor Hastie, and Rob Tibshirani.
\newblock Regularization paths for generalized linear models via coordinate
  descent.
\newblock \emph{Journal of statistical software}, 33\penalty0 (1):\penalty0 1,
  2010.

\bibitem[Fu(1998)]{fu1998penalized}
Wenjiang~J Fu.
\newblock Penalized regressions: the bridge versus the lasso.
\newblock \emph{Journal of computational and graphical statistics}, 7\penalty0
  (3):\penalty0 397--416, 1998.

\bibitem[Han and Liu(2013)]{han2013direct}
Fang Han and Han Liu.
\newblock A direct estimation of high dimensional stationary vector
  autoregressions.
\newblock \emph{arXiv preprint arXiv:1307.0293}, 2013.

\bibitem[Hesterberg et~al.(2008)Hesterberg, Choi, Meier, Fraley,
  et~al.]{hesterberg2008least}
Tim Hesterberg, Nam~Hee Choi, Lukas Meier, Chris Fraley, et~al.
\newblock Least angle and l1 penalized regression: A review.
\newblock \emph{Statistics Surveys}, 2:\penalty0 61--93, 2008.

\bibitem[Ivanov et~al.(2005)Ivanov, Kilian, et~al.]{ivanov2005practitioner}
Ventzislav Ivanov, Lutz Kilian, et~al.
\newblock A practitioner¡¯s guide to lag order selection for var impulse
  response analysis.
\newblock \emph{Studies in Nonlinear Dynamics and Econometrics}, 9\penalty0
  (1):\penalty0 1--34, 2005.

\bibitem[Kami{\'n}ski et~al.(2001)Kami{\'n}ski, Ding, Truccolo, and
  Bressler]{kaminski2001evaluating}
Maciej Kami{\'n}ski, Mingzhou Ding, Wilson~A Truccolo, and Steven~L Bressler.
\newblock Evaluating causal relations in neural systems: Granger causality,
  directed transfer function and statistical assessment of significance.
\newblock \emph{Biological cybernetics}, 85\penalty0 (2):\penalty0 145--157,
  2001.

\bibitem[Kirch(2007)]{kirch2007resampling}
Claudia Kirch.
\newblock Resampling in the frequency domain of time series to determine
  critical values for change-point tests.
\newblock \emph{Statistics \& Decisions}, 25\penalty0 (3/2007):\penalty0
  237--261, 2007.

\bibitem[Kolmogorov(1933)]{kolmogorov1933sulla}
AN~Kolmogorov.
\newblock Sulla determinazione empirica delle leggi di probabilita.
\newblock \emph{Giorn. Ist. Ital. Attuari}, 4:\penalty0 1--11, 1933.

\bibitem[Kreiss(1992)]{kreiss1992bootstrap}
Jens-Peter Kreiss.
\newblock Bootstrap procedures for ar (co)-processes.
\newblock In \emph{Bootstrapping and related techniques: proceedings of an
  international conference held in Trier, FRG, June 4-8, 1990}, volume 376,
  page 107. Springer-Verlag, 1992.

\bibitem[Liu et~al.(2013)Liu, Yu, et~al.]{liu2013asymptotic}
Hanzhong Liu, Bin Yu, et~al.
\newblock Asymptotic properties of lasso+ mls and lasso+ ridge in sparse
  high-dimensional linear regression.
\newblock \emph{Electronic Journal of Statistics}, 7:\penalty0 3124--3169,
  2013.

\bibitem[Mairal and Yu(2012)]{mairal2012complexity}
Julien Mairal and Bin Yu.
\newblock Complexity analysis of the lasso regularization path.
\newblock \emph{arXiv preprint arXiv:1205.0079}, 2012.

\bibitem[Ombao and Van~Bellegem(2008)]{ombao2008evolutionary}
Hernando Ombao and S{\'e}bastien Van~Bellegem.
\newblock Evolutionary coherence of nonstationary signals.
\newblock \emph{IEEE Transactions on Signal Processing}, 56\penalty0
  (6):\penalty0 2259--2266, 2008.

\bibitem[Paparoditis(2002)]{paparoditis2002frequency}
Efstathios Paparoditis.
\newblock Frequency domain bootstrap for time series.
\newblock In \emph{Empirical process techniques for dependent data}, pages
  365--381. Springer, 2002.

\bibitem[Politis et~al.(2003)]{politis2003impact}
Dimitris~N Politis et~al.
\newblock The impact of bootstrap methods on time series analysis.
\newblock \emph{Statistical Science}, 18\penalty0 (2):\penalty0 219--230, 2003.

\bibitem[Shao(2010)]{shao2010dependent}
Xiaofeng Shao.
\newblock The dependent wild bootstrap.
\newblock \emph{Journal of the American Statistical Association}, 105\penalty0
  (489):\penalty0 218--235, 2010.

\bibitem[Shumway and Stoffer(2006)]{shumway2006time}
Robert~H Shumway and David~S Stoffer.
\newblock \emph{Time series analysis and its applications: with R examples}.
\newblock Springer Science \& Business Media, 2006.

\bibitem[Tibshirani(1996)]{tibshirani1996regression}
Robert Tibshirani.
\newblock Regression shrinkage and selection via the lasso.
\newblock \emph{Journal of the Royal Statistical Society. Series B
  (Methodological)}, pages 267--288, 1996.

\bibitem[Zhao and Yu(2006)]{zhao2006model}
Peng Zhao and Bin Yu.
\newblock On model selection consistency of lasso.
\newblock \emph{Journal of Machine learning research}, 7\penalty0
  (Nov):\penalty0 2541--2563, 2006.

\bibitem[Zhao et~al.(2009)Zhao, Rocha, and Yu]{zhao2009composite}
Peng Zhao, Guilherme Rocha, and Bin Yu.
\newblock The composite absolute penalties family for grouped and hierarchical
  variable selection.
\newblock \emph{The Annals of Statistics}, pages 3468--3497, 2009.

\end{thebibliography}
